\documentclass[12pt]{article}

\usepackage[utf8]{inputenc}
\usepackage{cancel}
\usepackage{enumitem}
\usepackage{amssymb}
\usepackage{soul}
\usepackage{tabularx}
\usepackage{xcolor}
\usepackage{booktabs}
\usepackage{subcaption} 
\usepackage{colortbl}
\usepackage{authblk}
\DeclareUnicodeCharacter{00A0}{ }

\newcolumntype{C}{>{\centering\arraybackslash}X}
\usepackage[T1]{fontenc}
\usepackage{epsfig}
\usepackage{latexsym}
\usepackage{graphicx}
\usepackage{amsmath}
\usepackage{amsfonts}   
\usepackage{amssymb}    
\usepackage{float}
\usepackage{bm}
\usepackage{url}
\usepackage{hyperref} 
\usepackage[nodisplayskipstretch]{setspace}
\def\lsim{\raise0.3ex\hbox{$\;<$\kern-0.75em\raise-1.1ex\hbox{$\sim\;$}}}
\def\gsim{\raise0.3ex\hbox{$\;>$\kern-0.75em\raise-1.1ex\hbox{$\sim\;$}}}

\def    \beq            {\begin{equation}}
\def    \eeq            {\end{equation}}
\def    \bea           {\begin{eqnarray}}
\def    \eea           {\end{eqnarray}}

\def \mn{\mu\nu{\rm SSM}}

\def\g2{{\rm GeV}^2}

\def\sw2{sin^2 \theta_w}

\def\a^tau{\alpha_{\tau}}

\def\beq{\begin{equation}}
\def\eeq{\end{equation}}
\def\beqa{\begin{eqnarray}}
\def\eeqa{\end{eqnarray}}

\newcommand{\newc}{\newcommand}
\newc\BR{BR}
\newc{\akappa}{A_{\kappa} }
\newc\deltagmtwo{\delta (g-2)_{\mu}} 
\newc\deltaamu{\Delta a_{\mu}}

\def\anti{\overline}

\def\la{\lambda}

\newc{\haa}{BR\(h_1\to a_1 a_1\)}
\newc{\abb}{BR\(a_1\to b\anti{b}\)}
\newc{\hbb}{BR\(h_1\to b\anti{b}\)}
\newc{\abund}{\Omega h^2}
\newc\bsgamma{b\rightarrow s \gamma }
\newc\bxsgamma{\overline{B}\rightarrow X_{s}\gamma}
\newc\brbsgamma{\BR(\overline{B}\rightarrow X_s\gamma)}

\allowdisplaybreaks

\usepackage{array, makecell}
\usepackage{boldline}
\usepackage[nosort]{cite}

\title{\bf{
$U(1)'$ extensions of the $\mn$
}}

\author[a]{J.~A.~Aguilar--Saavedra\thanks{jaas@ugr.es}}
\author[b]{I.~Lara\thanks{inaki.lara@fuw.edu.pl}}
\author[c,d]{D.~E.~L\'opez-Fogliani\thanks{daniel.lopez@df.uba.ar}}
\author[e,f]{C.~Mu\~noz\thanks{c.munoz@uam.es}} 

  \affil[a]{Departamento de F\'{\i}sica Te\'orica y del Cosmos,  Universidad de Granada,  E-18071 Granada, Spain}
    \affil[b] {Faculty of Physics, University of Warsaw, Pasteura 5, 02-093 Warsaw, Poland}
  \affil[c]{Instituto de F\'isica de Buenos Aires UBA \& CONICET, Departamento de F\'isica, Facultad de Ciencia Exactas y Naturales, Universidad de Buenos Aires, 1428 Buenos Aires, Argentina}
  \affil[d]{Pontificia Universidad Cat\'olica Argentina, 1107 Buenos Aires, Argentina}
    \affil[e]{Departamento de F\'{\i}sica Te\'{o}rica, Universidad Aut\'{o}noma de Madrid (UAM), Campus de Cantoblanco, 28049 Madrid, Spain}
  \affil[f]{Instituto de F\'{\i}sica Te\'{o}rica (IFT) UAM-CSIC,  Campus de Cantoblanco, 28049 Madrid, Spain}

\date{}

\begin{document}

\maketitle

\begin{abstract}
In the $\mu\nu$SSM, the presence of 
$R$-parity violating couplings involving right-handed (RH) neutrinos solves simultaneously the 
$\mu$- and $\nu$-problems.
We explore extensions of the $\mu\nu$SSM adding a $U(1)'$ gauge group, which provides the RH neutrinos with a non-vanishing charge.
In these models, dubbed U$\mu\nu$SSM, the anomaly cancellation conditions impose 
the presence of exotic quarks in the spectrum that are vector-like under the standard model (SM) gauge group:
either three pairs $SU(2)$ quark singlets, or a pair of quark singlets together with a pair of quark doublets.
Several singlets under the SM group can also be present, with the $U(1)'$ charges making distinctions among them, and therefore allowing different types of couplings. Some of these singlets dynamically generate Majorana masses for the RH neutrinos, and others can be candidates for dark matter. 
The useful characteristics of models with $U(1)'$s are also present in U$\mu\nu$SSM models: baryon-number-violating operators as well as explicit Majorana masses and $\mu$ terms are forbidden, and the domain wall problem is avoided. 
The phenomenology of U$\mu\nu$SSM models is very rich.
We analyze the experimental constraints on their
parameter space, specially on the mass and mixing of the new $Z'$ boson. In addition to the exotic quarks, which can hadronize inside the detector or decay producing SM particles,
the U$\mu\nu$SSM models can also have new signals such as decays of the $Z'$ to sparticle pairs like right sneutrinos, charginos or neutralinos. Besides, $Z'$ and Higgs mediated annihilations and interactions with the visible sector of WIMP dark matter particles, can also be present.
\end{abstract}  

Keywords: Supersymmetry Phenomenology; Supersymmetric Standard Model; $Z'$ phenomenology

\clearpage 

\tableofcontents 

\section{Introduction}
\label{introduction}

The addition of right-handed (RH) neutrinos $\nu_{R}$ to the spectrum of the standard model (SM) provides the light neutrinos with masses.
In the framework of supersymmetry (SUSY), 
the $SU(3)\times SU(2)\times U(1)_Y$ gauge invariant superpotential containing the most general renormalizable couplings involving the three families of SM superfields and an arbitrary number of RH neutrino superfields,
$\hat\nu_{i'}^c$, is given by:
\bea
\label{Eq:superpotential10}
W &=&  
Y^e_{ij} \, \hat H_d\, \hat L_i \, \hat e_j^c
\ +
Y^d_{ij} \, \hat H_d\, \hat Q_i \, \hat d_j^c 
\ -
Y^u_{ij} \, \hat {H_u}\, \hat Q_i \, \hat u_j^c
\ +
\mu \, \hat H_u \, \hat H_d
\nonumber\\
&+&
\lambda_{ijk} \, \hat L_i  \, \hat L_j \, \hat e_k^c
+
\lambda'_{ijk} \, \hat L_i \, \hat Q_j \, \hat d_k^c
\ +
\lambda''_{ijk} \, \hat u_i^c \, \hat d_j^c \, \hat d_k^c
\ + \mu_i \, \hat H_u \, \hat L_i
\nonumber\\
&-&
Y^\nu_{ij'} \, \hat {H_u}\, \hat L_i \, \hat \nu^c_{j'}
\ + \lambda_{i'} \, \hat {H_u} \, \hat H_d \, \hat \nu^c_{i'}
\ +
\frac{1}{3}
\kappa{_{i'j'k'}} \, \hat \nu^c_{i'}\, \hat \nu^c_{j'}\, \hat \nu^c_{k'}
\ +
{\mathcal M}_{i'j'} \, \hat \nu^c_{i'}\, \hat \nu^c_{j'}
\ + t_{i'} \hat \nu^c_{i'}
\,,
\eea
where the summation convention is implied on repeated indexes, with 
$i,j,k=1,2,3$ the usual family indexes of the SM, and
$i'= 1,....n_{\nu^c}$, with $n_{\nu^c}$ the number of RH neutrinos. 
Our convention for the contraction of two $SU(2)$ doublets is e.g.
$\hat {H}_u \,  \hat H_d\equiv  \epsilon_{ab} \hat H^a_u \, \hat H^b_d$,
$a,b=1,2$ and $\epsilon_{ab}$ the totally antisymmetric tensor with $\epsilon_{12}=1$.

The four terms in the first line of Eq.~(\ref{Eq:superpotential10}) determine the superpotential of the minimal supersymmetric standard 
model (MSSM) (for reviews, see e.g. Refs.~\cite{Nilles:1983ge,Haber:1984rc,Martin:1997ns}).
The four terms in the second line are the conventional trilinear and bilinear
$R$-parity violating (RPV) couplings (for a review, see e.g. Ref.~\cite{Barbier:2004ez}). Finally, all the terms in the last line contain RH neutrinos, and are allowed by gauge invariance since these fields have vanishing hypercharges by construction, $y(\nu^c)=0$.
Of these terms, the first three are characteristic 
of the `$\mu$ from $\nu$' supersymmetric standard 
model ($\mu\nu$SSM)~\cite{LopezFogliani:2005yw} (for a recent review, see Ref.~\cite{Lopez-Fogliani:2020gzo}), the fourth term gives Majorana masses for RH neutrinos, and the fifth is a tadpole term for them with $t_{i'}$ of dimension mass squared.

The $\mn$ is a natural construction where
trilinear couplings involving RH neutrino superfields
are 
added to the usual quark and charged-lepton Yukawa couplings, solving dynamically
crucial theoretical problems of SUSY. It contains in the superpotential the following subset of terms of Eq.~(\ref{Eq:superpotential10})~\cite{LopezFogliani:2005yw,Escudero:2008jg,Ghosh:2017yeh}:
\bea
\label{Eq:superpotentialmunu}
W_{\text{$\mu\nu$SSM}} &=&  
Y^e_{ij} \, \hat H_d\, \hat L_i \, \hat e_j^c
\ +
Y^d_{ij} \, \hat H_d\, \hat Q_i \, \hat d_j^c 
\ -
Y^u_{ij} \, \hat {H_u}\, \hat Q_i \, \hat u_j^c
\nonumber\\
&+&
\lambda_{ijk} \, \hat L_i  \, \hat L_j \, \hat e_k^c
+
\lambda'_{ijk} \, \hat L_i \, \hat Q_j \, \hat d_k^c
\nonumber\\
&-&
Y^\nu_{ij'} \, \hat {H_u}\, \hat L_i \, \hat \nu^c_{j'}
\ +
\lambda_{i'} \, \hat {H_u} \, \hat H_d \, \hat \nu^c_{i'}
\ +
\frac{1}{3}
\kappa{_{i'j'k'}} \, \hat \nu^c_{i'}\, \hat \nu^c_{j'}\, \hat \nu^c_{k'}
\,,
\eea
In particular, the first term of the last line provides neutrino Yukawa couplings. It also effectively generates bilinear terms as the ones of the bilinear 
RPV model~\cite{Barbier:2004ez},
when
the right sneutrinos  
develop electroweak-scale vacuum expectation values (VEVs) after the
electroweak symmetry breaking (EWSB).
By the same mechanism, the RPV couplings 
$\lambda_{i'} \hat\nu_{i'}^c\hat H_u \hat H_d$ effectively
generate a $\mu$-term solving the so-called
$\mu$-problem~\cite{Kim:1983dt} (for a recent review, see Ref.~\cite{Bae:2019dgg}).
Besides, the RPV couplings 
$\kappa_{i'j'k'} \hat\nu_{i'}^c\hat\nu_{j'}^c\hat \nu_{k'}^c$,
effectively generate electroweak-scale Majorana masses 
instrumental 
in solving the $\nu$ problem through a EW-scale seesaw, i.e. the generation of correct neutrino masses 
and mixing angles in SUSY~\cite{LopezFogliani:2005yw,Escudero:2008jg,Ghosh:2008yh,Bartl:2009an,Fidalgo:2009dm,Ghosh:2010zi,Liebler:2011tp} with $Y^{\nu}\lsim 10^{-6}$.
In fact, the $\mn$ seesaw is a {\it generalized} electroweak seesaw, because it involves not only left-handed (LH) neutrinos $\nu_{L}$ with RH ones 
$\nu_{R}$ as in the usual seesaw, but also the neutralinos.
This fact favors the accommodation of neutrino {data~\cite{Capozzi:2017ipn,deSalas:2017kay,deSalas:2018bym,Esteban:2018azc}}, and occurs because of the $R$-parity violation
present in the $\mn$.
The presence of couplings of the type $\kappa_{i'j'k'}$ also forbids a global $U(1)$ symmetry in the superpotential, avoiding therefore the existence of a Goldstone boson.
Finally, note that the lepton-number-violating couplings 
$\lambda_{ijk} \hat L_i\hat L_j\hat e_k^c$ and 
$\lambda'_{ijk}\hat L_i \hat Q_j \hat d_k^c$ 
are naturally present in the superpotential once we impose the presence of the Yukawa couplings 
$Y^e_{ij} \hat H_d\hat L_i\hat e_j^c$ and 
$Y^d_{ij} \hat H_d\hat Q_i\hat d_j^c$,
since $\hat H_d$ and $\hat L_i$ have the same SM quantum numbers.

Studies of 
interesting signals of 
the $\mn$ at the Large Hadron Collider (LHC) 
were carried out in the literature, and can be found summarized in Ref.~\cite{Lopez-Fogliani:2020gzo}.
They were
focused on the three characteristic terms of the model, written in the last line
of $W_{\mn}$ above.
In addition to the enlarged Higgs
sector with sneutrinos, the intimate
connection between the lightest supersymmetric particle (LSP)
lifetime and the size of neutrino Yukawa couplings was also analyzed. Displaced vertices
and/or multileptons are some of the interesting signatures that can be
probed.

Despite these attractive signals and interesting properties of the $\mn$, there are interesting arguments from the theoretical viewpoint to try to extend the model.
First, we would like to have an explanation for the absence of the  baryon-number-violating couplings 
$\lambda''_{ijk} \hat u_i^c\hat d_j^c\hat d_k^c$, which together with the lepton-number-violating couplings would give rise
to fast proton decay. Similarly,
the absence of the bilinear terms
$\mu\hat H_u \hat H_d$, $\mu_i \hat H_u \hat L_i$ and 
${\mathcal M}_{i'j'} \hat\nu_{i'}^c\hat\nu_{j'}^c$ (and the linear term
$t_{i'} \hat\nu_{i'}^c$), which would reintroduce 
the $\mu$ problem and additional naturalness problems, must be explained. 
Finally, since the superpotential of the $\mn$ contains only trilinear couplings, it has a $Z_3$ discrete symmetry just like 
the next-to-minimal supersymmetric standard model (NMSSM) (for reviews, see e.g. Refs.~\cite{Maniatis:2009re,Ellwanger:2009dp}), and 
therefore one expects to have also a cosmological domain wall 
problem~\cite{Holdom:1983vk,Ellis:1986mq,Rai:1992xw,Abel:1995wk,Chung:2010cd} unless inflation at the weak scale is invoked.

In Ref.~\cite{Fidalgo:2011tm}, the strategy of adding an extra $U(1)'$ gauge symmetry
to the $\mn$
to explain the absence of the above terms in the superpotential, and to solve the potential domain wall problem, was adopted.
There, the SM gauge group was therefore extended to
$SU(3)\times SU(2)\times U(1)_Y\times U(1)'$ (see Ref.~\cite{Langacker:2008yv} for a review of this strategy in other models). 
 Generically, 
with the extra $U(1)'$ one is able to forbid not only the presence of the linear term in the superpotential, but also the above dangerous trilinear and bilinear terms\footnote{This avoids relying on string theory arguments or discrete symmetries to forbid them (see Ref.~\cite{Lopez-Fogliani:2017qzj} for a summary about these solutions in the $\mn$).}, since the fields that participate in them can be charged under this group making the terms not invariant under this symmetry.
Besides, the domain wall problem disappears once the discrete symmetry is 
embedded in the gauge 
symmetry~\cite{Lazarides:1982tw,Kibble:1982dd,Barr:1982bb}.\footnote{
Otherwise, it has to be solved with the presence 
of non-renormalizable operators in the superpotential.
These operators break explicitly the $Z_3$ symmetry lifting the degeneracy of the three original vacua, and they can be chosen small enough as not to alter the low-energy 
phenomenology~\cite{Holdom:1983vk,Ellis:1986mq,Rai:1992xw,Chung:2010cd}.  
But in the context of supergravity they can reintroduce in the superpotential the linear and bilinear terms forbidden by the $Z_3$ symmetry~\cite{Abel:1995wk}, and generate quadratic tadpole divergences~\cite{Ellwanger:1983mg,Bagger:1993ji,Jain:1994tk,Bagger:1995ay,Abel:1996cr},
although
these problems can be eliminated in models which possess a
$R$-symmetry in the non-renormalizable superpotential~\cite{Abel:1996cr,Panagiotakopoulos:1998yw}.}
A crucial characteristic of these models is the presence in their spectrum of exotic matter dictated by the anomaly cancellation conditions, such as extra quark representations or singlets under the SM gauge group.
For an alternative $U(1)'$ extension of the $\mn$ where this exotic matter is not present, see Ref.~\cite{Lozano:2018esg}. There, this result is obtained allowing non-universal $U(1)'$ charges for the SM fields, and non-holomorphic Yukawa couplings generating fermion masses at one loop by gluino or neutralino exchange.

In addition to the above arguments favoring extra $U(1)'$ charges for the fields, we would like to remark the special role of the RH neutrino when added to the SM spectrum, since it is the only field with no quantum numbers under the SM gauge group. 
In a sense,
the feeling of having a field with no charges is uneasy.
For example, in string constructions where extra $U(1)'$ groups arise 
naturally~\cite{Casas:1987us,Casas:1988se,Langacker:2008yv}, no ordinary fields appear that are singlets under the full gauge group.
We will assume therefore as natural to have a non-vanishing $U(1)'$ charge for the RH neutrinos.
In fact, forbidding the $\mu$ term $\mu\hat H_u \hat H_d$ and the bilinear term $\mu_i \hat H_u \hat L_i$ using the
$U(1)'$ symmetry as discussed above, necessarily implies that the RH neutrino must be charged for the terms $\lambda_{i'} \hat H_u \hat H_d\hat\nu_{i'}^c$
and $Y^\nu_{ij'} \hat {H_u} \hat L_i  \hat \nu^c_{j'}$ to be allowed.
Although now the couplings 
$\kappa_{i'j'k'}\, \hat\nu_{i'}^c\hat\nu_{j'}^c\hat \nu_{k'}^c$
are automatically forbidden in the superpotential, this fact does not imply that a Goldstone boson is reintroduced in the spectrum, since the $U(1)'$ symmetry is gauged and therefore the would-be Goldstone boson is eaten by the new $Z'$ gauge boson in the EWSB process.
Note that, as in the case of the $\mn$ with the SM gauge group, the terms in the second line of Eq.~(\ref{Eq:superpotentialmunu}) are now also allowed with the $U(1)'$ extension, since this symmetry cannot forbid them once we impose the simultaneous presence of 
$Y^\nu_{ij'}\hat {H_u}\hat L_i\hat \nu^c_{j'}$
and
$\lambda_{i'}\hat {H_u}\hat H_d\hat \nu^c_{i'}$ implying that 
$\hat H_d$ and $\hat L_i$ have the same quantum numbers.

Given the above discussion, our starting superpotential in what follows will be the one of 
Eq.~(\ref{Eq:superpotentialmunu}) without the last term.
This will give rise to the
$U(1)'$ extensions of the $\mn$, dubbed U$\mu\nu$SSM, which we will explore in this work.
Unlike Ref.~\cite{Fidalgo:2011tm},
we will not impose three families of RH neutrinos nor three families of exotic matter. 
This change will lead to significant modifications in the 
spectra of the RPV models that we will build,
even with the possible existence of candidates for stable dark matter (DM).

The assumption of three families of RH neutrinos can be motivated in principle as a replication of what happens with the other particles of the SM.
However, as already pointed out, the RH neutrino is the only singlet of the SM and therefore its way of arising from a more fundamental theory does not have to be the same as for the other particles. 
Actually, from the model-building viewpoint only one RH neutrino is sufficient to generate dynamically the 
$\mu$ term~\cite{LopezFogliani:2005yw,Bartl:2009an,Biekotter:2017xmf,Lara:2018rwv,Lara:2018zvf}, reproducing neutrino physics from loop corrections~\cite{Bartl:2009an}.
Two RH neutrinos are sufficient to reproduce at the tree level the neutrino mass differences and mixing angles, and it is also the minimal case 
that can yield
spontaneous CP violation in the neutrino sector \cite{Fidalgo:2009dm}.
Moreover, more than three RH neutrinos can also be used, without spoiling any of the useful properties present in the case of three families~\cite{Lopez-Fogliani:2017qzj}.
These arguments lead us to motivate 
U$\mu\nu$SSM models
where the number of RH neutrinos 
is a free parameter, to be fixed by the anomaly cancellation conditions. 
As we will show, models with three RH neutrinos or with a different number of them can be obtained.

{In connection with this strategy, we will also take into account that the $U(1)'$ charges can make distinctions among fields that are singlets under the SM gauge group. 
Thus we will allow for an arbitrary number of these singlets with different $U(1)'$ charges, and therefore with 
different allowed couplings. 
In particular,
some of them will behave as RH neutrinos ($\hat\nu^c$). Others ($\hat S$) will generate Majorana masses for the latter through couplings of the type
$\hat S\hat\nu^c\hat\nu^c$, which is helpful since
couplings $\hat\nu^c\hat\nu^c\hat \nu^c$ are now forbidden. 
Besides, other singlets under the SM gauge group ($\hat\xi$) will be candidates for DM through couplings $\hat\nu^c\hat\xi\hat\xi$.
}

The paper is organized as follows. Section~\ref{munuSSM} will be devoted to introduce the
U$\mu\nu$SSM superpotential.
In Section~\ref{Section:anomalies}, 
the anomaly cancellation conditions will be imposed to find the consistent models. 
We will see that a minimum number of exotic quarks is required, which
are vector-like under the SM gauge group but chiral under the $U(1)'$. Namely,
either three pairs of quark singlets of $SU(2)$, or a pair of quark singlets of $SU(2)$ together with a pair of quark doublets of $SU(2)$, must be present.
Optionally, singlets under the SM gauge group distinguished from the RH neutrinos by the $U(1)'$ charges can also be present. Some of them can be candidates for DM.
In Section~\ref{Section:models}, several U$\mu\nu$SSM models with different characteristics and matter content will be built. 
The scalar potential of these models will be studied in Section~\ref{scalarpotential}.
In Section~\ref{masses}, the masses and mass mixings of different sectors of the models will be analyzed, paying special attention to the mixing of the SM $Z$ boson and the $Z'$ boson associated to the $U(1)'$.
The experimental constraints on the parameter space of U$\mu\nu$SSM models will be discussed in Section~\ref{ephenomenology}, 
specially the limits on $Z'$ masses and $Z-Z'$ mixing from searches at the LHC and precision electroweak data.
In Section~\ref{phenomenology}, other phenomenological prospects of these models will be introduced, discussing briefly characteristic signals of the $Z'$ boson when decaying to sparticle pairs, as well as possible WIMP dark matter signals. Finally, our conclusions are left for Section~\ref{conclusions}.

\section{The superpotential of U$\mn$ models
}
\label{munuSSM} 

Based on the discussion of the Introduction, 
we consider the following starting superpotential:
\bea
\label{Eq:superpotentialUmunu}
W_{1} &=&  
Y^e_{ij} \, \hat H_d\, \hat L_i \, \hat e_j^c
\ +
Y^d_{ij} \, \hat H_d\, \hat Q_i \, \hat d_j^c 
\ -
Y^u_{ij} \, \hat {H_u}\, \hat Q_i \, \hat u_j^c
\nonumber\\
&+&
\lambda_{ijk} \, \hat L_i  \, \hat L_j \, \hat e_k^c
+
\lambda'_{ijk} \, \hat L_i \, \hat Q_j \, \hat d_k^c
\nonumber\\
&-&
Y^\nu_{ij'} \, \hat {H_u}\, \hat L_i \, \hat \nu^c_{j'}
\ +
\lambda_{i'} \, \hat {H_u} \, \hat H_d \, \hat \nu^c_{i'}
\,,
\eea
whose terms are invariant under the $SU(3)\times SU(2)\times U(1)_Y\times U(1)'$ gauge group, and we are assuming 
a non-vanishing $U(1)'$ charge of the RH neutrinos.
Defining in what follows $z(F)$ as the the $U(1)'$ charge of a field $F$, we have for the RH neutrinos
$z(\nu^c)\neq 0$.

In order to build U$\mu\nu$SSM models, let us note that the introduction of exotic matter with color charge in the spectrum is mandatory.
As we will discuss in the next section in detail,
the $[SU(3)]^2 - U(1)'$ anomaly cancellation condition implies the presence of these exotic quarks.
Their minimal representations are
singlets or doublets of $SU(2)$, denoted respectively by 
$\hat{\mathbb{K}}$ and $\hat{\mathbb{D}}$.
To avoid conflicts with experiments, they
must be sufficiently heavy to not have been detected.
To give them masses, one can couple them to the already present RH neutrino superfields whose scalar components acquire VEVs after the EWSB, 
with the general superpotential
\bea
\label{Eq:superpotential2}
W_2 =
 Y^{\mathbb{K}}_{i' I}  \, \hat \nu^c_{i'}  \hat{\mathbb{K}}_{I}  \hat{\mathbb{K}}^c_{I}
+
 Y^{\mathbb{D}}_{i' I'}  \, \nu^c_{i'} \hat{\mathbb{D}}_{I'}  \hat{\mathbb{D}}^c_{I'}\,,
 \label{exoticquarks}
\eea
where
$I=0,1,...,n_{\mathbb{K}}$ and $I'=0,1,...,n_{\mathbb{D}}$,
with $n_{\mathbb{K}}$ and $n_{\mathbb{D}}$ the number of $\hat{\mathbb{K}}$
and $\hat{\mathbb{D}}$ superfields, respectively.
Note that the vanishing hypercharge of the RH neutrinos, $y(\nu^c)=0$,
implies that the exotic quarks must have opposite hypercharges to be coupled to each other,
\bea
y({\mathbb{K}}_I^c) = - y({\mathbb{K}}_I)\,,\;\;\;
y({\mathbb{D}}^c_{I'}) = - y({\mathbb{D}}_{I'})\,.
\label{hypercharges}
\eea
i.e. they must be
vector-like pairs under the SM gauge group (as the Higgs doublets $H_u$ and $H_d$).
This is also helpful in reducing the sensitivity 
to precision electroweak 
constraints~\cite{Zyla:2020zbs}, and not altering the following anomaly cancellation conditions: $[SU(3)]^2 - U(1)_Y$, $[SU(2)]^2 - U(1)_Y$, $[\text{Gravity}]^2 - U(1)_Y$ and
$[U(1)_Y]^3$.
For the sake of generality, we allow 
$\mathbb{K}_I$ and $\mathbb{D}_{I'}$
to have arbitrary hypercharges in our analysis,
nevertheless we will also find solutions of the anomaly cancellation conditions 
where they have the same hypercharges as the SM quarks.

In fact, we will show in the next section that only two solutions of $[SU(3)]^2 - U(1)'$ anomaly cancellation condition are possible with this minimal type of exotic quarks.
One of them has 
$n_{\mathbb{K}} = 3$ and $n_{\mathbb{D}}=0$, and therefore the corresponding superpotential is
\bea
\label{Eq:superpotential2k}
W_2(\hat{\mathbb{K}}_j) =
Y^{\mathbb{K}}_{i' j}  \, \hat \nu^c_{i'}  \hat{\mathbb{K}}_{j}  \hat{\mathbb{K}}^c_{j}
\,,
\label{solution1}
\eea
with $j=1,2,3$ as the family indexes.
The other solution has $n_{\mathbb{K}} = 1$ and $n_{\mathbb{D}}=1$, and the superpotential is
\bea
\label{Eq:superpotential2kd}
W_2(\hat{\mathbb{K}},\hat{\mathbb{D}}) =
Y^{\mathbb{K}}_{i'}  \, \hat \nu^c_{i'}  \hat{\mathbb{K}}  \hat{\mathbb{K}}^c
+
Y^{\mathbb{D}}_{i'}  \, \nu^c_{i'} \hat{\mathbb{D}}  
\hat{\mathbb{D}}^c\,.
\label{superk}
\eea

{In Section~\ref{Section:models}, we will see that 
very simple realizations of U$\mn$ models can be constructed. There, apart from the exotic quarks, only two RH neutrinos are present.
In these cases, 
since in U$\mn$ models the couplings 
$\kappa{_{i'j'k'}} \hat \nu^c_{i'} \hat \nu^c_{j'} \hat \nu^c_{k'}$
are forbidden, the useful outcome of the
$\mn$ where Majorana masses are dynamically generated cannot be obtained.
Nevertheless, it was shown
in Ref.~\cite{Fidalgo:2011tm}, where three families of RH neutrinos were used, that they can acquire large
masses through the mixing with the extra gaugino and the Higgsinos.
This produces 
two RH neutrinos with electroweak-scale masses, whereas the third one combines with the
LH neutrinos to form a nearly massless Dirac particle. 
So the electroweak seesaw only works for two linear combinations of LH neutrinos, and at the tree level there are
four light Majorana states.
To account for neutrino data some of the entries of the
Yukawa matrix must be of the order of $Y^{\nu}\lsim 10^{-11}$.
At the end of the day, one obtains
two heavy RH neutrinos of the order of TeV and four light (three active and one
sterile) neutrinos. Actually, 
given the oscillation anomalies in LSND~\cite{Athanassopoulos:1996wc,Athanassopoulos:1996jb,Athanassopoulos:1997er,Athanassopoulos:1997pv}, and MiniBooNE~\cite{AguilarArevalo:2010wv} 
the extra light sterile neutrino might be welcome~\cite{Akhmedov:2010vy}. 
}

{Alternatively, we can avoid to tune the neutrino Yukawa couplings, assuming the existence of additional singlets under the SM gauge group. This also produces a different type of neutrino spectrum. As we will see in the next section, we have enough freedom in the constructions (fulfilling anomaly cancellation conditions) to allow the presence of $\hat S$ superfields with couplings
}
\bea
\label{Eq:superpotential3}
W_3 =
\kappa_{\alpha j' k'}  \, \hat S_{\alpha}  \, \hat \nu^c_{j'}  \, \hat \nu^c_{k'},
\label{couplingsnunu}
\eea
where $\alpha=0,1, ..., n_S$, with $n_S$ the number of $\hat S$ superfields. 
They are distinguished from the RH neutrinos by the value of the $U(1)'$ charge, and their VEVs dynamically generate Majorana masses for the RH neutrinos.
In this case, all RH neutrinos acquire masses of the order of TeV, and the light neutrinos have the appropriate masses with all neutrino Yukawa couplings 
$Y^{\nu}\lsim 10^{-6}$.

{Once the $\hat S$ superfields are present, we will show that it is helpful for the anomaly cancellation conditions to be fulfilled, to allow also the presence of other singlets under the SM gauge group $\hat N$ with different $U(1)'$ charges. 
They can be used
as an additional source of masses for the $\hat S$ fields through the couplings
\bea
\label{Eq:superpotential5}
W_4 =
\kappa'_{\alpha' \alpha \beta}  \, \hat N_{\alpha'} \hat S_{\alpha} \hat S_{\beta},
\eea
where $\alpha'=0,1,.....n_N$, with $n_N$ the number of $\hat N$ superfields.}

\begin{table}[t!]
\begin{center}
\begin{tabular}{|c|c|}
\hline
\text{Fields} ($F$)
&
$SU(3)\times SU(2)\times U(1)_Y\times U(1)'$
\tabularnewline
 \hline 
 $\hat{Q}_i$ 
&
$\left(3, 2, 1/6, z(Q)\right)$
 \tabularnewline
\hline 
 $\hat{u}^c_i$
&
$\left(3, 1, -2/3, z(u^c)\right)$
 \tabularnewline
 \hline 
 $\hat{d}^c_i$
 &
$\left(3, 1, 1/3, z(d^c)\right)$
 \tabularnewline
 \hline 
 $\hat{L}_i$
&
$\left(1, 2, -1/2, z(L)\right)$
 \tabularnewline
 \hline 
 $\hat{e}^c_i$
&
$\left(1, 1, 1, z(e^c)\right)$
 \tabularnewline
\hline 
$\hat{H}_d$
&
$\left(1, 2, -1/2, z(H_d)\right)$
 \tabularnewline
\hline 
$\hat{H}_u$
&
$\left(1, 2, 1/2, z(H_u)\right)$
 \tabularnewline
\hline 
  $\hat\nu^c_{i'}$
  &
$(1,1,0,z(\nu^c))$   
\tabularnewline
\hline 
  $\hat S_{\alpha}$
&
$(1,1,0,z(S))$   
\tabularnewline
\hline 
 $\hat N_{\alpha'}$
 &
$(1,1,0,z(N))$   
\tabularnewline
\hline 
 $\hat\xi_{\alpha''}$
 &
$(1,1,0,z(\xi))$  
\tabularnewline
\hline 
 $\hat{\mathbb{K}}_i $
&
$\left(3,1,y({\mathbb{K}}_i),z({\mathbb{K}}_i)\right)$
 \tabularnewline
\hline 
$\hat{\mathbb{K}}^c_i $
&
$(\bar 3,1,-y({\mathbb{K}}_i),z({\mathbb{K}}^c_i))$ 
\tabularnewline
\hline
\end{tabular}
\caption{Chiral superfields in U$\mn$ models and their quantum numbers,
for the solution in Eq.~(\ref{solution1}) with the presence of the
extra quark singlets $\hat{\mathbb{K}}_{i}$ and $\hat{\mathbb{K}}^c_{i}$, $i=1,2,3$.
The indexes $i',\alpha,\alpha'$ and $\alpha''$ correspond to the number of different singlet superfields. $z(F)$ denotes the $U(1)'$ charge of a field $F$.
For the alternative solution in Eq.~(\ref{superk}),
the last two representations must be replaced
by the four representations
 $\hat{\mathbb{K}}\left(3,1,y({\mathbb{K}}),z({\mathbb{K}})\right)$, 
$\hat{\mathbb{K}}^c (\bar 3,1,-y({\mathbb{K}}),z({\mathbb{K}}^c))$,  
$\hat{\mathbb{D}}\left(3,2,y({\mathbb{D}}),z({\mathbb{D}})\right)$ and
  $\hat{\mathbb{D}}^c (\bar 3,2,-y({\mathbb{D}}),z({\mathbb{D}}^c))$.
}
\label{table1}
\end{center}
\end{table}

{
Another interesting possibility that we can realize in U$\mn$ models, is the presence of other singlets under the SM gauge group which can be candidates for DM.
Specifically, we can ask for the presence 
of $\hat\xi$ superfields in the spectrum  with couplings
\bea
\label{Eq:superpotential4}
W_5 =
\kappa''_{i' \alpha'' \beta''}  \, \, \hat \nu^c_{i'} \, \hat \xi_{\alpha''} \, \hat \xi_{\beta''},
\eea
where $\alpha'', \beta''=0,1,.....n_{\xi}$, with $n_{\xi}$ the number 
of $\hat\xi$ superfields.
Remarkably, 
the scalar and fermionic components of these superfields are weakly interacting massive particles (WIMPs), and because of the $Z_2$ symmetry present in $W_5$
the lightest of them can be used as stable WIMP DM, even though we are working in the framework of RPV models.
In Section~\ref{Section:models},
we will build models with three RH neutrinos, DM candidates, and no more extra matter. Constructions with a different number of these fields, and including 
$\hat S$ and $\hat N$ fields will also be obtained.
}

Summarizing, to study the most interesting phenomenology of U$\mn$ models we will work in what follows either with
the superpotential
\bea
\label{Eq:superpotentialt}
W_{\text{U$\mu\nu$SSM}}(\hat{\mathbb{K}}_i)
=
W_1 + 
W_2(\hat{\mathbb{K}}_i)
+ W_3 + W_4 + W_5 
\,,
\eea
or with the superpotential
\bea
\label{Eq:superpotentialt2}
W_{\text{U$\mu\nu$SSM}}(\hat{\mathbb{K}},\hat{\mathbb{D}}) =
W_1 + 
W_2(\hat{\mathbb{K}},\hat{\mathbb{D}})
+ W_3 + W_4 + W_5 
\,,
\label{neww}
\eea
with the chiral fields shown in Table~\ref{table1}. As we will show in the next section, the cases of interest contain a pair
of Higgs doublets $H_d$ and $H_u$.

In the following, we will concentrate in these scenarios analyzing the solutions of the anomaly cancellation conditions, thus constraining further the spectrum.
{Note that if $z(F)$ is any particular solution 
(with $F$ all the fields of the spectrum), then the following linear combination is also solution~\cite{Lee:2007fw}:
\bea
\rho \, z(F) + \sigma \,  y(F).
\label{linear}
\eea
Here the $U(1)'$ charges are scaled by an arbitrary normalization factor
$\rho$, as well as `rotated' by hypercharge with another arbitrary coefficient $\sigma$.
In the case of the hypercharge contribution, this is possible because it fulfills automatically the SM anomaly cancellation conditions, as discussed in
Eq.~(\ref{hypercharges}), but also, by construction of the models, the other two necessary conditions,
$[U(1)_Y]^2 - U(1)'$ and $[U(1)']^2 - U(1)_Y$, as we will discuss in the next section.
This arbitrariness in the numerical values for the $U(1)'$ charges has no effect on the phenomenology of the models, since it can be reabsorbed in the definition of the gauge coupling.}


\section{$U(1)'$ charges and anomaly cancellation} 
\label{Section:anomalies}

The $U(1)'$ charges of the fields must fulfill the following conditions:
\bea
\label{zneutrino}
z(\nu^c) & \neq & 0\,,
\\ 
\label{zneutrino2}
z(H_u)& \neq & - z(H_d)\,,
\\
\label{proton}
z(u^c) & \neq & -2\ z(d^c)\,,
\\
\label{hdl}
z(H_d) & = & z(L)\,.
\eea
The first one was extensively discussed in the Introduction. The second condition arises from the first one since the effective $\mu$ term 
$\lambda_{i'}\hat {H_u}\hat H_d\hat \nu^c_{i'}$ must be present in the superpotential.
The third condition is in order
to avoid fast proton decay, forbidding the baryon-number-violating terms 
$\lambda''_{ijk}\hat u^c_i \hat d^c_j \hat d^c_k$.
Finally, the last condition arises from the simultaneous presence of 
$Y^\nu_{ij'}\hat {H_u}\hat L_i\hat \nu^c_{j'}$
and
$\lambda_{i'}\hat {H_u}\hat H_d\hat \nu^c_{i'}$.
Now, we ask that the rest of the terms in 
$W_{\text{U}{\mn}}$
are also invariant under
the $U(1)'$ symmetry, obtaining more conditions for the $U(1)'$ charges of the fields:
\bea
Y^d: &\; \; \; \; &  z(H_d) + z(Q) + z(d^c) =0,
\label{2.8}
\\ 
Y^e: &\; \; \; \; & 2 \, z(H_d) + z(e^c) =0,
\label{2.9}
\\  
Y^u: &\; \; \; \; & z(H_u) + z(Q) + z(u^c) =0,
\label{2.10}\\  
\lambda: &\; \; \; \; & z(H_u) + z(H_d) + z(\nu^c)  =0,
\label{2.11}\\ 
\kappa: &\; \; \; \; & z(S) + 2 \, z(\nu^c) =0,  
\label{2.12}\\  
\kappa': &\; \; \; \; & z(N) + 2 \, z(S) =0, 
\label{2.13}\\ 
\kappa'': &\; \; \; \; &  \, z(\nu^c) + 2 \, z(\xi)  =0, 
\label{2.14}\\ 
Y^{\mathbb{K}}: &\; \; \; \; & z(\nu^c) +  \, z(\mathbb{K}_I) + z(\mathbb{K}_I^c) =0, 
\label{2.15}\\
Y^{\mathbb{D}}: &\; \; \; \; & z(\nu^c) +  \, z(\mathbb{D}_{I'}) + z(\mathbb{D}^c_{I'}) =0.
\label{Eq:charges}
\eea
Using the above equations, we can express 
$7+n_{\mathbb{K}}+n_{\mathbb{D}}$ charges in terms of the others. We choose as independent charges,  
$z(Q)$, $z(\nu^c)$, $z(H_d)$, $z(\mathbb{K}_I)$ and $z(\mathbb{D}_{I'})$.

Armed with these results,
let us now impose the six anomaly cancellation conditions associated to the $U(1)'$ gauge symmetry.

\begin{enumerate}

\item
In the absence of exotic fields, taking into account Eqs.~(\ref{2.8}),~(\ref{2.10}) and~(\ref{2.11}), the $[SU(3)]^2 - U(1)'$ condition, 
$\sum z =0$ (where the sum extends over all color triplet fermions), 
gives rise to the constraint
\bea
3\left[2z(Q)+z(u^c)+z(d^c)\right]=3z(\nu^c)=0\ .
\label{ja1}
\eea
If we were studying the $U(1)_Y$ of the SM, we would have to
substitute $z(\nu^c)=0$ by the condition for the hypercharge $y(\nu^c)=0$, which is trivially fulfilled as expected. However, the 
$U(1)'$ charge of the RH neutrino $z(\nu^c)$ is different from zero, and therefore to cancel the anomaly it is mandatory to add
exotic quarks. Assuming the presence of the quarks associated to the superpotential of Eq.~(\ref{Eq:superpotential2}), it is straightforward to see that
the constraint 
\bea
(3-n_{\mathbb{K}}-2n_{\mathbb{D}})z(\nu^c)=0\ ,
\label{ja2}
\eea
must be fulfilled.
The equation $n_{\mathbb{K}}+2n_{\mathbb{D}}=3$ has two solutions. The one with
\bea
n_{\mathbb{K}} = 3\;, \;\;\;\ n_{\mathbb{D}}=0\,,
\label{exotic1} 
\eea
implies the presence of three SM vector-like quark pairs, singlets of $SU(2)$,
$\hat{\mathbb{K}}_{i}$ and $\hat{\mathbb{K}}^c_{i}$, $i=1,2,3$.
The second solution with
\bea
n_{\mathbb{K}} = 1\;, \;\;\;\ n_{\mathbb{D}}=1\,,
\label{exotic2} 
\eea
implies the presence of a SM vector-like quark pair, singlet of $SU(2)$,
$\hat{\mathbb{K}}, \hat{\mathbb{K}}^c$, together with a SM vector-like quark pair, doublet of $SU(2)$,
$\hat{\mathbb{D}}, \hat{\mathbb{D}}^c$. It is worth emphasizing here that both solutions have the same number of degrees of freedom. 
The corresponding superpotentials are written in 
Eqs.~(\ref{Eq:superpotential2k}) and~(\ref{Eq:superpotential2kd}), respectively. The representations of these exotic quarks are also shown in
Table~\ref{table1}.

\item
Let us now consider the $[\text{Gravity}]^2 - U(1)'$ anomaly cancellation condition.
Following similar arguments as in the previous case, one obtains the constraint
\bea
 \left(-9 + 2n_H + 3 n_{\mathbb{K}}+6n_{\mathbb{D}}
-n_{\nu^c} +2n_S -4n_N  +\frac{1}{2}n_{\xi}\right)z(\nu^c)=0\ , 
\label{ja3}
\eea
{where
$n_H$ is the number of SUSY Higgs families,
e.g. $n_H=1$ denotes the usual two Higgs doublets of the MSSM, $H_u$ and $H_d$}.
For both solutions in Eqs.~(\ref{exotic1}) and~(\ref{exotic2}), $3 n_{\mathbb{K}}+6n_{\mathbb{D}} = 9$, and therefore this constraint turns out to 
be 
\bea
n_{\nu^c}=  2 \, (n_H + n_S - 2 n_N )+ \frac{1}{2}{n_{\xi}}\,.
\label{gravity1} 
\eea

\item
From the cancellation of the $[SU(2)]^2 - U(1)'$ anomaly, one obtains for $z(Q)$ the constraint
\bea
z(Q) = \frac{1}{9} \left[(n_H + 3n_{\mathbb{D}} )\ z(\nu^c)- 3 \, z(H_d) \right]\,.
\label{doublet}
\eea
Since we will use this constraint in the computation of the rest of the anomaly cancellation conditions, our list of independent charges is reduced in the case of solution~(\ref{exotic1}) to
\bea
z(\nu^c),\;\;\; z(H_d),\;\;\; z(\mathbb{K}_i)\,,
\label{independent1}
\eea
and in the case of solution~(\ref{exotic2}) to
\bea
z(\nu^c),\;\;\; z(H_d),\;\;\; z(\mathbb{K}),\;\;\; z(\mathbb{D})\,.
\label{independent2}
\eea

Note that if $n_H=3$, the constraint in Eq.~(\ref{doublet}) implies  
$z(\nu^c) - z(H_d) - 3z(Q) = - n_{\mathbb{D}} z(\nu^c)$.
This result, together with Eqs.~(\ref{2.8}),~(\ref{2.10}) and~(\ref{2.11}), give rise to the relation
$z(u^c)+2\ z(d^c)= -n_{\mathbb{D}} z(\nu^c)$.
For solution~(\ref{exotic1}) with $n_{\mathbb{D}}=0$, this relation does not fulfill the constraint~(\ref{proton}), and therefore the baryon-number-violating couplings
$\lambda''_{ijk}\, \hat u^c_i \hat d^c_j \hat d^c_k$ would be allowed.

\item
The $[U(1)_Y]^2 - U(1)'$ anomaly cancellation condition
gives rise to the result 
\bea
\left[\frac{1}{2}(-8+2n_H+3n_{\mathbb{D}})
+3 \sum^{n_{\mathbb{K}}}_{I=1} \ y({\mathbb{K}}_I)^2 
+6n_{\mathbb{D}}\ y({\mathbb{D}})^2\right]z(\nu^c)=0. 
\label{ja4}
\eea
For solution~(\ref{exotic1}) with the exotic quarks
$\hat{\mathbb{K}}_{1,2,3}$ and $\hat{\mathbb{K}}^c_{1,2,3}$, 
the latter constraint can be written as
\bea
\sum^3_{i=1} y({\mathbb{K}_i})^2 
=
\frac{4-n_H}{3} \,.
\label{Eq:uno}
\eea
The case $n_H=3$ is excluded by the requirement of proton stability, as discussed before. For $n_H>4$ we obtain a complex value for $y({\mathbb{K}_i})$.
For $n_H=4$ one needs $y(\mathbb{K}_i)=0$, but then the $[U(1)']^2 - U(1)$ anomaly cancellation condition (see below) has solution only for
$z(\nu^c) =0$.
Thus all cases with $n_H > 1$ are excluded in our scenario, except $n_H=2$.
{To avoid the potential problem of flavour-changing neutral currents (FCNC) facing all models with an extended Higgs sector (see e.g. Ref.~\cite{Escudero:2005hk} and references therein), we use as our case of interest 
$n_H = 1$}. Therefore, Eq.~(\ref{Eq:uno}) is written as
\bea
\sum^3_{i=1} y(\mathbb{K}_i)^2   = 1\,.
\label{Eq:unos}
\eea
For solution~(\ref{exotic2})
with the exotic quarks 
$\hat{\mathbb{K}}$, $\hat{\mathbb{K}}^c$ and
$\hat{\mathbb{D}}$, $\hat{\mathbb{D}}^c$, the constraint from the anomaly cancellation condition (\ref{ja4}) can be written as
\bea
6y({\mathbb{K}})^2 + 12y({\mathbb{D}})^2 
=
5-2n_H \,.
\label{Eq:unoss}
\eea
For $n_H>2$ we obtain a complex value for the hypercharge. Nevertheless, for our case of interest $n_H=1$:
\bea
2 y(\mathbb{K})^2  + 4 y(\mathbb{D})^2 = 1\,.
\label{Eq:unosss}
\eea

Interesting examples of hypercharges that satisfy either (\ref{Eq:unos}) or (\ref{Eq:unosss}) will be given in the next section. 

\item
The equation associated to $[U(1)']^2 - U(1)_Y$ is given by
\bea\label{quinta1}
&&
\Big\{6\sum^{n_\mathbb{K}}_{I=1} y(\mathbb{K}_I)  z(\mathbb{K}_I)  
+12n_{\mathbb{D}}\ y({\mathbb{D}}) z({\mathbb{D}})
+
\left(16-4n_H-6n_{\mathbb{D}}\right) z(H_d)
\nonumber\\
&&
+
\left[6-\frac{7}{3}n_H - 4 n_{\mathbb{D}}
+3\sum^{n_\mathbb{K}}_{I=1} y(\mathbb{K}_I)  
+ 6n_{\mathbb{D}}\ y({\mathbb{D}})\right]  z(\nu^c)
\Big\} 
z(\nu^c)
 = 0\,.
\eea

\item
Finally, the equation associated to $[U(1)']^3$ is given by 
\bea
&&
\Big\{
9\sum_{I=1}^{n_\mathbb{K}} z(\mathbb{K}_I)^2 + 18 \, n_{\mathbb{D}} z(\mathbb{D})^2 - 3 \left(16 -4 \, n_H  -  6 n_{\mathbb{D}} \, \right) z(H_d)^2   \nonumber \\
&&
+ 
\left[9\sum_{I=1}^{n_\mathbb{K}} z(\mathbb{K}_I) + 
18 \, n_{\mathbb{D}} \, z(\mathbb{D}) \, 
- 2(18-7 \, n_H  -12\, n_{\mathbb{D}}) \,  z(H_d) \right] z(\nu^c) \nonumber \\
&& 
+ 
 \left[
 \left(5-\frac{1}{3} n_H - 2 n_{\mathbb{D}}\right) n_H 
+ 6 n_{\mathbb{D}}  
+ J
 \right] z(\nu^c)^2 
\Big\} 
z(\nu^c) = 0\,,
\label{uprimacubo}
\eea 
where we have defined
\bea
J = - n_{\nu^c}  + 8 n_S  -  64 n_N  +
 \frac{1}{8} n_{\xi}.
 \label{Eq:F}
\eea

It is helpful to note that fixing the values of $n_H$, $n_\mathbb{D}$ and $n_\mathbb{K}$, the only equations with a dependence in the remaining number of fields are (\ref{gravity1}) and (\ref{uprimacubo}).
Using  
(\ref{gravity1}), we can rewrite 
$J$ as
\bea
J = - 2 n_H  + 6 n_S  -  60 n_N  - \frac{3}{8} n_{\xi}.
\label{Eq:F-v2}
\eea
From this result, it is straightforward to realize that in our constructions the following transformations in the number of fields do not modify the values of the allowed charges:
\bea
& n_{\nu^c} & \rightarrow n_{\nu^c} + 10 n_1 + 16 n_2, \nonumber \\
& n_S & \rightarrow n_S + n_1 + 10 n_2, \nonumber \\
& n_N & \rightarrow n_N + n_2, \nonumber \\
& n_{\xi} & \rightarrow n_{\xi} + 16 n_1,
\label{transformations}
\eea
where $n_1,n_2 = 0,1,2,...$
This implies that a given solution for the $U(1)'$ charges of a given number of fields applies also for other constructions with the number of fields related by
Eqs.~(\ref{transformations}).

\end{enumerate}

\section{Building U$\mn$ models} 
\label{Section:models}

In order to construct different models we have to fix the values of the
$U(1)'$ charges.
Let us choose $z(\nu^c)=\frac{1}{4}$, then conditions 
(\ref{2.12}$-$\ref{2.14})
automatically determine the charges of the rest of singlets under the SM gauge group $S$, $N$, 
and $\xi$. If all of them are present, their 
charges are shown in Table~\ref{table:general}. 
With this result,
we can study in the next subsections the solutions of exotic quarks~(\ref{exotic1}) and~(\ref{exotic2}) with superpotentials~(\ref{Eq:superpotential2k}) and~(\ref{Eq:superpotential2kd}), respectively.
Because of previous arguments, we work in what follows with one family of Higgses
($H_u + H_d$), i.e. $n_H=1$.


 \begin{table}[t!]
\begin{center}
\begin{tabular}{|c|c|}
\hline  
\;  $z(\nu^c)$ \;  &  \; \;  $\frac{1}{4}$  \; \; 
\tabularnewline
\hline 
\;   $ z(S)$ \;   & \ $- \frac{1}{2} $   \; \; 
\tabularnewline
\hline 
\;   $z(\xi)$  \; & \ $- \frac{1}{8}$   \; \;  
\tabularnewline
\hline 
\;   $ z(N)$ \; & \ \ \ $ 1$  \; \; 
\tabularnewline
\hline 
\end{tabular}
\caption{Values of the $U(1)'$ charges of the singlets under the SM gauge group after normalization, valid for all 
U$\mu\nu$SSM models.
}
\label{table:general}
\end{center}
\end{table}

%
%

\subsection{Solutions with exotic quarks $\hat{\mathbb{K}}_{1,2,3}$ and $\hat{\mathbb{K}}^c_{1,2,3}$}

\label{primera}


The solution of exotic quarks~(\ref{exotic1}) contains SM vector-like pairs, singlets of $SU(2)$, 
$\hat{\mathbb{K}}_{i}$ and $\hat{\mathbb{K}}^c_{i}$, with $i=1,2,3$.
Since we have fixed above the value of $z(\nu^c)$ our independent charges are now $z(H_d)$ and $z(\mathbb{K}_i)$,
as we can see from the discussion in Eq.~(\ref{independent1}).
{For simplicity we start with the leptophobic example $z(H_d)=0$, since 
constraints~(\ref{hdl}) and~(\ref{2.9}) imply $z(L) = 0$ and $z(e^c) = 0$.
Using in addition 
constraints~(\ref{2.8}), (\ref{2.10}), (\ref{2.11})
and the anomaly cancellation condition~(\ref{doublet}), we obtain the charges of the rest of the
SM matter shown in Scenario 1 of Table~\ref{table:NK3-general}.}
For scenarios with non-leptophobic charges, i.e. $z(H_d)\neq 0$, we use e.g. $z(H_d)=\frac{1}{2}, -\frac{1}{2}$, as shown in Scenario 2 and 3 of Table~\ref{table:NK3-general}, respectively.


\begin{table}[t!]
\begin{center}
\begin{tabular}{|c|c|c|c|}
\hline  
  &  {\bf Scenario 1}   &  {\bf Scenario 2}   &  {\bf Scenario 3} 
\tabularnewline
\hline 
$z(L)  $  & \ \ $0$ & \ \ $\frac{1}{2}$ & $-\frac{1}{2}$
\tabularnewline
\hline 
$z(e^c)$  & \ \ $0$ & $-1$ & \ \ 1
\tabularnewline
\hline 
$z(H_u) $ & $- \frac{1}{4}$ & $- \frac{3}{4}$ & \ \ $ \frac{1}{4}$ 
\tabularnewline
\hline 
$z(H_d)$ & \ \ 0 & \ \ $\frac{1}{2} $ &  $ -\frac{1}{2}$ 
\tabularnewline
\hline 
 $z(Q) $ & \ \ $\frac{1}{36}$ & $-\frac{5}{36}$  & \ \ $\frac{7}{36}$
\tabularnewline
\hline 
$ z(u^c)$ &\  \ $\frac{2}{9} $ & \ \ $\frac{8}{9} $ & $ -\frac{4}{9} $
\tabularnewline
\hline 
$z(d^c)$ & $- \frac{1}{36}$ & $ - \frac{13}{36}$ & \ \  $\frac{11}{36}$ 
\tabularnewline
\hline 
\end{tabular}
\caption{Values of the $U(1)'$ charges for the SM matter in
U$\mu\nu$SSM models
using solution~(\ref{exotic1}) with the exotic quarks
$\hat{\mathbb{K}}_{1,2,3}$ and $\hat{\mathbb{K}}^c_{1,2,3}$,
and fixing  $z(H_d)= 0,\frac{1}{2},-\frac{1}{2}$. 
These values are independent of the solutions for $z(F)$, with $F$ denoting the exotic quarks.
}
\label{table:NK3-general}
\end{center}
\end{table}

On the other hand,
the gravitational anomaly cancellation condition (\ref{gravity1})
determines the number of singlets under the SM gauge group that we can use consistently.
Since we are assuming $n_H=1$, we can see that the minimal situation with only RH neutrinos as singlets implies that their number must be $n_{\nu^c}=2$.
Using now solutions for the hypercharges 
from the anomaly cancellation condition (\ref{Eq:unos}), 
we can find the allowed $U(1)'$ charges for the exotic quarks ${\mathbb{K}}_{i}$ solving conditions~(\ref{quinta1}) and~(\ref{uprimacubo}).

Examples of rational hypercharges fulfilling condition (\ref{Eq:unos}) are 
\bea
\left(|y(\mathbb{K}_1)|,|y(\mathbb{K}_2)|,|y(\mathbb{K}_3)|\right) =
\left(\frac{1}{3},\frac{2}{3},\frac{2}{3}\right),\;\;\;
\left(\frac{1}{9},\frac{4}{9},\frac{8}{9}\right). 
\label{examples}
\eea
The first example 
has an interesting interpretation because the exotic quarks can have the same hypercharges 
as the ordinary quarks. In particular, for the values 
shown in Table~\ref{table:ND1-11}, one can use for 
$\mathbb{K}^c_1$ and
$\mathbb{K}^c_{2,3}$ the following notation 
by similarity to the SM one:
$\left(y(\mathbb{K}^c_1\equiv \mathbb{B}^c),y(\mathbb{K} ^c_2\equiv 
\mathbb{T}^c_1),y(\mathbb{K}^c_3\equiv \mathbb{T}^c_2)\right)
=
\left(\frac{1}{3},-\frac{2}{3},-\frac{2}{3}\right)$.


\begin{table}[t!]
\begin{center}
\begin{tabular}{|c|c|}
\hline 
\;  $y({\mathbb{K}_1})$ \; & \; $ -  \frac{1}{3}$ \; \; 
\tabularnewline
\hline 
\;  $y({\mathbb{K}}_2)$ \; & \; \ \  $\frac{2}{3} $ \; \; 
\tabularnewline
\hline 
\;  $y({\mathbb{K}}_3)$ \; & \; \ \ $\frac{2}{3}  $ \; \;  
\tabularnewline
\hline 
\end{tabular}
\caption{An example of
values of the hypercharges for the exotic quarks 
$\mathbb{K}_i$ using condition (\ref{Eq:unos}).
As discussed
in Eq.~(\ref{hypercharges}),
$y({\mathbb{K}}_i^c) = - y({\mathbb{K}}_i)$.
These values are independent of the solutions for $z(F)$, with $F$ denoting the exotic quarks.
}
\label{table:ND1-11}
\end{center}
\end{table}

\begin{table}[t!]
\begin{center}
\begin{tabular}{|c|c|c|c|c|}
\hline
 {\bf Scenario 1}  
&  Solution 1  &  Solution 2 &  Solution 3 &  Solution 4
\tabularnewline
\hline 
$z({\mathbb{K}}_1) $ &  $- \frac{1}{27}$ &  $- \frac{1}{9}$ &  $- \frac{1}{54}$ & $- \frac{7}{54}$ 
\tabularnewline
\hline 
$z({\mathbb{K}}_2) $ &  $- \frac{7}{27}$ &  $- \frac{5}{18}$ &  $- \frac{13}{54}$ &  $- \frac{29}{108}$ 
\tabularnewline
\hline 
$z({\mathbb{K}}_3) $ &  $- \frac{19}{108}$ &  $- \frac{7}{36}$ &  $- \frac{5}{27}$ &  $- \frac{23}{108}$ 
\tabularnewline 
\hline 
$z({\mathbb{K}}^c_1) $ & $- \frac{23}{108}$ & $- \frac{5}{36}$ & $- \frac{25}{108}$ & $- \frac{13}{108}$   
\tabularnewline
\hline 
$z({\mathbb{K}}^c_2) $ & \ \ $\frac{1}{108}$ & \ \ $\frac{1}{36}$ & $- \frac{1}{108}$ & \ \ $ \frac{1}{54}$
\tabularnewline
\hline 
$z({\mathbb{K}}^c_3) $ & $- \frac{2}{27}$  &  $ -\frac{1}{18}$  &  $ -\frac{7}{108}$  & $- \frac{1}{27}$ 
\tabularnewline
\hline
\end{tabular}
\caption{Values of the $U(1)'$ charges for the exotic quarks in 
U$\mu\nu$SSM models
using solution~(\ref{exotic1}) with the hypercharges of
Table~\ref{table:ND1-11}.
For each 
column,
the $U(1)'$ charges of the rest of the fields are given in 
Table~\ref{table:general} and Scenario 1 of Table~\ref{table:NK3-general}.
The minimal number of singlets under the SM gauge group consistent with these charges is 
${\bf n_{\nu^c}=2}$. 
}
\label{table:NK3_NR2}
\end{center}
\end{table}

\begin{table}[t!]
\begin{center}
\begin{tabular}{|c|c|c|c|c|}
\hline   
 {\bf Scenario 1}  
&  Solution 1  &  Solution 2 &  Solution 3 &  Solution 4
\tabularnewline
\hline 
 $z({\mathbb{K}}_1) $ 
& \ \ $ \frac{1}{108}$  
& $- \frac{17}{108}$  
&  $- \frac{1}{36}$ 
&  $- \frac{13}{108}$ 
\tabularnewline
\hline 
$z({\mathbb{K}}_2)  $ &  $- \frac{26}{108}$ &  $- \frac{61}{216}$ &  $- \frac{5}{18}$ &  $- \frac{65}{216}$ \tabularnewline
\hline 
$z({\mathbb{K}}_3)  $ &  $- \frac{37}{216}$ &  $- \frac{23}{108}$ &  $- \frac{11}{72}$ &  $- \frac{19}{108}$ \tabularnewline 
\hline 
$z({\mathbb{K}}^c_1)  $ & $- \frac{28}{108}$ & $- \frac{5}{54}$ & $- \frac{2}{9}$ & $- \frac{7}{54}$   \tabularnewline
\hline 
$z({\mathbb{K}}^c_2)  $ & $- \frac{1}{108}$ & \ \ $\frac{7}{216}$ & \ \  $ \frac{1}{36}$ & \ \ $\frac{11}{216}$\tabularnewline
\hline 
$z({\mathbb{K}}^c_3)  $ & $ - \frac{17}{216}$  & $- \frac{1}{27}$  & $- \frac{7}{72}$  & $ - \frac{2}{27}$ \tabularnewline
\hline
\end{tabular}
\caption{The same as in Table~\ref{table:NK3_NR2}, but with the minimal number of singlets under the SM gauge group ${\bf n_{\nu^c}=3}$ and ${\bf n_{\xi}=2}$. 
}
\label{table:NK3_NR3-NDM2}
\end{center}
\end{table}

Using the example of Table~\ref{table:ND1-11}, with the above strategy 
we arrive in Scenario 1 to 
the four rational solutions for
the
$U(1)'$ charges of the exotic quarks ${\mathbb{K}}_{i}$ shown in Table~\ref{table:NK3_NR2}.
Note that the values of $z(\mathbb{K}_i^c)$ are automatically fixed by
the constraint~(\ref{2.15}).
As explained, the minimal number of singlets under the SM gauge group is $n_{\nu^c}=2$,
$n_S=n_N=n_{\xi}=0$. Nevertheless, following the discussion in Eq.~(\ref{transformations}) solutions with a larger number of fields but the same charges, exist. For example, for $n_1=1$ and $n_2=0$ the number of fields is
$n_{\nu^c}=12$, $n_S=1$, $n_N=0$, and $n_{\xi}=16$. This is also true for all the tables shown in this section.

Following the same strategy for the minimal cases $n_{1,2}=0$, we can find other solutions with an even number of RH neutrinos.
First, in Table~\ref{table:NK3_NReven-2} of the Appendix we show models with the same number of RH neutrinos as those in Table~\ref{table:NK3_NR2}, obtained
introducing other singlets under the SM gauge group in the construction. In particular, we impose $n_S=2$, $n_N=1$, $n_{\xi}=0$. 
In general, we can see from condition~(\ref{gravity1}) that there is always an even number of RH neutrinos if 
$n_{\xi}=0$.
Several models with $n_{\nu^c} = 4, 6$, and different numbers of $S$ and $N$ fields, are shown in 
Tables~\ref{table:NK3_NReven-4} and~\ref{table:NK3_NReven-6}, 
respectively.
Solutions with the same number of RH neutrinos but without the $N$ field turn out to be complex for the examples discussed here with $z(H_d)=0$ (and with $z(H_d)=\pm 1/2$ to be discussed below).

\begin{table}[b!]
\begin{center}
\begin{tabular}{|c|c|c|c|c|}
\hline  
 {\bf Scenario 2}  
&  Solution 1  &  Solution 2 &  Solution 3 &  Solution 4
\tabularnewline
\hline 
 $z({\mathbb{K}}_1)$ 
& \ \  $\frac{37}{108}$  
& \ \ $\frac{19}{108}$ 
 & \ \ $ \frac{11}{36}$  
& \ \ $ \frac{23}{108}$  
\tabularnewline
\hline 
$z({\mathbb{K}}_2)$ & $- \frac{49}{54}$ & $- \frac{205}{216}$ & $-  \frac{17}{18}$  & $-  \frac{209}{216}$ \tabularnewline
\hline 
$z({\mathbb{K}}_3)$ &  $-\frac{181}{216}$ & $- \frac{95}{108}$ & $- \frac{59}{72}$  & $- \frac{91}{108}$ \tabularnewline
\hline 
$z({\mathbb{K}}^c_1)$ &   $-\frac{16}{27}$ &- $\frac{23}{54}$ & $- \frac{5}{9}$ & $- \frac{25}{54}$\tabularnewline
\hline 
$z({\mathbb{K}}^c_2)$ & \ \   $\frac{71}{108}$& \ \   $ \frac{151}{216}$ &  \ \  $ \frac{25}{36}$ & \ \   $ \frac{155}{216}$\tabularnewline
\hline 
$z({\mathbb{K}}^c_3)$ &  \ \   $\frac{127}{216}$ & \ \   $\frac{17}{27}$ & \ \   $ \frac{41}{72}$ &  \ \  $ \frac{16}{27}$ \tabularnewline
\hline 
\end{tabular}
\caption{
Values of the $U(1)'$ charges for the exotic quarks in 
U$\mu\nu$SSM models
using solution~(\ref{exotic1}) with the hypercharges
of
Table~\ref{table:ND1-11}.
For each column,
the $U(1)'$ charges of the rest of the fields are given in 
Table~\ref{table:general} and Scenario 2 of Table~\ref{table:NK3-general}.
The minimal number of singlets under the SM gauge group consistent with these charges is 
${\bf n_{\nu^c}=3}$ and
${\bf n_{\xi}= 2}$.
}
\label{table:NK3_NR3-Nolepto-DM}
\end{center}
\end{table}

\begin{table}[t!]
\begin{center}
\begin{tabular}{|c|c|c|c|c|}
\hline  
 {\bf Scenario 3}  
&  Solution 1  &  Solution 2 &  Solution 3 &  Solution 4
\tabularnewline
\hline 
$z({\mathbb{K}}_1)$
&  $- \frac{35}{108}$  
&  $- \frac{53}{108}$
 &  $ - \frac{13}{36}$
&  $- \frac{49}{108}$  
\tabularnewline
\hline 
$z({\mathbb{K}}_2)$ & \ \ $\frac{23}{54}$ & \ \  $ \frac{83}{216}$ &  \ \ $ \frac{7}{18}$  &  \ \ $ \frac{79}{216}$ \tabularnewline
\hline 
$z({\mathbb{K}}_3)$ &  \ \  $\frac{107}{216}$ & \ \  $\frac{49}{108}$ & \ \  $ \frac{37}{72}$  &  \ \ $ \frac{53}{108}$ \tabularnewline
\hline 
$z({\mathbb{K}}^c_1)$ &  \ \   $\frac{2}{27}$ &  \ \ $\frac{13}{54}$ &  \ \ $ \frac{1}{9}$ &  \ \ $ \frac{11}{54}$\tabularnewline
\hline 
$z({\mathbb{K}}^c_2)$ & $- \frac{73}{108}$& $- \frac{137}{216}$ & $- \frac{23}{36}$ & $ - \frac{133}{216}$\tabularnewline
\hline 
$z({\mathbb{K}}^c_3)$ &  $- \frac{161}{216}$ & $- \frac{19}{27}$ & $- \frac{55}{72}$ & $- \frac{20}{27}$ \tabularnewline
\hline 
\end{tabular}
\caption{
Values of the $U(1)'$ charges for the exotic quarks in 
U$\mu\nu$SSM models
using solution~(\ref{exotic1}) with the hypercharges
of
Table~\ref{table:ND1-11}.
For each column,
the $U(1)'$ charges of the rest of the fields are given in
Table~\ref{table:general} and Scenario 3 of Table~\ref{table:NK3-general}.
The minimal number of singlets under the SM gauge group consistent with these charges is 
${\bf n_{\nu^c}=3}$ and
${\bf n_{\xi}= 2}$.
}
\label{table:NK3_NR3-Nolepto-DM-Ze1-DM}
\end{center}
\end{table}

To find solutions with an odd number of RH neutrinos we need to impose $n_{\xi}=2 + 4n$, with $n=0,1,2,...$ Actually, the presence of fields of the type $\xi$ is also interesting because the
lightest one has the capability of being a DM candidate, 
as discussed in Eq.~(\ref{Eq:superpotential4}). 
Working with $n_{\xi}=2$, we show in Table \ref{table:NK3_NR3-NDM2} four minimal models with $n_{\nu^c}=3$.
Of course, this table contains also models with $S$ and $N$ fields once we allow
$n_{1,2}\neq 0$.
In Tables~\ref{table:NK3_NR-Odd-1},~\ref{table:NK3_NR-Odd-3}, and~\ref{table:NK3_NR-Odd-5}, we show other models with $n_{\nu^c} = 1,3,5$, respectively.
Solutions with five RH neutrinos but without the $N$ field turn out to be complex for the examples with $z(H_d)=0$ (and with $z(H_d)=\pm 1/2$ to be discussed below).

For Scenario 2 with 
$z(H_d)=\frac{1}{2}$, 
we give in Tables~\ref{table:NK3_NR3-Nolepto-DM}~and~\ref{table:NK3_NR3-Nolepto} examples of solutions with the same number of singlets under the SM gauge group as those in
Tables~\ref{table:NK3_NR3-NDM2}~and~\ref{table:NK3_NR-Odd-3}.  
For Scenario 3 with $z(H_d)=- \frac{1}{2}$, 
we give in Tables~\ref{table:NK3_NR3-Nolepto-DM-Ze1-DM}~and~\ref{table:NK3_NR3-Nolepto-Ze1} again examples of solutions with the same number of singlets under the SM gauge group as those in
Tables~\ref{table:NK3_NR3-NDM2}~and~\ref{table:NK3_NR-Odd-3}.

{Finally, it is worth noting that for the hypercharges of Table~\ref{table:ND1-11}
used in these solutions for the $z(\mathbb{K}_i)$ charges, since they fulfill $y(\mathbb{K}_2)=y(\mathbb{K}_3)$ seemingly new solutions can be generated through the redefinitions $\mathbb{K}_2 \leftrightarrow  \mathbb{K}_3$, $\mathbb{K}_2^c \leftrightarrow  \mathbb{K}_3^c$, because they also fulfill 
straightforwardly
conditions~(\ref{quinta1}) and~(\ref{uprimacubo}).
However, it is trivial to realize that they correspond to a simple renaming of the fields, and are therefore equivalent.
The same comment applies in general, for any values of the hypercharges, for the solutions generated through the redefinitions 
$y(\mathbb{K}_i)\leftrightarrow -y(\mathbb{K}_i)$, $\mathbb{K}_i \leftrightarrow \mathbb{K}_i^c $.
}

%
%

\subsection{Solutions with exotic quarks $\hat{\mathbb{K}}$, $\hat{\mathbb{K}}^c$
and $\hat{\mathbb{D}}, \hat{\mathbb{D}}^c$}

\label{segunda}

\begin{table}[t!]
\begin{center}
\begin{tabular}{|c|c|c|c|c|}
\hline  
  &  {\bf Scenario 4}   &  {\bf Scenario 5}   &  {\bf Scenario 6} & {\bf Scenario 7} 
\tabularnewline
\hline 
$z(L)  $  & \ \ $0$ & \ \ $\frac{1}{2}$ & $-\frac{1}{2}$ & $-\frac{1}{4}$ 
\tabularnewline
\hline 
$z(e^c)$  & \ \ $0$ & $-1$ & \ \ 1& $\frac{1}{2}$ 
\tabularnewline
\hline 
$z(H_u) $ & $- \frac{1}{4}$ & $- \frac{3}{4}$ & \ \ $ \frac{1}{4}$ & 0
\tabularnewline
\hline 
$z(H_d)$ & \ \ 0 & \ \ $\frac{1}{2} $ &  $ -\frac{1}{2}$ & $-\frac{1}{4}$ 
\tabularnewline
\hline 
 $z(Q) $ & \ \ $\frac{1}{9}$ & $-\frac{1}{18}$  & \ \ $\frac{5}{18}$& \ \ $\frac{7}{36}$
\tabularnewline
\hline 
$ z(u^c)$ &\  \ $\frac{5}{36} $ & \ \ $\frac{29}{36} $ & $ -\frac{19}{36} $ & $ -\frac{7}{36} $
\tabularnewline
\hline 
$z(d^c)$ & $- \frac{1}{9}$ & $ - \frac{4}{9}$ & \ \  $\frac{2}{9}$ & \ \  $\frac{1}{18}$ 
\tabularnewline
\hline 
\end{tabular}
\caption{Values of the $U(1)'$ charges for the SM matter in
U$\mu\nu$SSM models
using solution~(\ref{exotic1}) with the exotic quarks
$\hat{\mathbb{K}}_{1,2,3}$ and $\hat{\mathbb{K}}^c_{1,2,3}$,
and fixing  $z(H_d)= 0,\frac{1}{2},-\frac{1}{2}$. 
These values are independent of the solutions for $z(F)$, with $F$ denoting the exotic quarks.
}
\label{table:NK1-ND1-general-lepto}
\end{center}
\end{table}

Let us now move to the solution for the exotic quarks in Eq.~(\ref{exotic2}), where one SM vector-like pair of singlets of $SU(2)$, $\hat{\mathbb{K}}, \hat{\mathbb{K}}^c$, and doublets $\hat{\mathbb{D}}, \hat{\mathbb{D}}^c$, are present. After fixing
$z(\nu^c)$ as in Table~\ref{table:general}, the independent charges in this case are $z(H_d)$, $z(\mathbb{K})$ and $z(\mathbb{D})$. 
If we choose again as a first example the simple case $z(H_d)=0$, then the
new charges of the SM matter are shown in Scenario 4 of  Table~\ref{table:NK1-ND1-general-lepto}.
For scenarios with non-leptophobic charges, we use $z(H_d)=\frac{1}{2}, -\frac{1}{2},-\frac{1}{4}$, as shown in scenarios 5, 6 and 7 of Table~\ref{table:NK1-ND1-general-lepto}, respectively.

On the other hand,
examples of rational hypercharges fulfilling condition~(\ref{Eq:unosss}) are:
\bea
\left(|y(\mathbb{K})|,|y(\mathbb{D})|\right) =
\left(0,\frac{1}{2}\right),\;\;\;
\left(\frac{2}{3},\frac{1}{6}\right). 
\label{examples2}
\eea
In Table~\ref{table:ND1-12}, we show the two cases that we will use in this subsection.
For the second case with $y(\mathbb{K})=\frac{2}{3}$ and $y(\mathbb{D})=\frac{1}{6}$, again by similarity with the SM
one can use for the exotic quarks $\mathbb{K}^c$ and $\mathbb{D}$ the following notation:
$\left(y(\mathbb{K}^c\equiv \mathbb{T}^c),y(\mathbb{D}\equiv \mathbb{Q})\right)
=
\left(-\frac{2}{3},\frac{1}{6}\right)$.

{It is worth noticing here that the possible existence of this extra type of quarks $\mathbb{D}\equiv \mathbb{Q}$ with the SM hypercharge $1/6$, was previously proposed in the context of the $\mn$ \cite{Lopez-Fogliani:2017qzj}. The argument used there was the reinterpretation of the two Higgs doublets as a fourth family of vector-like lepton superfields, with the vector-like quark doublet $\mathbb{D}$ as part of this family. In the present context of U$\mn$ models containing an extra $U(1)'$, two possibilities arise for these exotic quarks with the same hypercharges as the ordinary quarks:
the exotic quarks have different $U(1)'$ charges from the ordinary quarks, or they have the same ones and couple therefore with them. We will show below a model with the latter characteristic. 
}

\begin{table}[t!]
\begin{center}
\begin{tabular}{|c|c|c|}
\hline 
&  Case 1  &    Case 2 
\tabularnewline
\hline 
\; $y({\mathbb{K}})$ \; & $0$ &      $\frac{2}{3}$
\tabularnewline
\hline 
\;  $y({\mathbb{D}})$ \; & \ $\frac{1}{2} $  & $\frac{1}{6} $
\tabularnewline
\hline
\end{tabular}
\caption{Two examples of
values of the hypercharges for the exotic quarks 
$\mathbb{K}$ and $\mathbb{D}$ 
using condition~(\ref{Eq:unosss}).
As discussed
in Eq.~(\ref{hypercharges}),
$y({\mathbb{K}}_i^c) = - y({\mathbb{K}}_i)$ 
and $y({\mathbb{D}}_i^c) = - y({\mathbb{D}}_i)$.
These values are independent of the solutions for $z(F)$, with $F$ denoting the exotic quarks.
}
\label{table:ND1-12}
\end{center}
\end{table}


As in the case of 
the previous subsection, many models can be built fulfilling all the anomaly cancellation conditions. 
In the context of Scenario 4,
We show in Table~\ref{table:ND1-1} one of those models with the same number of singlets under the SM gauge group as those in Table~\ref{table:NK3_NR2}. The hypercharges correspond to those of the Case 1 of Table~\ref{table:ND1-12}.
Other models with the same number of singlets under the SM gauge group are shown in Table~\ref{table:ND1-2}, imposing 
the hypercharges of Case 2 of Table~\ref{table:ND1-12}.
In all the cases, the values of $z(\mathbb{K}^c)$ and $z(\mathbb{D}^c)$ are automatically fixed by
the constraints~(\ref{2.15}) and~(\ref{Eq:charges}), respectively.

\begin{table}[t!]
\begin{center}
\begin{tabular}{|c|c|c|c|}
\hline  
{\bf Scenario 4} &
\tabularnewline
\hline 
$z({\mathbb{K}}) $ & $-  \frac{1}{9}$ \,
\tabularnewline
\hline 
  $z({\mathbb{K}}^{c})$ & $- \frac{5}{36} $  \,
\tabularnewline
\hline 
 $z({\mathbb{D}})$ &  $- \frac{1}{9}$  \,
 \tabularnewline
\hline 
 $z({\mathbb{D}}^{c})$ &  $- \frac{5}{36}$  \,
\tabularnewline
\hline 
\end{tabular}
\caption{
Values of the $U(1)'$ charges for the exotic quarks in 
U$\mu\nu$SSM models
using solution~(\ref{exotic2}) with the hypercharges shown in Case 1
of Table~\ref{table:ND1-12}.
The $U(1)'$ charges of the rest of the fields are given in  
Table~\ref{table:general} and Scenario 4 of Table~\ref{table:NK1-ND1-general-lepto}.
The minimal number of singlets under the SM gauge group consistent with these charges is 
${\bf n_{\nu^c}=2}$.
}
\label{table:ND1-1}
\end{center}
\end{table}


\begin{table}[t!]
\begin{center}
\begin{tabular}{|c|c|c|}
\hline  
{\bf Scenario 4} & Solution 1 & Solution 2
\tabularnewline
\hline 
\;\;\;\ $z({\mathbb{K}})$ \;\;\;&  \, \,  \,  
$ - \frac{1}{9} $ \, \, \,  & \, \, \, $- \frac{11}{108} $ \, \, \, \tabularnewline
\hline 
\;\;\;$z({\mathbb{K}}^{c})$ \;\;\; &  \, \,  \, $- \frac{5}{36} $ \, \, \,  & \, \, \, $-  \frac{4}{27} $ \, \, \, \tabularnewline
\hline 
\;\;\; $z({\mathbb{D}})$ \;\;\;  &  \, \,  \,    $- \frac{1}{9} $      \, \, \, & \, \, \, $- \frac{7}{54} $ \, \, \, \tabularnewline
\hline 
\;\;\; $z({\mathbb{D}}^{c})$ \;\;\; &  \, \,  \, $- \frac{5}{36} $ \, \, \, & \, \, \, $- \frac{13}{108} $ \, \, \, \tabularnewline
\hline 
\end{tabular}
\caption{
The same as in Table~\ref{table:ND1-1}, but for 
the hypercharges shown in Case 2 of Table~\ref{table:ND1-12}.
}
\label{table:ND1-2}
\end{center}
\end{table}

For Scenario 5 with $z(H_d)=\frac{1}{2}$, 
we show in 
Tables~\ref{table:ND1-1-nolepto-Zeneg}~and~\ref{table:ND1-3-Zeneg} 
models with similar characteristics to those 
given above for Scenario 4 with $z(H_d)=0$.
We do the same for scenario 6 with
$z(H_d)=-\frac{1}{2}$, 
showing the models in
Tables~\ref{table:ND1-1-nolepto-Ze1}~and~\ref{table:ND1-3-Ze1}.

Finally, we studied Scenario 7 with $z(H_d)=-\frac{1}{4}$. 
The corresponding models can be found in
Tables~\ref{table:ND1-1-nolepto-Ze05} and~\ref{table:ND1-3-Ze05}.
{It is remarkable that the left column of Table~\ref{table:ND1-3-Ze05} corresponds
to a concrete model with the characteristic discussed above: the SM field $Q$ and the
extra quark doublet $\mathbb{D}$
have all equal gauge charges, since also $z(Q)=z(\mathbb{D})=7/36$. Therefore, terms are allowed in the superpotential 
of Eq.~(\ref{neww}) where $\hat{Q}$ and 
$\hat{\mathbb{D}}\equiv \hat Q_4$ are interchanged, i.e. we must
add 
the extra terms
\bea
W^{\text{extra}}_{\text{U$\mu\nu$SSM}}
=  
\lambda'_{i4k} \, \hat L_i \, \hat Q_{4}\, \hat d_k^c
+
Y^d_{4k} \, \hat H_d\, \hat Q_4 \, \hat d_k^c 
\ -
Y^u_{4k} \, \hat {H_u}\, \hat Q_4 \, \hat{u_k^c}
+
Y^{\mathbb{Q}}_{j4k'}  \,  \hat{Q}_j \, 
\hat{Q}^c_4 \, \hat{\nu^c}_{k'}\,.
\label{extra}
\eea
This model is a $U(1)'$ extension of the scenario proposed in
Ref.~\cite{Lopez-Fogliani:2017qzj} with a
fourth family of vector-like quark doublet representation, which due to anomaly cancellation has in addition 
the pair of singlets of $SU(2)$, $\hat{\mathbb{K}}$, $\hat{\mathbb{K}}^c$.
Compared to non-SUSY models, this model produces a variety of
new decay modes for the vector-like quarks involving the extra scalars present in SUSY, as discussed in
Ref.~\cite{Aguilar-Saavedra:2017giu}.
}

{Similar to the discussion above for the other solution of exotic quarks,
the solutions that can be generated here through the redefinitions
$y(\mathbb{K})\leftrightarrow -y(\mathbb{K})$,
$y(\mathbb{D})\leftrightarrow -y(\mathbb{D})$, 
$\mathbb{K} \leftrightarrow \mathbb{K}^c $, $\mathbb{D} \leftrightarrow \mathbb{D}^c$, are also equivalent.}

{Let us finally remark that rational solutions for the charges have been used so far in the discussion. This is expected if the $U(1)'$ is embedded in a simple group, however in general there is no reason for this to hold. As examples of this possibility, we show in Tables~\ref{table:ND1-2-irrational} and~\ref{table:ND1-3-irrational} models with irrational charges which have the same number of SM singlets as those in 
Tables~\ref{table:NK3_NR3-NDM2} and~\ref{table:NK3_NR-Odd-3}, respectively.  
They correspond to solution~(\ref{exotic2}) for the exotic quarks, 
hypercharges as in Case 2 of Table~\ref{table:ND1-12}
and Scenario 4 with $z(H_d)=0$. With these characteristics we have already shown the models in Table~\ref{table:ND1-2} with $n_{\nu^c} = 2$. However,
for $n_{\nu^c} = 3$ and the other singlets as in 
Tables~\ref{table:ND1-2-irrational} and~\ref{table:ND1-3-irrational}, it is not possible to find rational solutions. The same happens for $z(H_d)=\pm 1/2$ and also with hypercharges as those of Case 2 of 
Table~\ref{table:ND1-12}.
Nevertheless, we have checked that irrational solutions can be found in all these cases.

Summarizing, in this section we have built explicit models that clearly show that the expected phenomenology can be very diverse, with a variety of exotic quarks, extra singlets and couplings, in addition to the extra $Z'$.
The next sections are devoted to discuss their impact on the experimental analyses. 

\section{The scalar potential}
\label{scalarpotential}


The study of the scalar potential will allow us to discuss the spontaneous breaking of the gauge symmetry of U$\mu\nu$SSM models, $SU(3)\times SU(2)\times U(1)_Y \times U(1)' \to SU(3)\times U(1)_{\text{em}}$. The neutral scalar potential is the sum of three contributions: F-terms, D-terms and soft terms.
Working in the framework of a typical low-energy SUSY, and taking first into account the superpotential in Eq.~(\ref{Eq:superpotentialt}), the soft terms are given by: 
\bea
-\mathcal{L}_{\text{soft}}  =&&
\epsilon_{ab} \left(
T^e_{ij} \, H_d^a  \, \widetilde L^b_{iL}  \, \widetilde e_{jR}^* +
T^d_{ij} \, H_d^a\,   \widetilde Q^b_{iL} \, \widetilde d_{jR}^{*} 
+
T^u_{ij} \,  H_u^b \widetilde Q^a_{iL} \widetilde u_{jR}^*
+ \text{h.c.}
\right)
\nonumber \\
&+&
\epsilon_{ab} 
\left(T^{\lambda}_{ijk}\, \widetilde L_{iL}^a \, \widetilde L_{jL}^b \, \widetilde e_{kR}^*
+ 
T^{\lambda'}_{ijk} \, \widetilde L_{iL}^a \, \widetilde Q_{jL}^{b} \,  \widetilde d_{kR}^{*} 
\
+ \text{h.c.}
\right)
\nonumber\\
&+& 
\epsilon_{ab} \left(
T^{\nu}_{ij'} \, H_u^b \, \widetilde L^a_{iL} \widetilde \nu_{j'R}^*
- 
T^{\lambda}_{i'} \, \widetilde \nu_{i'R}^*
\, H_d^a  H_u^b
\
+ \text{h.c.}\right)
 \nonumber\\
&+&  
\left(T^{\kappa}_{\alpha j' k'}  {S}_{\alpha} 
\widetilde{\nu}_{j'R}^*
\widetilde{\nu}_{k'R}^* 
+  T^{\kappa'}_{\alpha' \alpha \beta} {N}_{\alpha'} {S}_{\alpha} {S}_{\beta}    + T^{\kappa''}_{i' \alpha'' \beta''} 
\widetilde{\nu}_{i'R}^*
 {\xi}_{\alpha''} {\xi}_{\beta''} + \text{h.c.}  \right) 
 \nonumber \\ 
&+& 
\left(T^{\mathbb{K}}_{i' j} 
\widetilde{\nu}_{i'R}^* \widetilde{\mathbb{K}}_j \widetilde{\mathbb{K}}_j^c + \text{h.c.}\right)
\nonumber\\
&+&
m_{\widetilde{\mathbb{K}}_{ij}}^2 \widetilde{\mathbb{K}}_i^* 
\widetilde{\mathbb{K}}_j  
+ m_{\widetilde{\mathbb{K}}^c_{i j}}^2  
\widetilde{\mathbb{K}}^{c*}_i 
\widetilde{\mathbb{K}}^{c}_j
\nonumber \\ 
&+&
m_{S_{\alpha\beta}}^2  {S}_{\alpha}^*  {S}_{\beta} + 
m_{N_{\alpha ' \beta '}}^2 {N}_{\alpha '}^* {N}_{\beta '} + 
m_{\xi_{\alpha '' \beta ''}}^2 {\xi}_{\alpha ''}^* {\xi}_{\beta ''}
+
m^2_{\widetilde{\nu}_{i'j'}}
\widetilde{\nu}_{i'R}^*
\widetilde{\nu}_{j'R} 
\nonumber \\
&+& 
m^2_{\widetilde{Q}_{ij}}
\widetilde{Q}_{iL}^{a*}
\widetilde{Q}^a_{jL}
+m^2_{\widetilde{u}_{ij}} \widetilde{u}_{iR}^*
\widetilde u_{jR}
+ m^2_{\widetilde{d}_{ij} } \widetilde{d}_{iR}^*
\widetilde d_{jR}
\nonumber\\
&+&
m^2_{\widetilde{L}_{ij}}
\widetilde{L}_{iL}^{a*}  
\widetilde{L}^a_{jL}
+
m^2_{\widetilde{e}_{ij}}  \widetilde{e}_{iR}^*
\widetilde e_{jR}
+ 
m_{H_d}^2 {H^a_d}^*
H^a_d + m_{H_u}^2 {H^a_u}^*
H^a_u
\nonumber \\
&+& 
\frac{1}{2}\, \left(M_3\, {\widetilde g}\, {\widetilde g}
+
M_2\, {\widetilde{W}}\, {\widetilde{W}}
+ 
M_1\, {\widetilde B}^0 \, {\widetilde B}^0 
+ 
M'_1\ {\widetilde B}^{0'} {\widetilde B}^{0'}  
+ \text{h.c.} \right)
\,.
\label{Eq:softW3K}
\eea
In case of following the result based on the breaking of supergravity, that 
all the soft trilinear parameters are proportional to their corresponding couplings
in the superpotential (for a review, see e.g. Ref.~\cite{Brignole:1997dp}), one can write
\bea
T^{e}_{ij} &=& A^{e}_{ij} Y^{e}_{ij}\ , \;\;\;\;\;\;\;\;\;\;\;\;
T^{d}_{ij} = A^{d}_{ij} Y^{d}_{ij}\ , \;\;\;\;\;\;\;\;\;\;\;\;\;\;\;
T^{u}_{ij'} = A^{u}_{ij'} Y^{u}_{ij'}\ ,
\nonumber
\\
T^{\nu}_{ij'} &=& A^{\nu}_{ij'} Y^{\nu}_{ij'}\ , \;\;\;\;\;\;\;\;\;\;\;
T^{\lambda}_{i'}= A^{\lambda}_{i'}\lambda_{i'}\ , \;\;\;\;\;\;\;\;\;\;\;\;\;\;\;
T^{\mathbb{K}}_{i'j}= A^{\mathbb{K}}_{i'j'} Y^{\mathbb{K}}_{i'j}
\ ,
\nonumber
\\
T^{\kappa}_{\alpha j'k'} &=& A^{\kappa}_{\alpha j'k'} \kappa_{\alpha j'k'}\ , \;\;
T^{\kappa'}_{\alpha' \alpha\beta} = A^{\kappa'}_{\alpha' \alpha\beta} 
\kappa'_{\alpha' \alpha\beta}\ , \;\;
T^{\kappa''}_{i' \alpha'' \beta''}= A^{\kappa''}_{i' \alpha'' \beta''} \kappa''_{i' \alpha'' \beta''}
\ ,
\label{tmunu2}
\eea
where $A$ is of the order of one TeV, and the summation convention on repeated indexes does not apply.

For the superpotential in Eq.~(\ref{Eq:superpotentialt2}), the fifth and sixth lines in
Eq.~(\ref{Eq:softW3K}) must be replaced by:
\bea
&+& 
\left(T^{\mathbb{D}}_{i'} \, 
\widetilde \nu_{i'R}^* \, \widetilde{\mathbb{D}} \, \widetilde{\mathbb{D}}^c + 
T^{\mathbb{K}}_{i'} \, \widetilde \nu_{i'R}^* \, \widetilde{\mathbb{K}} \, \widetilde{\mathbb{K}}^c + \text{h.c.} \right)   
\nonumber \\
&+&  
m_{\widetilde{\mathbb{D}}}^2 \,  \widetilde{\mathbb{D}}^* \, \widetilde{\mathbb{D}} + m_{\widetilde{\mathbb{D}}^c}^{2} \, \widetilde{\mathbb{D}}^{c*} \, \widetilde{\mathbb{D}}^c  + m_{\widetilde{\mathbb{K}}}^2 \, \widetilde{\mathbb{K}}^{*} \, \widetilde{\mathbb{K}} + m_{\widetilde{\mathbb{K}}^c}^{2} \, \widetilde{\mathbb{K}}^{c*} \, \widetilde{\mathbb{K}}^c \,,
\label{soft2}
\eea
with
\bea
T^{\mathbb{D}}_{i'}= A^{\mathbb{D}}_{i'} Y^{\mathbb{D}}_{i'}\ , \;\;\;\;\;\;\;\;\;
T^{\mathbb{K}}_{i'}= A^{\mathbb{K}}_{i'} Y^{\mathbb{K}}_{i'}
\ .
\label{tmunu3}
\eea

It is worth remarking that because of the $Z_2$ symmetry present in the
 superpotential of Eq.~(\ref{Eq:superpotential4}) containing
the singlet superfields under the SM gauge group of type $\hat\xi$, the latter can only appear 
in pairs in the Lagrangian.
As a consequence, 
if we consider the presence of only one of those superfields 
or several of them without mixing, it is straightforward 
to realize 
that a vanishing VEV for the scalar component $\xi$,  $\langle\xi\rangle = 0$, is a solution of the minimization equations. Thus the $Z_2$ symmetry is not broken spontaneously, but the other  
neutral scalars develop in general the following VEVs: 
\bea
 \langle H_d^0 \rangle &=& \frac{v_d}{\sqrt{2}}, \;\;\;\;  \langle H_u^0 \rangle = \frac{v_u}{\sqrt{2}}, \;\;\;\; \langle {\widetilde \nu}_{iL} \rangle= \frac{v_{iL}}{\sqrt{2}}, 
\nonumber \\
 \langle {\widetilde \nu}_{i'R}\rangle &=& \frac{v_{i'R}}{\sqrt{2}}, \;\;\;\; 
\langle S_{\alpha} \rangle = \frac{v_{\alpha S}}{\sqrt{2}}, \;\;\;\; \langle N_{\alpha'} \rangle= \frac{v_{\alpha' N}}{\sqrt{2}} \,. 
\label{vevs}
\eea
This is a simplifying assumption that will not essentially modify the following discussion, but can be helpful to have either the bosonic or the fermionic components of 
$\hat\xi$ as WIMP DM, as we will discuss below. In order to study the phenomenology associated to
superpotentials~(\ref{Eq:superpotentialt}) and~(\ref{Eq:superpotentialt2}), it is enough to consider the simple case of no mixing between generations,
as well as in the corresponding soft terms 
(\ref{Eq:softW3K}) and (\ref{soft2}), respectively, and to assume that only one generation of sneutrinos and of additional singlets under the SM gauge group get VEVs: 
${v_L}/{\sqrt{2}}$, ${v_R}/{\sqrt{2}}$, ${v_S}/{\sqrt{2}}$, ${v_N}/{\sqrt{2}}$. 
We will use this assumption in what follows.
The extension of the analysis to all generations is straightforward, and the conclusions are similar.
The expression of the tree-level neutral scalar potential is then given by:
\begin{eqnarray} \label{VEV potencial extra u1}
\langle V^0 \rangle 
&= &
 \frac{1}{4} \, \left\{ \,
\frac{1}{8}g_Z^2
\left(|v_d|^2+|v_L|^2-|v_u|^2\right)^2 
\right.
\nonumber \\ 
&+ &
\frac{1}{2}  g^2_{Z'}\left[z(H_d)|v_d|^2 + z(H_u) |v_u|^2 + z(L) |v_L|^2 + z(\nu ^c) |v_R|^2  + z(S) |v_S|^2  + z(N) |v_N|^2\right]^2 \nonumber \\ 
&+ &
|Y^\nu|^2\left(|v_u|^2|v_R|^2+|v_u|^2|v_L|^2 + |v_L |^2 |v_R|^2\right) \nonumber \\
&+ &
|\lambda |^2\left(|v_d|^2|v_u|^2+| v_R |^2|v_u|^2+| v_R |^2|v_d|^2\right) \nonumber \\
&+ &
|\kappa |^2 \left(4 \, | v_R |^2 |v_S|^2 + | v_R |^4\right) \nonumber \\
&+ &
|\kappa' |^2 \left(4 \, |v_S|^2 |v_N|^2 + |v_S|^4\right) \nonumber \\
&+ &
\left(-\lambda  Y^{\nu*} v_d v_L^* |v_u|^2-\lambda  Y^{\nu*} v_d v_L^* | v_R |^2 + \text{h.c.}\right) \, \nonumber \\
&+ &
\left.
{2} \left(\kappa Y^{\nu*} v_u^* v_L^* v_R v_S {-\lambda^*} \kappa   v_u^*  v_d^*  v_S v_R + 
{\kappa^*} \kappa' (v_R^*)^2 v_N v_S + \text{h.c.}\right)\right\} 
\nonumber \\ 
&+ &
\frac{1}{2} \, \left(m_{\widetilde{L}}^2 |v_L|^2 +
m_{\widetilde{\nu}}^2 |v_R|^2 + m_{N}^2|v_N|^2 + m_{S }^2|v_S|^2 
+ m_{H_d}^2|v_d|^2+m_{H_u}^2|v_u|^2\right) 
\nonumber \\
&+ & \frac{1}{2 \, \sqrt{2}} \, 
\left(T^\nu
v_u v_L v_R - 
T^{\lambda}
v_R v_d v_u
+ 
T^{\kappa} v_S v_R^2
+
T^{\kappa'}
v_S^2 v_N  + \text{h.c.} \right),
\end{eqnarray}
where $g$ and $g'$ are the $SU(2)$ and $U(1)_Y$ gauge couplings estimated at the $m_Z$ scale by
$e=g\sin\theta_W=g'\cos\theta_W$,
$g_{Z'}$ is the $U(1)'$ gauge coupling, and we have defined 
\bea
g_Z^2\equiv g^2 + g'^2.
\eea

Assuming CP conservation for simplicity, 
the six minimization conditions with respect to $v_d$, $v_u$, 
$v_R$, $v_S$, $v_N$ and $v_L$, are respectively: 
\begin{eqnarray}
&& \frac{1}{4}
g_Z ^2
(v_d^2 + v^2_L -v_u^2)v_d  \nonumber \\
&+& 
g_{Z'}^2
\left[z(H_d) v_d^2 + z(H_u) v_u^2 + z(L) v_L^2 + z(\nu ^c) v_R^2  + z(S) v_S^2 + z(N) v_N^2\right] z(H_d) v_d \nonumber \\
&+& \lambda ^2 v_d (v_u^2 + v_R^2) - \lambda  Y^\nu v_L {v_u}^2  - \lambda  Y^\nu v_L v_R^2  
{-2} \lambda \kappa v_S v_u v_R \nonumber \\
&+& 2 \, m_{H_d}^2 v_d - \sqrt{2} 
T^{\lambda}
v_R v_u = 0\,, 
\end{eqnarray}
\begin{eqnarray}
& -& \frac{1}{4}
g_Z^2
(v_d^2 + v_L^2-v_u^2) v_u \nonumber \\
&+& 
g_{Z'}^2
\left[z(H_d) v_d^2 + z(H_u) v_u^2 + z(L) v_L^2 + z(\nu ^c) v_R^2  + z(S) v_S^2 + z(N) v_N^2\right] z(H_u) v_u \nonumber \\
& +& {Y^\nu}^2 v_u (v_L^2 + v_R^2) + \lambda^2 v_u ( v_d^2 + v_R^2) - 2 \lambda  Y^\nu v_d v_L v_u 
{+2} \kappa Y^{\nu} v_L v_R v_S {-2} \lambda \kappa v_d v_S v_R   \nonumber \\
&+& 2 \, m_{H_u}^2 v_u + \sqrt{2} 
T^{\nu}
v_L v_R - \sqrt{2} 
T^{\lambda}
v_R v_d = 0\,,
\end{eqnarray}
\begin{eqnarray}
&& 
g_{Z'}^2
\left[z(H_d) v_d^2 + z(H_u) v_u^2 + z(L) v_L^2 + z(\nu ^c) v_R^2  + z(S) v_S^2 + z(N) v_N^2\right] {z(\nu ^c)}
v_R \nonumber \\
& +& \lambda ^2 v_R (v_d^2 + v_u^2)
+ \kappa^2 (4 v_R v_S^2 + 2 v_R^3  )     
+ {Y^\nu}^2 v_R (v_u^2 + v_L^2)  
  \nonumber \\
&{+}& 
\kappa Y^{\nu} v_u v_L v_S - 2\lambda  Y^\nu v_d v_L v_R 
{-} \lambda \kappa v_u v_d v_S + 4 \kappa \kappa' v_R v_N v_S 
\nonumber \\
&+& 2 \, 
m_{\widetilde{\nu}}^2
\, v_R + \sqrt{2} 
T^{\nu}
v_u v_L  - \sqrt{2}
T^{\lambda}
v_d v_u + {2}\sqrt{2} 
T^{\kappa}
v_S v_R  = 0\,, 
\label{rightsneutrinovev}
\end{eqnarray}
\begin{eqnarray}
&&
g_{Z'}^2
\left[z(H_d) v_d^2 + z(H_u) v_u^2 + z(L) v_L^2 + z(\nu ^c) v_R^2  + z(S) v_S^2 + z(N) v_N^2\right] {z(S)}
v_S  \nonumber \\
&+&  {4} \ \kappa^2  {v_R^2} v_S + {\kappa'^2}  ( {4} v_N^2 v_S + 2 v_S^3  )  {+2} \kappa Y^\nu v_u v_L v_R {-2} \lambda \kappa v_u v_d v_R + 2 \, \kappa \kappa' v_R^2 v_N  \nonumber \\
&+& 2 \, m_{S}^2 \, v_S + \sqrt{2} 
T^{\kappa}
v_R^2  + 2 \, \sqrt{2} 
T^{\kappa'}
 v_S \, v_N 
= 0\,,
\label{ese}
\end{eqnarray}
\begin{eqnarray}
&& 
g_{Z'}^2
\left[z(H_d) v_d^2 + z(H_u) v_u^2 + z(L) v_L^2 + z(\nu ^c) v_R^2  + z(S) v_S^2 + z(N) v_N^2\right]  z(N) v_N  \nonumber \\
&+& {4 \ \kappa'^2}   v_S^2 v_N  + 2 \, \kappa \kappa' v_R^2 v_S + \, {2} m_{N}^2 v_N+ \sqrt{2} 
T^{\kappa'}
v_S^{2} 
= 0\,, 
\label{ene}
\end{eqnarray}

\begin{eqnarray}
&& \frac{1}{4}
g_{Z}^2
(v_d^2 + v_L^2-v_u^2) v_L  \nonumber \\
&+& 
g_{Z'}^2
\left[z(H_d) v_d^2 + z(H_u) v_u^2 + z(L) v_L^2 + z(\nu ^c) v_R^2  + z(S) v_S^2 + z(N) v_N^2\right]{z(L)} v_L \nonumber \\
&+& {Y^\nu}^2 {v_L} (v_u^2 + v_R^2) -\lambda  Y^\nu v_d v_u^2 - \lambda  Y^\nu v_d v_R^2 + 2\kappa Y^{\nu} v_u v_R v_S
 \nonumber \\
&+& 2 \, m_{\tilde L} ^2 v_L + \sqrt{2} \, 
T^{\nu}
v_u v_R=0\,.
\label{ecuaciones minimo}
\end{eqnarray}
Note that apart from the presence of the singlets under the SM gauge group $S$ and $N$, these equations are similar to the minimization conditions for the $\mn$, 
where correct EWSB is known to take place~\cite{LopezFogliani:2005yw,Escudero:2008jg,Ghosh:2008yh,Bartl:2009an}. As in that model,
the scale of the soft terms is in the ballpark of one TeV and they induce the EWSB in the U$\mn$.
It is also worth noticing that whereas $v_R$, $v_S$ and $v_N$ are naturally of the order of TeV,
as it happens in the $\mn$ $v_L\sim 10^{-4}$ GeV~\cite{LopezFogliani:2005yw}. This small value
of $v_L$ from its minimization equation is because of the proportional contributions to $Y^{\nu}$.
These contributions enter through the F-terms and soft terms in the scalar potential
(assuming 
$T^{\nu}=A^{\nu}Y^{\nu}$ as in Eq.~(\ref{tmunu2})),
and 
are small
due to the generalized electroweak seesaw discussed in the Introduction that
determines $Y^{\nu}\lsim 10^{-6}$.
A simple estimation gives $v_{iL}\lsim m_{{\mathcal{D}_i}}$, with
$m_{{\mathcal{D}_{i}}}= Y^{\nu}_{i} {v_u}/{\sqrt 2}$
the Dirac masses for neutrinos.
The smallness of the left sneutrino VEVs for a correct description of the neutrino sector,
compatible with current data, has been shown in Refs.~\cite{LopezFogliani:2005yw,Escudero:2008jg,Ghosh:2008yh,Bartl:2009an,Fidalgo:2009dm,Ghosh:2010zi}.


\section{Masses and mass mixings}
\label{masses}

{The Higgses and left sneutrinos have non-vanishing $U(1)'$ charges, producing therefore the mixing of the SM $Z$ boson and the $Z'$ boson associated to the $U(1)'$.
The relevant terms of the covariant derivative appearing in the Lagrangian of U$\mn$ models are:}
\begin{eqnarray}
D_\mu = \partial_\mu - i g_Z T_3 Z_\mu -i g_{Z'} z(F) Z'_\mu,
\label{derivative}
\end{eqnarray}
{where $T_3$ is the third component of the isospin.
After EWSB, Higgses, left sneutrinos and singlet scalars under the SM gauge group charged under $U(1)'$ acquire VEVs as discussed
in Eq.~(\ref{vevs}), giving rise to the following
mass-squared matrix:
}
\begin{eqnarray}
\left(
\begin{array}{cc}
m_{ZZ}^2 & m_{ZZ'}^2 \\
m_{ZZ'}^2 & m_{Z'Z'}^2\ 
\end{array}
\right),
\label{MZZ'2x2}
\end{eqnarray}
where the entries are functions of the VEVs, gauge coupling constants and 
$U(1)'$ charges
 \begin{eqnarray}
m_{ZZ}^2 &=& \frac{1}{4}
g_{Z}^2 v^2,
\nonumber \\
m_{Z'Z'}^2 &=& g_{Z'}^2 \left[ z(H_u)^2 v_u^2 + z(H_d)^2 v_d^2 + 
z(L)^2 \sum_i v_{iL}^2 \right.
\nonumber\\
&+&
\left.
z(\nu ^c)^2 \sum_{i'}v_{i'R}^2 +
 z(S)^2 \sum_{\alpha} v^2_{{\alpha}S} 
+ z(N)^2 \sum_{\alpha'} v^2_{{\alpha'}N} \right], 
\nonumber \\
m_{ZZ'}^2 &=& \frac{1}{2}g_{Z'}
g_Z
\left[z(H_u) v_u^2-z(H_d) v_d^2 -z(L)\sum_i v_{iL}^2 \right],
\label{entries}
\end{eqnarray}
and we have not included the effect of a potential kinetic mixing which is usually negligible~\cite{Langacker:2008yv}.
Here
$v^2 \equiv v_d^2 + v_u^2 + 
\sum_i v^2_{iL}
=({2 m_W}/{g})^2\approx$ (246 GeV)$^2$.

As in the SM, 
this matrix can be diagonalized by a rotation of the fields $Z$ and $Z'$ around the mixing angle $\theta_{\text{mix}}$.
Then, the eigenvectors $Z_1$ and $Z_2$ are a combination of $Z$ and $Z'$, and the mixing angle is related to the entries of the mass matrix~(\ref{MZZ'2x2}) as
\begin{equation}
\tan 2 \theta_{\text{mix}} = \frac{2 m_{ZZ'}^2}{M^2_{Z'Z'} -   M^2_{ZZ}} = \frac{g_{Z'} g_Z v^2} { (M^2_{Z'Z'} -   M^2_{ZZ}) } z_\text{mix},
\label{tanz}
\end{equation}
where
\begin{equation}
z_\text{mix} = z(H_u) \sin^2 \beta - z(H_d) \cos^2 \beta = \frac{z(H_u) \tan^2 \beta - z(H_d)}{\tan^2 \beta + 1}.
\label{ec:zmix}
\end{equation}
Here we have defined
$\tan\beta= {v_u}/{v_d}$, and 
since $v_{iL} \ll v_d, v_u$, we have also used 
$v^2 \approx v_d^2 + v_u^2$.
As we can see, in the limit of large $\tan\beta$, $z_{\text{mix}}\to z(H_u)$.

There exist strong experimental constraints on how large $\theta_{\text{mix}}$ can be, {mainly stemming from precise measurements of the $Z$ boson couplings to fermions performed at the $Z$ pole at LEP (see for example Ref.~\cite{Erler:2009jh}).} In a realistic scenario, the mixing angle is very small
and therefore $m_{Z_1}\approx m_{ZZ}\approx m_Z$, $m_{Z_2}\approx m_{Z'Z'}\approx m_{Z'}$ and $Z_1\approx Z$, $Z_2\approx Z'$.
Thus, $\theta_{\text{mix}}$ can be approximated by:
\begin{equation}
\theta_{\text{mix}} \approx \frac{g_{Z'} g_Z v^2} { 2(m^2_{Z'} -   m^2_{Z}) } z_\text{mix}.
\label{tanzz}
\end{equation}

\vspace{0.25cm}

\noindent
Let us
focus now our attention on the neutralino sector. 
In U$\mu\nu$SSM models, because of the $R$-parity violation the neutralinos, including the extra gaugino,
mix with LH and RH neutrinos, and with the other singlets under the SM gauge group. 
Of course, now we have to be sure that one eigenvalue of this matrix is very small, reproducing the experimental results on neutrino masses.

Working in the basis of 2-component spinors,\footnote{For a description of the notation used, see Appendix B of Ref.~\cite{Ghosh:2017yeh}.}
the neutral fermions have the flavor composition 
${\psi^{0}}^T=({(\nu_{L})^c}^*\ 
\widetilde Z'\ 
\widetilde B^0\ 
\widetilde W^{0}\ \widetilde H_{d}^0\ \widetilde H_{u}^0\ \nu^*_{R}\ 
\widetilde S\ \widetilde N)$,
and one obtains the mass terms in the Lagrangian, 
$-\frac{1}{2} {\psi^{0}}^T 
{m}_{\psi^0} 
\psi^0 + \mathrm{h.c.}$,
with ${m}_{\psi^0}$ a $9\times 9$ (symmetric) neutrino/singlino/neutralino mass matrix
\begin{equation}
{m}_{\psi^0} 
=\left(\begin{array}{cc}
 0 & m^{T}\\
m & {\mathcal M} \end{array}\right).
\label{matrizse}
\end{equation}
In this matrix,
$m$ is a $8\times 1$ submatrix containing the mixing of the LH neutrino with the neutralinos, the RH neutrino and the extra singlinos $\widetilde S$ and $\widetilde N$:
\begin{eqnarray}
m^T=\left(\ g_{Z'} z(L) v_L \  \ -\frac{1}{2}g' v_L \  \ \frac{1}{2}g v_L \  \ 0 \  \ \frac{Y^\nu v_R}{\sqrt{2}} \  \ \frac{Y^\nu v_u}{\sqrt{2}} \  \ 0 \  \ 0 \ \right).
\end{eqnarray}
${\mathcal M}$ is a $8\times 8$ submatrix 
containing the mixing of the neutralinos with the RH neutrino and the extra singlinos $\widetilde S$ and $\widetilde N$:
\begin{eqnarray}
\scriptsize
{\mathcal M}=\left( \begin{array}{cccccccc}  M'_1 & 0 & 0 & g_{Z'} z(H_d) v_d &  g_{Z'} z(H_u) v_u &  g_{Z'} z(\nu ^c) v_R &  g_{Z'} z(S) v_S &  g_{Z'} z(N) v_N 
\\
0 & M_1 & 0 & -\frac{g'}{2}v_d & \frac{g'}{2}v_u & 0 & 0 & 0 
\\
0 & 0 & M_2 & \frac{g}{2}v_d & - \frac{g}{2}v_u & 0 & 0 & 0 
\\
g_{Z'} z(H_d) v_d & -\frac{g'}{2}v_d & \frac{g }{2}v_d & 0 & 
-\frac{\lambda}{\sqrt 2}v_R & -\frac{\lambda}{\sqrt 2}v_u & 0 & 0 
\\ 
g_{Z'} z(H_u) v_u & \frac{g'}{2}v_u & -\frac{ g}{2}v_u & 
-\frac{\lambda}{\sqrt 2}v_R & 0 & 
-\frac{\lambda}{\sqrt 2} v_d + \frac{Y^{\nu}}{\sqrt 2} v_L & 0 & 0
\\
g_{Z'} z(\nu^c) v_R & 0 & 0 & -\frac{\lambda}{\sqrt 2}v_u & 
-\frac{\lambda}{\sqrt 2} v_d + \frac{Y^{\nu}}{\sqrt 2} v_L &
{\sqrt{2}}\kappa v_S &
{\sqrt{2}}\kappa v_R
& 0
\\
g_{Z'} z(S) v_S & 0 & 0 & 0 & 0 & 
{\sqrt{2}}\kappa v_R
& \sqrt{2} {\kappa'} v_N & \sqrt{2} {\kappa'} v_S 
\\
g_{Z'} z(N) v_N & 0 & 0 & 0 & 0 & 0 & \sqrt{2} {\kappa'} v_S & 0
\\
\end{array} \right)
\nonumber 
\label{Matriz Neutralinos}
\end{eqnarray}
\vspace*{-1.3cm}

\bea
\eea

It is relevant to note that, similarly to the $\mn$, the neutral fermion mass matrix ${m}_{\psi^0}$  has the structure of a generalized electroweak seesaw. 
Note in this respect that the entries of ${\mathcal M}$
are of the order of about one TeV. 
On the contrary, the entries of the matrix $m$ are much smaller
being determined mainly by the
neutrino Yukawa coupling and the left sneutrino VEV, 
which are very small 
as discussed above, $Y^\nu\lsim 10^{-6}$, $v_L\lsim 10^{-4}$ GeV.
Thus, as expected we obtain a very small mass for the light neutrino (actually, three small masses in the case of including in the analysis the three generations).
This is what happens in some of the models of Sec.~\ref{Section:models}, 
such as e.g. the one of Table~\ref{table:NK3_NR-Odd-3}, where Majorana masses are generated by the present couplings of the singlet under the SM gauge group $S$ (see the discussion in Eq.~(\ref{couplingsnunu})).

{
However, other constructions can give rise to a different phenomenology. For example, the model 
of Table~\ref{table:NK3_NR3-NDM2} leads to a situation similar to that of Refs.~\cite{Ghosh:2010hy,Barger:2010iv,Fidalgo:2011tm}, where four light particles are present because of the absence of Majorana masses for the RH neutrinos (see the discussion below Eq.~(\ref{superk})).
Thus, this particular extension predicts the existence of two heavy RH neutrinos of the order of TeV, and four light (three active and one sterile) neutrinos.
}


\vspace{0.25cm}

\noindent
{Finally, we have also performed an estimation of the tree-level mass of the 
{SM-like Higgs} in these models. Let us remember that neglecting the
{mixing of the SM-like Higgs with the right sneutrinos, and the} small neutrino Yukawa coupling effects, the expression of the 
mass in the $\mu\nu$SSM is similar to the one of the NMSSM once we define 
${\bm \la}\equiv \sqrt{\sum_{i'} \lambda^2_{i'}}$.
This is given by the following tree-level expression \cite{Escudero:2008jg}:
\begin{eqnarray}
m_h^2 = 
m^2_Z \cos^2 2\beta + 
({v}/{\sqrt 2})^2\
{\bm \lambda}^2 \sin^2 2\beta
\,.
\end{eqnarray}
This mass receives a positive contribution from the $U(1)'$ sector \cite{Morrissey:2005uz},
in such a way 
that the tree-level formula for the mass of the SM-like Higgs in 
U$\mu\nu$SSM models is given by:
\begin{eqnarray}
m_h^2 =
m^2_Z \cos^2 2\beta + 
({v}/{\sqrt 2})^2\
{\bm \lambda}^2 \sin^2 2\beta
 + g_{Z'} v^2 \left[z(H_u) \cos^2 \beta+ z(H_d) \sin^2 \beta\right]^2\,.
\label{upper new}
\end{eqnarray}
Thus, the addition of the $U(1)'$ gauge group
to the $\mn$ has also the interesting feature of increasing the Higgs mass, relaxing therefore the constraints on SUSY spectra.
Note that effects lowering (raising) the tree-level mass appear when the SM-like Higgs mixes with heavier (lighter) right sneutrinos.
}

\section{Present bounds}
\label{ephenomenology}

Limits on exotic quarks/squarks and $Z'$ masses and couplings arise from direct searches at colliders. In addition, limits on $Z-Z'$ mixing stem from precision electroweak data. In this section, we will apply them to extract bounds on the parameter space of U$\mn$ models.


\subsection{Constraints from the LHC}

{
As discussed in previous sections, the presence of exotic quarks/squarks in the spectrum of U$\mn$ models is mandatory.
This type of particles can be produced at the LHC, and in the cases when they do not couple to ordinary quarks it is sensible to assume that they will hadronize
inside the detector into color-singlet states, known in the literature as R-hadrons. Thus, bound states of exotic {quarks/squarks} combined with SM quarks can be produced at the LHC (for a review, see e.g. Ref.~\cite{Kang:2007ib}). Unless these R-hadrons decay via non-renormalizable operators in specific constructions, we expect them to be stable and therefore {the current bounds at the LHC on their exotic constituents are of about 1.2 TeV~\cite{Aaboud:2019trc}}. In case the new quarks couple to SM quarks, their production (singly or in pairs) and decay gives various signals with multiple $b$ and top quarks, see Ref.~\cite{Aguilar-Saavedra:2017giu}.
As we know from the discussion of Section~\ref{munuSSM}, the VEVs of the right sneutrinos  $v_{i'R}$ are crucial to determine the masses of this exotic matter of U$\mn$ models (see Eq.~(\ref{exoticquarks})). 
Thus, values of $v_{i'R}$ of the order of TeV or larger, obtained from the minimization of the scalar potential (see Eq.~(\ref{rightsneutrinovev})), can fulfill the current bounds.
}

\vspace{0.25cm}

\noindent
Let us now discuss the limits on $Z'$ masses.
A $Z'$ can be discovered at the LHC through the Drell-Yan production including the following SM final states:
$Z' \to \ell \ell$~\cite{Aad:2019fac} (with $\ell = e, \mu$), $Z' \to jj$~\cite{Sirunyan:2019vgj} (with $j$ a light quark), $Z' \to t \bar t$~\cite{Aad:2020kop}, $Z' \to W W$~\cite{Aad:2019fbh} and $Z' \to Z h$~\cite{CMS:2020qrs}.
In this subsection, we will study how these limits translate into bounds on the $Z'$ mass 
of U$\mn$ models. 

For this analysis, we need to know first the $Z'$ decay widths 
(for a review, see e.g. Ref.~\cite{Leike:1998wr}).
{One
can obtain them in the case of Dirac fermions $f$ using the following
neutral current interactions:}
\begin{equation}
\mathcal{L}^{Z'}_{\text{NC}} = g_{Z'} Z'_{\mu} \sum_f \bar{f} \gamma^{\mu}  
\left[z(f)P_L - z(f^c) P_R \right] f, 
\end{equation}
where
$z(f)$ ($-z(f^c)$) are the $U(1)'$ charges for the left (right) chiral fermions.
Then, one is able to obtain the following $Z'$ decay widths: 
\begin{equation}
 \Gamma_{Z' \to f \bar{f}}= C_f \, m_{Z'} \, \frac{g^2_{Z'} }{{24}\pi} \left[z(f)^2 + z(f^c)^2\right],
 \label{fermions}
\end{equation}
where $C_f$ is the color factor (1 for color singlets and 3 for triplets), and the fermion masses have been neglected (formulas including fermion mass effects can be 
found in Ref.~\cite{Kang:2004bz}).
{In this approximation, one finds for the $Z'$ decays to SM fermions:}

\begin{eqnarray}
\sum_f \Gamma_{Z' \to f \bar{f}} & = & m_{Z'} \, \frac{g^2_{Z'} }{{24}\pi} 
\left\{9 \left[ z(Q)^2 + z(u^c)^2 \right] + 9 \left[ z(Q)^2   + z(d^c)^2\right]
+ 3 \left[ z(L)^2 + z(e^c)^2\right] \right.
\nonumber\\
&+& \left. 3  z(L)^2    \right\},
 \label{totalwidth}
\end{eqnarray}
{where the last term $3 z(L)^2$ corresponds to the contribution of the LH neutrinos. Although they are in fact mixed with RH neutrinos and neutralinos in the mass matrix discussed in Eq.~(\ref{matrizse}), in a good approximation they are almost pure LH neutrinos and the formula (\ref{fermions}) can be use without the contribution $z(\nu^c)$.
}

{Other $Z'$ decays to SM particles are present, although typically with small branching ratios. A pure $Z'$ has no couplings to $W$ bosons, unlike the $Z$ in the SM. However, as discussed in 
Sec.~\ref{masses} the $Z$ and the $Z'$ are mixed. 
Thus, the $Z$ component contained in the mass eigenstates $Z_{1,2}$ interacts with $W$ bosons.} The partial decay width of the $Z_2\approx Z'$ to a $W^+W^-$ pair can be approximated as
\begin{equation}
 \Gamma_{Z' \to W^{+} W^-} = \, m_{Z'} \, \frac{
 g^2_ Z\theta^2_{\text{mix}} }{192 \pi} \left( \frac{m_{Z'}}{m_{Z}} \right)^4
= m_{Z'} \frac{g_{Z'}^2}{48 \pi}  z_\text{mix}^2 .
 \label{ws}
\end{equation}
As expected, it is suppressed by the square of the small mixing angle $\theta_{\text{mix}}\propto m_Z^2/m_{Z'}^2$ (see Eq.~(\ref{tanzz})), which 
compensates the huge factor $(m_{Z'}/m_Z)^4$. For $Z'$ much heavier than the weak and Higgs bosons, we have in the alignment limit 
\begin{equation}
 \Gamma_{Z' \to Z h}=  \Gamma_{Z' \to W^{+} W^-}.
 \label{zh}
\end{equation}
Therefore, these two channels contribute each with $z_\text{mix}^2/2$ to the sum in Eq.~(\ref{totalwidth}), when computing the total decay width.

The presence of additional channels depends on the spectrum of the models. For example, the decay into $hA$, i.e.
the neutral CP-even SM-like Higgs 
and a CP-odd Higgs (both with doublet-like composition) is allowed if the latter is 
light enough. 
In U$\mn$ models, similar to the case of the $\mn$ there are seven neutral pseudoscalar states
from the mixing between Higgses and sneutrinos. However, the three left sneutrinos are almost decoupled, and we are left with the doublet-like pseudoscalar and the three right sneutrinos
contributing to this channel. The latter states can be even lighter than the SM-like 
Higgs~\cite{Kpatcha:2019qsz,Lopez-Fogliani:2020gzo}.
{The predictions of these scenarios could be constrained with the ATLAS searches~\cite{Aaboud:2017ecz}.}

Besides, $Z'$ decays into SUSY partners might be kinematically allowed, such as decays into
Majorana fermions or sfermions~\cite{Barger:1987xw,Haber:1984rc,Kang:2004bz}.
Concerning the latter decays, if the right sneutrinos are light as discussed above, $Z'$ decays to a right sneutrino pair can be sizeable. Also the left sneutrinos can be light, as discussed in Refs.~\cite{Ghosh:2017yeh,Lara:2018rwv,Kpatcha:2019gmq}, and therefore decays to a left sneutrino pair can be interesting to consider.

In addition, other particles present in U$\mn$ models might allow other $Z'$ decays, such as decays into the
exotic quarks or into the fermionic partners of the 
singlets under the SM gauge group.
However, we expect their masses to be large enough as not to contribute significantly to the analysis.

The $Z'$ production cross section depends on the $Z'$ mass and coupling, as well as on the $\text{U}(1)'$ charges of the quarks. On the other hand, the branching ratios into the SM final states depend on other details of the model.
The presence of other decay modes involving new particles as those discussed above reduces the branching ratio into SM final states, relaxing the constraints from those searches. Thus, we will conservatively ignore these other possible decay modes in order to discuss the limits,
and therefore we will work with the total decay width 
$\Gamma_{Z'}$
given by:
\begin{equation}
\Gamma_{Z'} = \sum_f \Gamma_{Z' \to f \bar{f}} +  \Gamma_{Z' \to W^{+} W^-} +  \Gamma_{Z'\to Z h}.
\label{gamma}
\end{equation}
The signals involving $Z'$ decay into new particles and their observability will be discussed in the next section.

Since the signals studied here depend only on the $U(1)'$ charges of the SM particles, our analysis will be focused on the 
seven scenarios built in 
Tables~\ref{table:NK3-general} and~\ref{table:NK1-ND1-general-lepto}
of
Sec.~\ref{Section:models} with different values for these charges.

The branching ratio into SM final states depends on $\tan \beta$, via the widths $\Gamma(Z' \to W^+ W^-)$ and $\Gamma(Z' \to Zh)$, which are proportional to the factor $z_\text{mix}^2$,
see Eq.~(\ref{ws}). For reference, we will fix $\tan \beta = 2$. For other values, the mixing angle scales with the factor
\begin{equation}
k_{\theta_{\text{mix}}} \equiv \frac{5}{4z(H_u) - z(H_d)}  \frac{z(H_u) \tan^2 \beta - z(H_d)}{\tan^2 \beta + 1} \,,
\label{ec:Mfac}
\end{equation}
shown in Fig.~\ref{fig:Mfac} for the different scenarios. 
In the limit of large $\tan\beta$, $k_{\theta_{\text{mix}}}\to 5z(H_u)/(4z(H_u) - z(H_d))$.
The widths $\Gamma(Z' \to W W)$ and $\Gamma(Z' \to Zh)$ scale with $k^2_{\theta_{\text{mix}}}$. Since these two decay modes are subdominant the $Z'$ branching ratios into fermionic final states practically are independent of $\beta$, and the BRs into $WW$ and $Zh$ approximately scale with $k^2_{\theta_{\text{mix}}}$.

\begin{figure}[t!]
\begin{center}
\includegraphics[height=5.4cm,clip=]{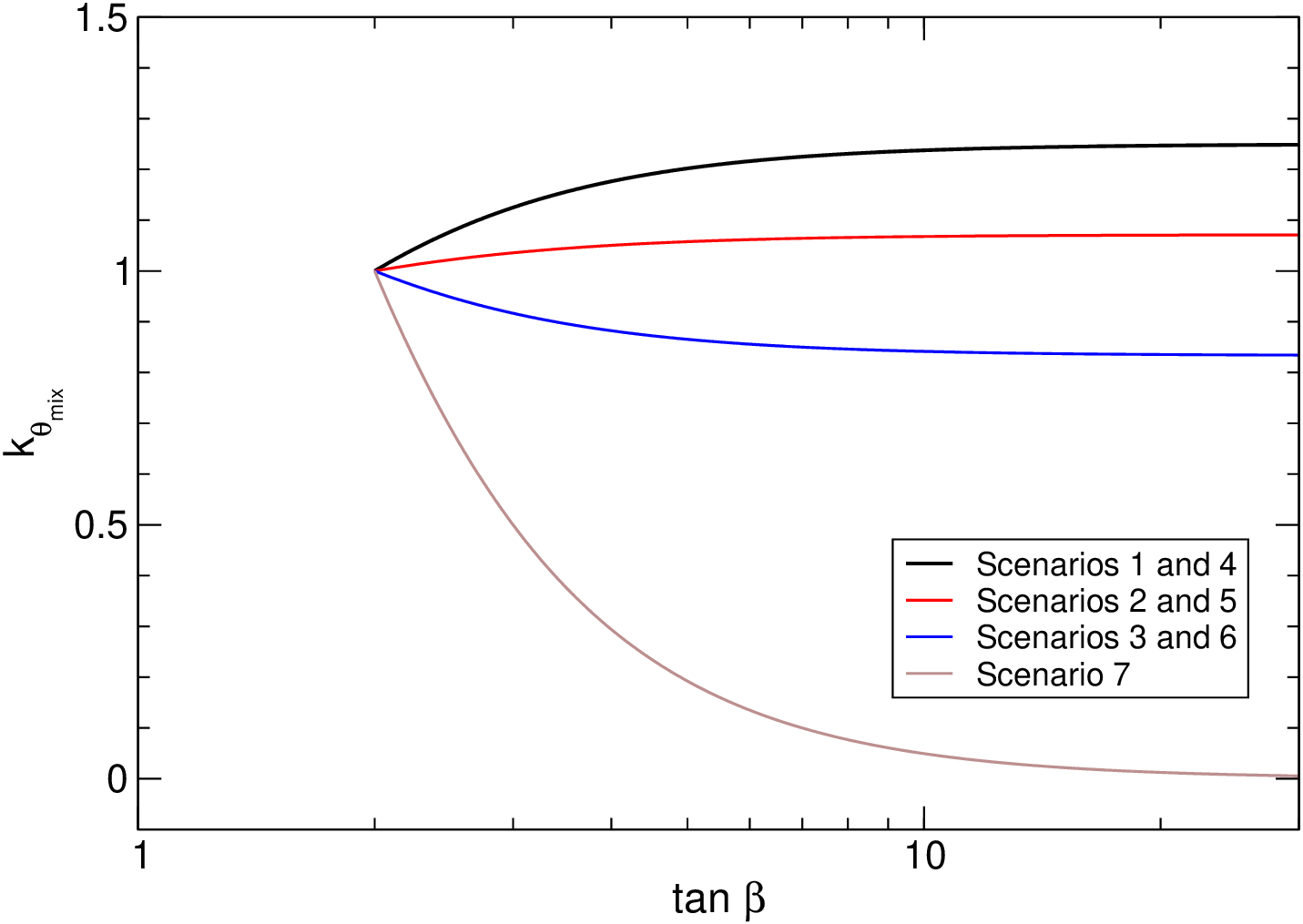} 
\caption{Scale factor for the mixing angle $\theta_{\text{mix}}$, defined in Eq.~(\ref{ec:Mfac}), as a function of $\tan \beta$ {in logarithmic scale}.}
\label{fig:Mfac}
\end{center}
\end{figure}

\begin{figure}[t!]
\begin{center}
\includegraphics[height=5.4cm,clip=]{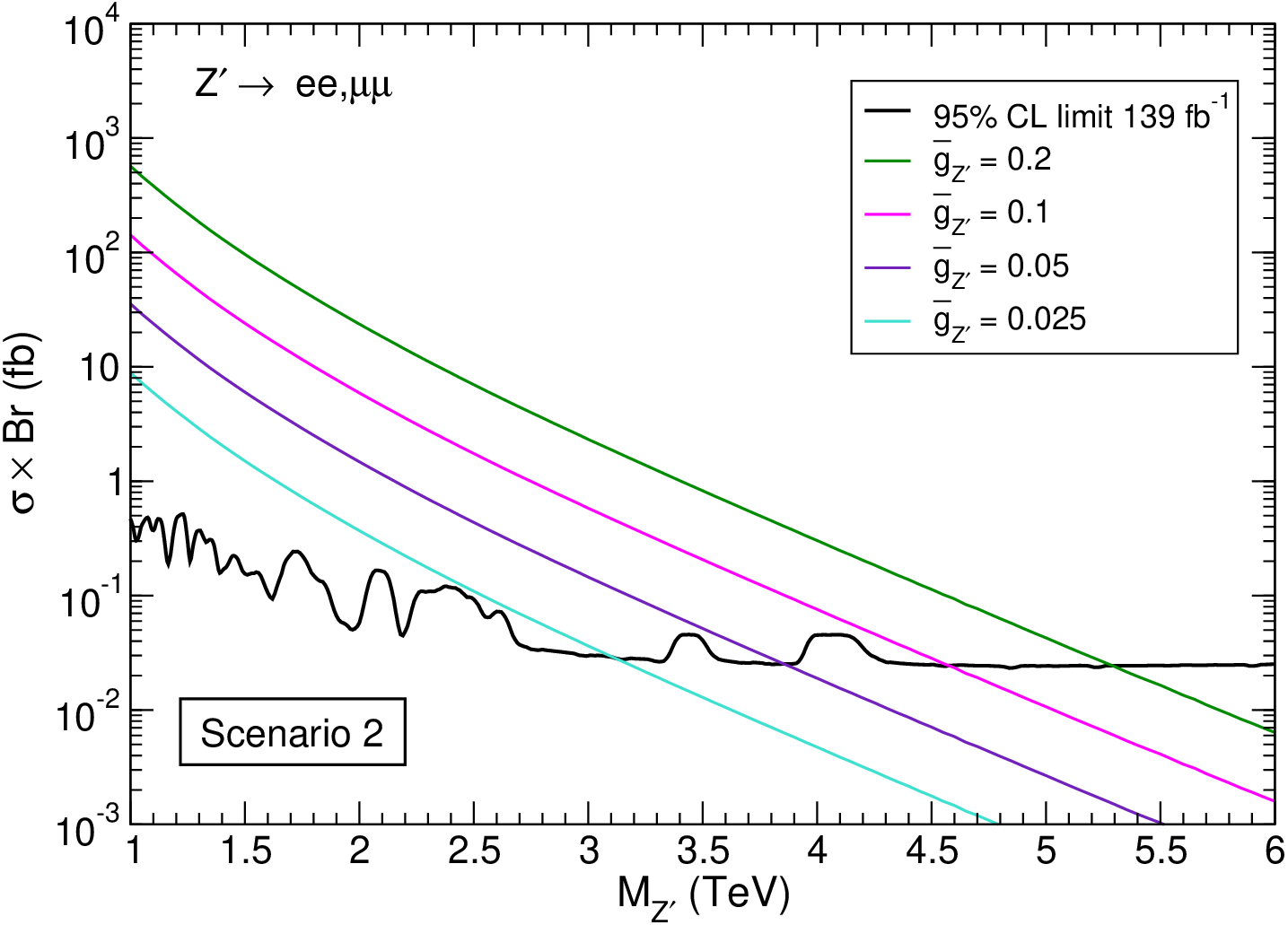} 
\caption{Limits on $Z'$ bosons arising from the search in Ref.~\cite{Aad:2019fac} in the dilepton final state. Together, we show the cross section predictions for Scenario 2
of Table~\ref{table:NK3-general}, with $\bar g_{Z'} \equiv g_{Z'}$ by definition. For the rest of scenarios the cross sections can be reinterpreted according to (\ref{ec:dicc}) and Table~\ref{tab:dicc2}.}
\label{fig:lim2}
\end{center}
\end{figure}

Searches for $Z' \to \ell \ell$ provide the strongest constraints except for scenarios 1 and 4 where the $Z'$ is leptophobic. The 95\% confidence level (CL) upper limit on $\sigma(pp \to Z') \times \text{Br}(Z' \to \ell \ell)$ from Ref.~\cite{Aad:2019fac} is presented in Fig.~\ref{fig:lim2}. We also show the cross section prediction for Scenario 2 for couplings $\bar g_{Z'} \equiv g_{Z'} = 0,025, 0,05, 0,1, 0.2$. 
For other scenarios the cross section predictions in Fig.~\ref{fig:lim2} can easily be reinterpreted by using the modified coupling
\begin{equation}
\bar g_{Z'} = g_{Z'} \times k_0 \left( 1 +  k_1 \frac{m_{Z'}}{\text{1 TeV}} \right) \,,
\label{ec:dicc}
\end{equation}
with $k_0$ and $k_1$ given in Table~\ref{tab:dicc2}. Note that because $k_1 \ll 1$ the mass depencence of this rescaling is very weak.

\begin{table}[t!]
\begin{center}
\begin{tabular}{|c|cc|cc|cc|cc|}
\hline
& \multicolumn{2}{c|}{Scenario 3}  & \multicolumn{2}{c|}{Scenario 5} & \multicolumn{2}{c|}{Scenario 6} & \multicolumn{2}{c|}{Scenario 7} \\
\hline 
& $k_0$ & $k_1$ & $k_0$ & $k_1$ & $k_0$ & $k_1$ & $k_0$ & $k_1$   \\
\hline
$\ell^+ \ell^-$    & 0.80 & -0.014 & 0.97 & -0.004  & 0.87 & -0.008 & 0.44 & -0.020 \\
\hline
\end{tabular}
\caption{Numerical factors used to compute the modified coupling $\bar g_{Z'}$ to reinterpret the cross section predictions in Fig.~\ref{fig:lim2}. }
\label{tab:dicc2}
\end{center}
\end{table}

Limits on the $Z'$ boson mass and couplings for Scenario 1 from dijet~\cite{Sirunyan:2018xlo,Sirunyan:2019vgj}, $t \bar t$~\cite{Aaboud:2018mjh,Aad:2020kop}, diboson~\cite{Aad:2019fbh} and $Zh$~\cite{CMS:2020qrs} resonance searches are presented in Fig.~\ref{fig:lim1}. We show the cross sections times BR for the corresponding final state (shown in the upper left corner of the plots) corresponding to couplings $\bar g_{Z'} \equiv g_{Z'} = 0.2, 0.4, 0.8$, as a function of the $Z'$ mass. For the remaining scenarios the cross section predictions can be obtained by using (\ref{ec:dicc}) with the values of $k_0$, $k_1$ given in Table~\ref{tab:dicc1}. It is found that in most cases this coupling rescaling is almost independent of the $Z'$ mass in the range of interest, and for the rest the dependence is quite weak.  In all cases the strongest constraints result from the diboson and $Zh$ final states. As discussed above, for other values of $\tan \beta$ the cross section times BR approximately scales with $k_\theta^2$.

\begin{figure}[t!]
\begin{center}
\begin{tabular}{ccc}
\includegraphics[height=5.4cm,clip=]{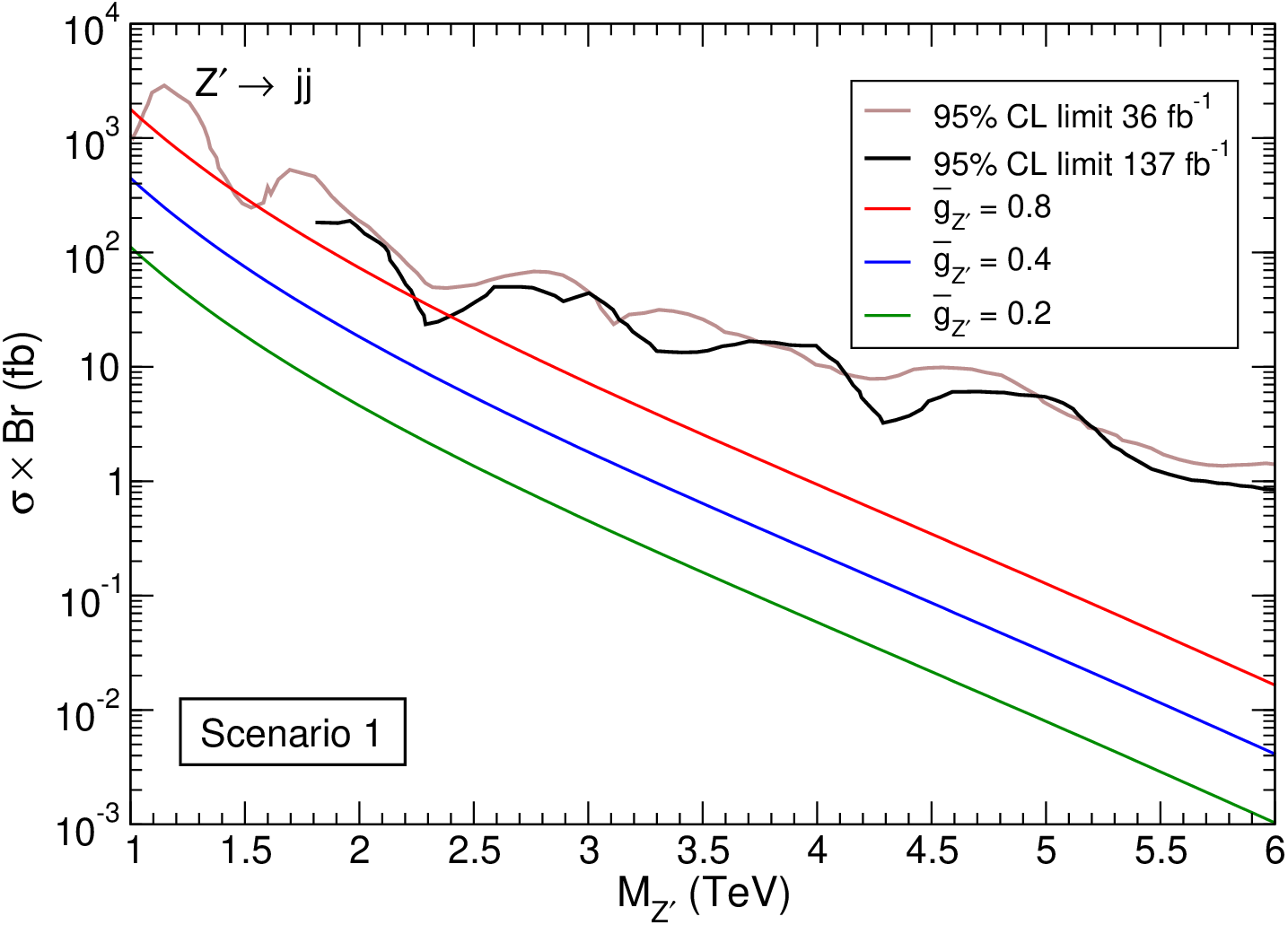} 
& & \includegraphics[height=5.4cm,clip=]{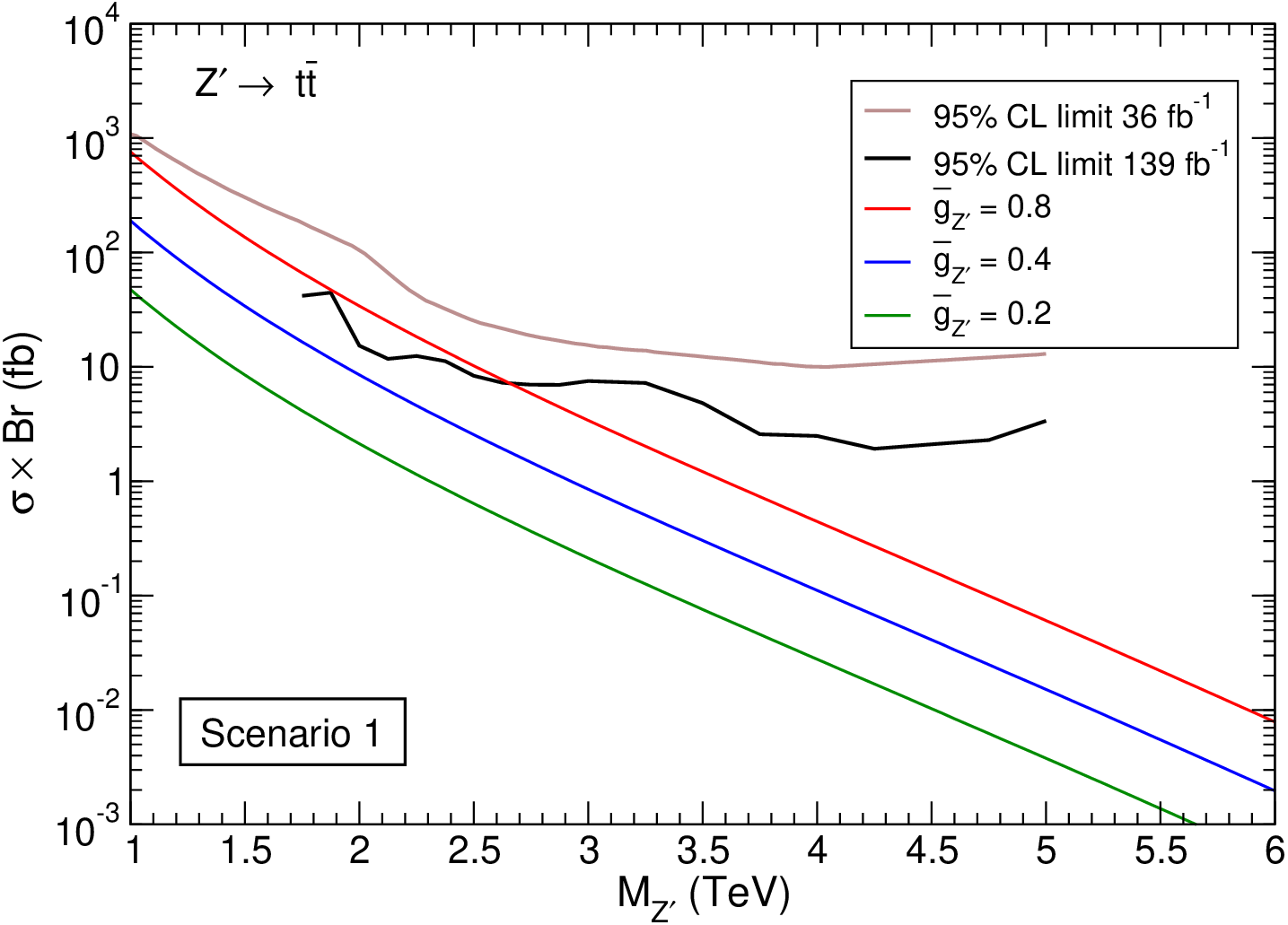} 
\\
\includegraphics[height=5.4cm,clip=]{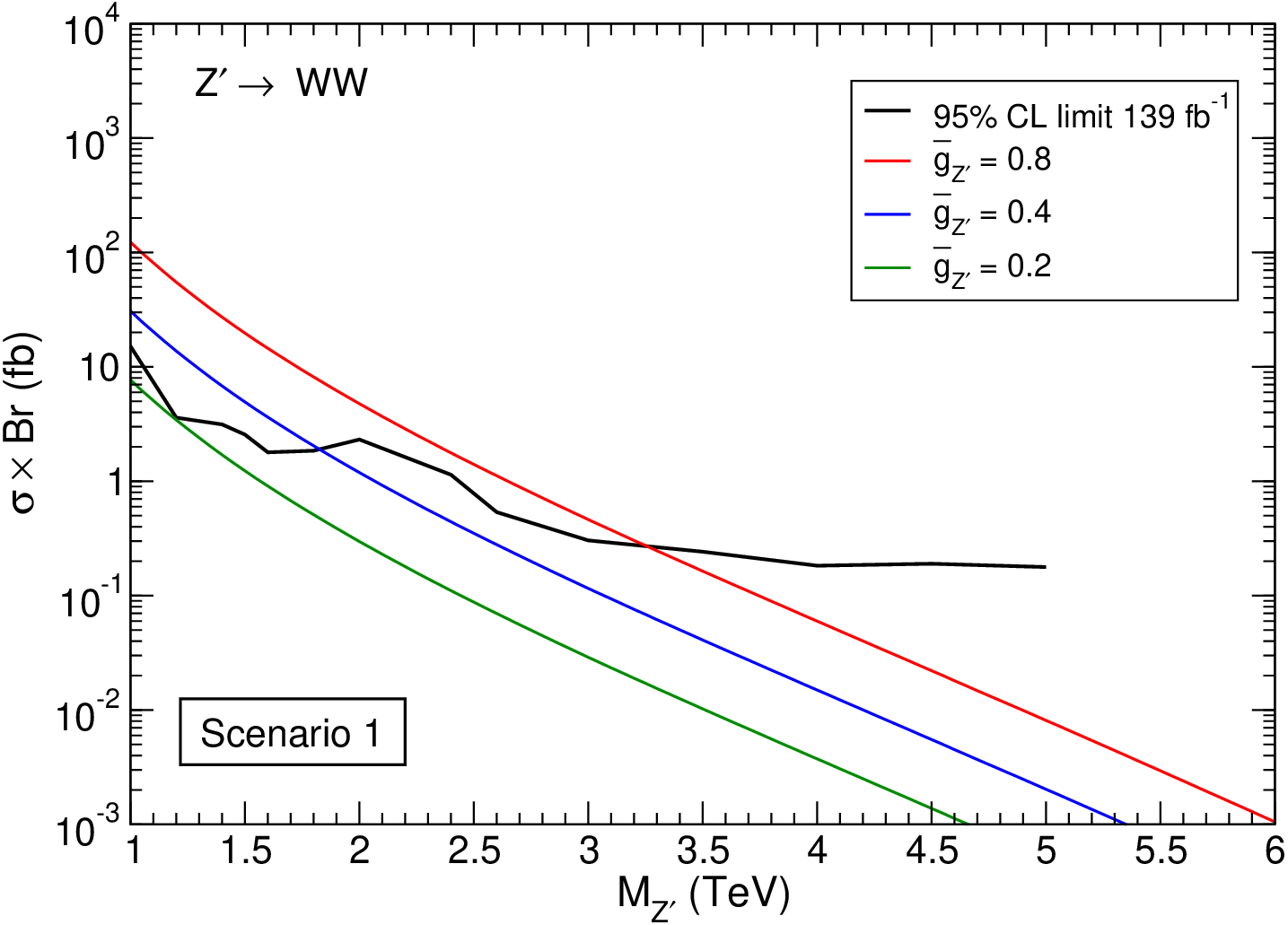} 
& & \includegraphics[height=5.4cm,clip=]{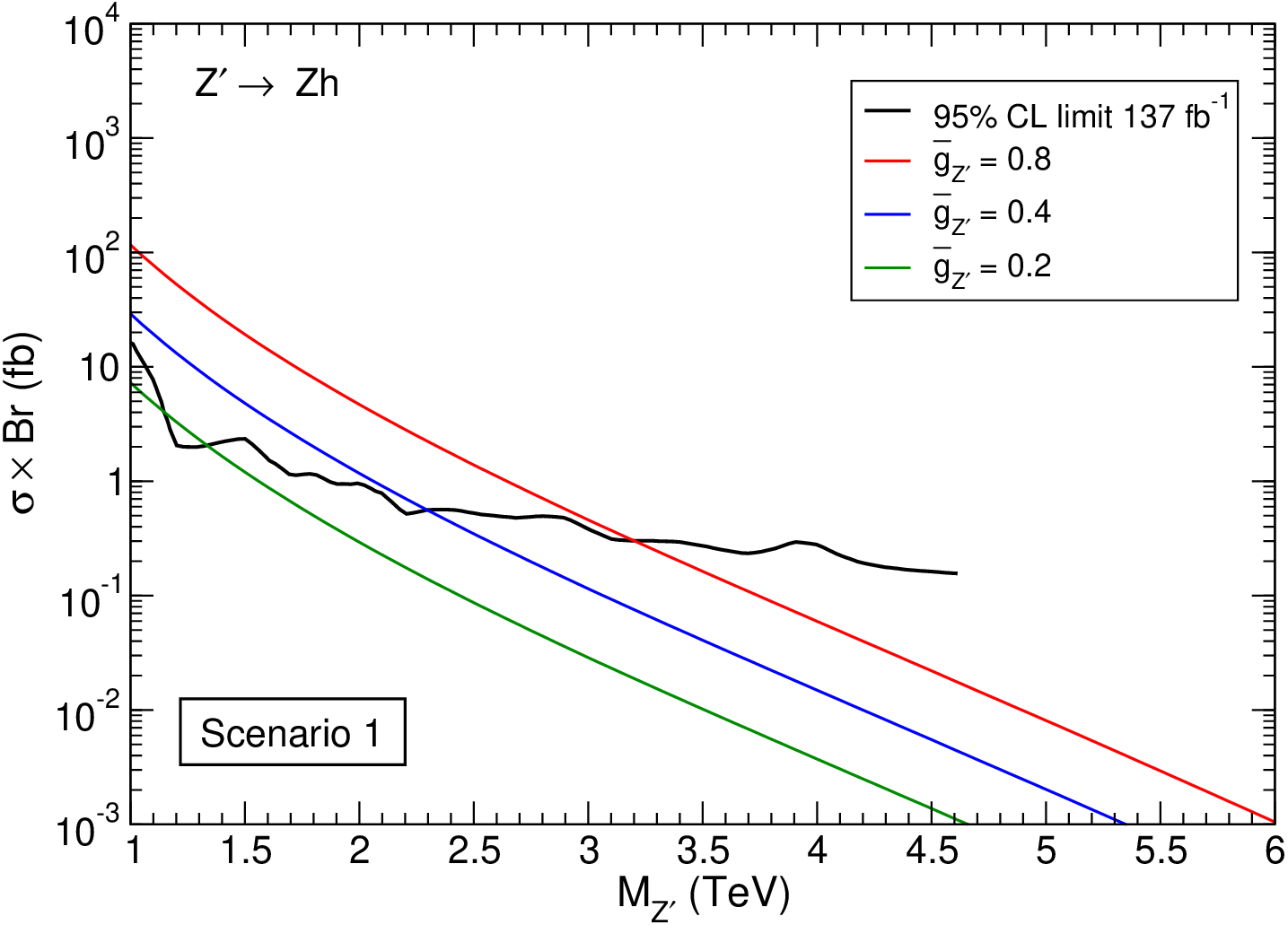} 
\end{tabular}
\caption{Limits on $Z'$ bosons arising from searches in several final states: dijets~\cite{Sirunyan:2018xlo,Sirunyan:2019vgj} (top left), $t \bar t$~\cite{Aaboud:2018mjh,Aad:2020kop} (top right), $WW$~\cite{Aad:2019fbh} (bottom left) and $Zh$~\cite{CMS:2020qrs} (bottom right). Together, we show the cross section predictions for Scenario 1, with $\bar g_{Z'} \equiv g_{Z'}$ by definition. For the rest of scenarios the cross sections can be reinterpreted according to (\ref{ec:dicc}) and Table~\ref{tab:dicc1}.}
\label{fig:lim1}
\end{center}
\end{figure}
\begin{table}[t!]
\begin{center}
\begin{tabular}{|c|cc|cc|cc|cc|cc|cc|}
\hline
& \multicolumn{2}{c|}{Scenario 2} & \multicolumn{2}{c|}{Scenario 3} & \multicolumn{2}{c|}{Scenario 4} & \multicolumn{2}{c|}{Scenario 5} & \multicolumn{2}{c|}{Scenario 6} & \multicolumn{2}{c|}{Scenario 7} \\
\hline
& $k_0$ & $k_1$ & $k_0$ & $k_1$ & $k_0$ & $k_1$ & $k_0$ & $k_1$ & $k_0$ & $k_1$ & $k_0$ & $k_1$  \\
\hline
$jj$                  & 3.6 & 0  & 1.8 & 0  & 1.09 & -0.019 & 3.3 & 0 & 2.2 & 0 & 1.1 & -0.013
\\
\hline
$t \bar t$            & 3.3 & 0  & 1.3 & 0  & 0.75 & -0.018 & 2.8 & 0 & 1.8 & 0 & 0.86 & -0.015
\\
\hline
$WW$, $Zh$     & 2.5 & 0  & 0.93 & 0& 0.95 & -0.019 & 2.7 & 0 & 0.34 & 0 & 0.17 & -0.013 \\
\hline
\end{tabular}
\caption{Numerical factors used to compute the modified coupling $\bar g_{Z'}$ to reinterpret the cross section predictions in Fig.~\ref{fig:lim1}. For the  entries marked with vanishing $k_1$ the scaling is practically independent of the mass within the range of interest. }
\label{tab:dicc1}
\end{center}
\end{table}

\vspace{0.25cm}

The direct limits on $Z'$ masses and mixings have implications on the VEVs of the singlets that are required to generate the $Z'$ mass. Since $m_{ZZ'}$ in 
Eq.~(\ref{entries}) approximately equals the  $Z'$ mass, we can write
\bea
\sqrt{
\sum_{i'}v_{i'R}^2 +
4 \sum_{\alpha} v^2_{{\alpha}S} +
16 \sum_{\alpha'} v^2_{{\alpha'}N} 
}
\approx \frac{4m_{Z'}}{g_{Z'}}, 
\label{entries2}
\eea
where we have neglected the small contributions from the Higgs and left sneutrino VEVs, and used for the $U(1)'$ charges of the singlets under the SM gauge group the result of Table~\ref{table:general}. The sum within the square root is easily dominated by the last two terms (specially the last one), for example under either of these conditions allowed by the minimization equations (\ref{rightsneutrinovev}$-$\ref{ene}):
\begin{itemize}
\item[(i)] all VEVs are of similar order,
$v_{i'R}\sim v_{{\alpha}S}\sim v_{{\alpha'}N}  \equiv
v_N$,
and 
the number of each type of singlets is similar, 
$n_{\nu^c} \sim n_S  \sim n_N\equiv n_{N}$ (as typically obtained in the models
of Appendix~\ref{SubAppendix:solution1}); 
\item[(ii)] there is a hierarchy in the VEVs $v_{i'R} \ll v_{{\alpha}S}\sim v_{{\alpha'}N} \equiv v_{N}$.
\end{itemize}
In such case, the number of each type of singlets is
\bea
n_{N} \sim \left( \frac{m_{Z'}}{g_{Z'} 
v_N
} \right)^2.
\label{entries4}
\eea
For leptophobic scenarios $v_N$ 
is naturally at the TeV scale.  
Let us select for Scenario 1 the masses $M_{Z'} \simeq 1.2, 2.5, 3.3$ TeV, 
and $g_{Z'}\simeq 0.2, 0.4, 0.8$ as the corresponding upper limits for the gauge couplings allowed for these masses by direct searches. The above equation implies that e.g. for VEVs of the order of 3 TeV one needs at least a number of singlets $n_N \sim 4, 4, 2$ to give the $Z'$ boson its mass.
These results are similar for Scenario 4. Therefore in these cases, unlike the non-leptophobic scenarios discussed below, the number of singlets under the SM gauge group can be small with all VEVs of the order of TeV or below. In particular, note that $v_{i'R}$ can be much smaller than $v_{{\alpha'}N}$, thus allowing for scalars at the electroweak scale with a TeV-scale $Z'$ boson. 

Notice also that in models where neither of the above conditions (i), (ii) hold, the VEVs and/or the number of singlets required are larger. An example are the scenarios without singlets of the type $N$ and $S$, as those shown in Tables of  
Sec.~\ref{Section:models} and Appendix~\ref{SubAppendix:solution2}. Then, Eq.~(\ref{entries4}) must be replaced by
\bea
n_{\nu^c} \sim \left( \frac{4m_{Z'}}{g_{Z'} 
v_R
} \right)^2,
\label{entries5}
\eea
implying that a large number of RH neutrinos is required for the masses and couplings discussed above. Alternatively, if $v_R\sim 10$ TeV, then $n_{\nu^c} \sim 4, 4, 2$ is sufficient. 
Let us point out that one can obtain this VEV hierarchy without increasing an order of magnitude the values of the soft terms. As can be straightforwardly deduced from the minimization equations, 
it is in fact sufficient to decrease an order of magnitude the $\lambda, \kappa, Y^\nu$ couplings, i.e. to values $\lambda, \kappa\sim 0.1$ and $Y^\nu\lsim 10^{-8}$. This is because the relevant quantities from the superpotential are the products
$\lambda v_{R}$, $\kappa v_{R}$ and  $Y^\nu v_R$.

The stringent limits on the mass and mixing of general (non leptophobic) $Z'$ bosons imply that the VEVs of (at least some of) the scalar singlets are well above the TeV scale, unless there are a large number of singlets present.
Selecting for Scenario 2 
the masses $M_{Z'}\simeq 2.5, 3.9, 4.5, 5.3$ TeV 
and the corresponding $g_{Z'}\simeq 0.025, 0.05, 0.1, 0.2$ upper limits for the gauge couplings allowed 
by direct searches, Eq.~(\ref{entries4}) implies that 
e.g. for a VEV of the order of 5 TeV one needs at least a number of $N$ singlets $n_N \sim 400, 244, 81, 28$, respectively, to give the $Z'$ boson its mass.
Slightly more fields are required for the other (non-leptophobic) scenarios of Table~\ref{tab:dicc2}.\footnote{Although this number of singlets may seem too big, let us remind the situation in string constructions where typically there is a large number of singlet fields. For example, in Ref.~\cite{Casas:1988hb} where a SM model-like construction from orbifols was obtained, the number of singlets at low energy is 33. Also, this number is as large as 196 in the D-brane constructions discussed in Ref.~\cite{Uranga:933469}.}

As mentioned above, alternatively we can reduce the number of fields required
allowing a 
hierarchy between the VEVs of the singlets under the SM gauge group. 
For example, with $v_{N}\gsim 10$ TeV, Eq.~(\ref{entries4}) can be fulfilled with 
$n_{N}\sim$ 10 for the cases of masses and couplings discussed above,
whereas $v_R$ stays in the order of one TeV.
To obtain this hierarchy without increasing an order of magnitude the soft terms associated to the fields with larger VEVs, $S$ and $N$,
it is sufficient to decrease an order of magnitude the corresponding couplings to values $\kappa,\kappa' \sim 0.1$, since now the relevant quantities from the superpotential are the products
$\kappa v_{\alpha S}$ and $\kappa' v_{\alpha' N}$.

\subsection{Indirect limits}

The $Z-Z'$ mixing angle $\theta_{\text{mix}}$ is constrained by the precise measurements of fermion couplings, in particular at LEP, and by the relation between the $W$ and $Z$ boson masses. 
As we will see, for $Z'$ boson masses of several TeV these limits are basically ineffective compared to those from direct searches.
For illustration we will select a reference mass $M_{Z'} = 2.5$ TeV as example. For the coupling $g_{Z'}$ we use the 95\% upper limit allowed for this $Z'$ mass by direct searches, as discussed in the previous subsection, for each scenario. Table \ref{tab:mix} collects these values of $g_{Z'}$, the factor $z_\text{mix}$ defined in Eq.~(\ref{ec:zmix}) for $\tan \beta = 2$, and the resulting $\theta_{\text{mix}}$.

\begin{table}[b!]
\begin{center}
\begin{tabular}{|c|c|c|c|c|}
\hline
Scenario & $g_{Z'}$ & $z_\text{mix}$ & $\theta_{\text{mix}}$ & $\delta_\text{mix}$
\\ \hline
 1 & 0.5 & $-0.2$ & $-3.6 \times 10^{-4}$ & $9.7 \times 10^{-5}$
\\ \hline
 2 & 0.024 & $-0.7$ & $-6.0 \times 10^{-5}$ & $2.8 \times 10^{-6}$
\\ \hline
3 & 0.019 & 0.3 & $2.0 \times 10^{-5}$ & $3.1 \times 10^{-7}$
\\ \hline
 4 & 0.45 & $-0.2$ & $-3.2 \times 10^{-4}$ & $7.9 \times 10^{-5}$
\\ \hline
 5 & 0.023 & $-0.7$ & $-5.8 \times 10^{-5}$ & $2.5 \times 10^{-6}$
\\ \hline
6 & 0.020 & 0.3 & $2.2 \times 10^{-5}$ & $3.7 \times 10^{-7}$
\\ \hline
 7 & 0.010 & 0.05 & $1.8 \times 10^{-7}$ & $2.4 \times 10^{-9}$
\\ \hline
\end{tabular}
\caption{Summary of 95\% CL upper limits on $Z'$ coupling for $m_{Z'} = 2.5$ TeV (second column) and $z_\text{mix}$ factors (third column) in each scenario. The resulting $Z-Z'$ mixing angle and relative contribution to the $Z$ boson mass are given in the fourth and fifth columns, respectively.}
\label{tab:mix}
\end{center}
\end{table}

The mass of the $Z_1$ ($\approx Z$) gauge boson
resulting from the diagonalisation of the $2 \times 2$ matrix (\ref{MZZ'2x2}) is given by
\begin{equation}
m_{Z}^2 \approx m_{ZZ}^2 (1 - \delta_\text{mix}) \,, 
\label{ec:MZ1}
\end{equation}
where $m_{ZZ}^2 = (m_W/c_W)^2$ at the tree level, with $c_W\equiv \cos\theta_W$, and the correction due to mixing is
\begin{equation}
\delta_\text{mix} = 2 \left(\frac{g_{Z'}}{g_Z}\right) 
\theta_\text{mix}\ z_\text{mix} \,.
\label{ec:deltaM}
\end{equation}
The numerical value of this correction for the examples selected is given in the last column of Table~\ref{tab:mix}. The impact of this correction can be assessed by the parameter~\cite{Zyla:2020zbs}
\begin{equation}
\rho_0 \equiv \frac{m_W^2}{m_Z^2 c_W^2 \hat \rho} \,,
\end{equation}
where $\hat \rho$ takes into account radiative corrections to the SM tree-level relation $m_W = m_Z c_W$. As a result of a global fit, $\rho_0 = 1.00038 \pm 0.00020$~\cite{Zyla:2020zbs}. From Eq.~(\ref{ec:MZ1}),
we have $\rho_0 = 1+ \delta_\text{mix}$, therefore the values of $\delta_\text{mix}$ in Table~\ref{tab:mix} are well in agreement with experimental data.

\begin{table}[t!]
\begin{center}
\begin{tabular}{|c|c|c|c|c|}
\hline
& SM prediction & Measurement & Scenario 1 & Scenario 4
\\ \hline
$R_b$ & $0.21581 \pm 0.00002$ & $0.21629 \pm 0.00066$ & $-4.9 \times 10^{-6}$ & $-1.0 \times 10^{-5}$ 
\\ \hline
$R_c$ & $017221 \pm 0.00003$ & $0.1721 \pm 0.0030$ & $1.1 \times 10^{-5}$ & $1.3 \times 10^{-5}$ 
\\ \hline
$A_\text{FB}^b$ & $0.1030 \pm 0.0002$ & $0.0996 \pm 0.0016$ & $-7.2 \times 10^{-7}$ & $-2.3 \times 10^{-6}$
\\ \hline
$A_\text{FB}^c$ & $0.0736 \pm 0.0002$ & $0.0707 \pm 0.0035$ & $-1.0 \times 10^{-5}$ & $-3.5 \times 10^{-6}$
\\ \hline
\end{tabular}
\caption{SM predictions and experimental values of selected $Z$-pole observables, as well as $Z-Z'$ mixing contributions arising in scenarios 1 and 4 (see the text).}
\label{tab:RbRc}
\end{center}
\end{table}

The mixing also induces modifications of the fermion couplings with respect to the predictions of the $\text{SU}(2) \times \text{U}(1)_Y$ theory. Writing the
SM
neutral current Lagrangian as
\begin{equation}
\mathcal{L}_\text{NC}^Z = g_Z Z_\mu \sum_f \bar f \gamma^\mu \left(c_L^f P_L + c_R^f P_R\right) f \,,
\end{equation}
the corrections are $c_{L,R}^f \to c_{L,R}^f + \delta c_{L,R}^f$, with
\bea
 \delta c_L^f = -
\frac{\delta_\text{mix}}{2z_\text{mix}}
\, 
z(f) 
\ , \;\;\;\;\;\;\;\;\;
 \delta c_R^f = 
\frac{\delta_\text{mix}}{2z_\text{mix}}
\, z(f^c) \,.
\label{ec:deltac}
\eea
For charged leptons the $Z-Z'$ mixing contributions to the vector and axial couplings $c_V^f = (c_L^f + c_R^f)/2$ and $c_A^f = (c_L^f - c_R^f)/2$, respectively, are $|\delta c^f_{V,A}| \leq 1.4 \times 10^{-6}$, i.e. three orders of magnitude below the current precision~\cite{Zyla:2020zbs}. (Note that for the leptophobic scenarios where $g_{Z'}$ can be larger the corrections to lepton couplings identically vanish.) For the quarks the corrections are of the order of $10^{-5}$. The impact of these contributions in $Z$-pole observables $R_b$, $R_c$, $A_\text{FB}^b$, $A_\text{FB}^c$, can be investigated by writing~\cite{Aguilar-Saavedra:2013qpa}
\begin{align}
& R_b = R_b^\text{SM} \; [ 1 - 1.81 \delta c_L^b + 0.33 \delta c_R^b - 0.42 \delta c_L^c + 0.19 \delta c_R^c ] \,, \notag \\
& A_\text{FB}^b = A_\text{FB}^{b,\text{SM}} \;
[ 1 - 0.16 \delta c_L^b - 0.88 \delta c_R^b ] \,, \notag \\
& R_c = R_c^\text{SM} \; [ 1 + 0.50 \delta c_L^b - 0.091 \delta c_R^b + 2.0 \delta c_L^c - 0.89 \delta c_R^c ] \,, \notag \\
& A_\text{FB}^c = A_\text{FB}^{c,\text{SM}} \; 
[ 1 + 1.2 \delta c_L^c + 2.7 \delta c_R^c ] \,,
\label{ec:RbRc}
\end{align}
where we label the SM predictions with the `SM' superscript. These expansions only assume that the corrections to the $b$ and $c$ couplings given by (\ref{ec:deltac}) are small, as is our case. We present in Table~\ref{tab:RbRc} the SM predictions and experimental measurements for these quantities~\cite{Zyla:2020zbs}, as well as the corrections in scenarios 1 and 4, with the input values summarised in Table~\ref{tab:mix}. As it can be observed, the contributions arising from the $Z-Z'$ mixing are in all cases much smaller than the experimental and theoretical uncertainties.


\section{Phenomenological prospects}
\label{phenomenology}

\noindent 
Once we have shown in the previous section that U$\mu\nu$SSM models are
phenomenologically viable, with reasonable lower bounds on the $Z'$ mass that depend on the parameter space analyzed, let us now discuss other type of signals that could be relevant for its detection.

\vspace{0.25cm}

\noindent 
The $Z'$ boson can produce quite characteristic signals 
when decaying to sparticle pairs. In this discussion we focus on leptophobic $Z'$ bosons with $z(H_d) = 0$, because otherwise the $Z' \to e^+ e^-$ and $Z' \to \mu^+ \mu^-$ signals that are already seeked for at the LHC would be the most visible ones.
The decays into weakly-interacting SUSY particles pose a special interest for  because in some regions of the parameter space they could produce the most prominent signals. We can have the following ones:

\begin{itemize}

\item $Z' \to \tilde \nu_{i'R} \tilde \nu_{j'R}$, with $\tilde \nu_{i'R}$ and $\tilde 
\nu_{j'R}$ decaying into lighter $\tilde \nu^c$ or weak bosons, which ultimately decay into quarks. These signals produce two or more massive multi-pronged jets \cite{Aguilar-Saavedra:2017zuc} that can be pinpointed with 
suitable tools \cite{Aguilar-Saavedra:2020uhm}. {On the other hand, the direct production of these neutral particles is quite suppressed by their small coupling to the SM particles}. The potential of a generic search to detect massive dijet signals in this context will be presented elsewhere \cite{inprep}.

\item $Z' \to  \tilde \chi_1^+ \tilde \chi_1^- ,\, \tilde \chi_2^0 \tilde \chi_2^0$, with the MSSM-like decays $\tilde \chi_1^\pm \to \ell^\pm \nu \tilde \chi_1^0$, $\tilde \chi_2^0 \to \ell^+ \ell^- \tilde \chi_1^0$ and RPV decay of the lightest neutralino $\tilde \chi_1^0 \to \nu \nu$ yielding missing energy. This type of signals has been studied in the context of DM models, with neutral heavy leptons $N_{1,2}$ and a charged lepton $E_1^\pm$ in the place of $\tilde \chi_{1,2}$ and $\tilde \chi_1^\pm$, respectively. From the results in Ref.~\cite{Aguilar-Saavedra:2019iil}, one can see that there are regions in the parameter space where the $Z'$-mediated signals are more significant than the direct production of $\tilde \chi_{1,2}^\pm$ and $\tilde \chi_2^0$ pairs.

\item $Z' \to  \tilde \chi_1^+ \tilde \chi_1^- ,\, \tilde \chi_2^0 \tilde \chi_2^0$ with the same decays $\tilde \chi_1^\pm \to \ell^\pm \nu  \tilde \chi_1^0$, $\tilde \chi_2^0 \to \ell^+ \ell^- \tilde \chi_1^0$ and RPV decays $\tilde \chi_1^0 \to \ell^+ \ell^- \nu, q \bar q' \ell$ with a displaced vertex \cite{Kpatcha:2019pve}. Depending on the $Z'$ mass, the final state particles may be more or less boosted and collimated. When the $Z'$ is much heavier than its decay products, the characteristic signature is a pair of complex objects, each consisting of two collimated leptons with an emerging lepton pair or jet. The possibility to identify these objects, and the sensitivity to these signals, either in the direct production or in the decay of a $Z'$ boson, deserve further investigation. 
\end{itemize}

\vspace{0.25cm}

\noindent
On the other hand, the $\hat\xi$ sector, when it is present in U$\mn$ models, can also give rise to characteristic signals. Let us discuss them briefly.
As mentioned in Sec.~\ref{munuSSM}, the superpotentials~(\ref{Eq:superpotentialt}) and~(\ref{Eq:superpotentialt2}) contain a $Z_2$ symmetry. If it is not spontaneously broken, one can have either the bosonic or the fermionic components
of the superfield $\hat\xi$ as potentially interesting WIMP DM candidates.
In order for this type of candidates to be absolute stable the $Z_2$ symmetry must be exact at the non-renormalizable level, which depends on the specific construction used. 

Defining for the bosonic component $\xi$ the scalar ($\xi^\mathcal{R}$) and pseudoscalar ($\xi^\mathcal{I}$) fields as  
\begin{equation}
\xi = \frac{1}{\sqrt{2}} ( \xi^\mathcal{R} + i \xi^\mathcal{I})\,,
\label{chibosoncomponents}
\end{equation}
their masses squared are given by:
\bea
m^2_{\xi^\mathcal{R}} 
&=& 
{\frac{1}{2}  g^2_{Z'}\left[z(H_d)|v_d|^2 + z(H_u) |v_u|^2 + z(L) |v_L|^2 + z(\nu ^c) |v_R|^2  + z(S) |v_S|^2  + z(N) |v_N|^2\right] }
\nonumber\\
&+& 
m^2_{\xi} + 
m^2_{\widetilde \xi}
+\,
 \left( \lambda {\kappa''} v_u v_d + Y^{\nu} \kappa'' v_u v_L + {2} \kappa \kappa''  v_R v_S + \sqrt{2} 
T^{\kappa''}
 v_R \right)\,,
\\
m^2_{\xi^\mathcal{I}} 
&=& 
{\frac{1}{2}  g^2_{Z'}\left[z(H_d)|v_d|^2 + z(H_u) |v_u|^2 + z(L) |v_L|^2 + z(\nu ^c) |v_R|^2  + z(S) |v_S|^2  + z(N) |v_N|^2\right] }
\nonumber\\
&+& 
m^2_{\xi} + 
m^2_{\widetilde \xi}
-\, \left( \lambda {\kappa''}
 v_u v_d + Y^{\nu} \kappa'' v_u v_L +  {2}\kappa \kappa''  v_R v_S + \sqrt{2} 
T^{\kappa''}
 v_R \right)\,,
\label{eq:chimass2}
\eea
where $m_{\xi}$ is the soft mass and $m_{\widetilde \xi}$ is the mass of the fermionic component $\widetilde \xi$:
\bea
m_{\widetilde \xi} =  {\sqrt 2} \kappa'' v_R\,.
\eea
The lightest of these fields above is the WIMP DM candidate, which has in general $Z'$ and Higgs mediated (though the mixing with the scalar and pseudocalar RH sneutrino) annihilations and interactions with the visible sector.  
This interesting subject will be discussed in another 
occasion~\cite{pierre:2020xxx}.


\section{Conclusions}
\label{conclusions}

In this work we have built $U(1)'$ extensions of the $\mn$,
dubbed U$\mn$. We have shown that the extension of the $\mn$ with a $U(1)'$ symmetry is very well motivated from the theoretical viewpoint in order to forbid dangerous couplings. Besides, these models give rise
to new and interesting phenomenological properties. In addition to a new $Z'$ gauge boson, exotic quarks and singlets under the SM gauge group are present in the spectrum due to the anomaly cancellation conditions. 
The singlets can be distinguished by the $U(1)'$ charges. Some of them behave as the RH neutrinos of the $\mn$, dynamically generating the $\mu$-term.
Others can generate electroweak-scale Majorana masses for the RH neutrinos solving straightforwardly the $\nu$-problem, and, alternatively, in some models light sterile neutrinos can be present. Finally, other type of singlets can be used as DM candidates.
{The new quarks have typically different $U(1)'$ charges from the SM quarks, in which case they are expected to hadronize inside the detector into color-singlet states. But in some models
they can have the same charges as the SM quarks, in which case their production and decay gives final states with multiple top and bottom quarks, and gauge and Higgs bosons.}

The phenomenology of U$\mn$ models is very rich. 
For non-leptophobic scenarios the $Z'\to \ell\ell$ decays are very clean, and the null results of current searches provide the strongest constraints on the mass and mixing of $Z'$ bosons, with for example a lower bound on the $Z'$ mass around 4.5 TeV for an $U(1)'$ gauge coupling of about 0.1. In the case of leptophobic scenarios these limits are weaker, and the strongest constraints result from the diboson and $Zh$ final states.
The lower bound on the $Z'$ mass is around 3 TeV for an $U(1)'$ gauge coupling of the order of the weak coupling $g$. We have also analyzed the limits on $Z-Z'$ mixing from precision electroweak data, but they turn out to be ineffective compared to the previous ones from direct searches. 

Finally, we have sketched novel signals that could be explored in forthcoming publications, produced in the decays of the $Z'$ to sparticle pairs such as right sneutrinos, charginos or neutralinos. In particular, if the $Z'$ boson is significantly heavier than its decay products, the experimental signature contains complex boosted objects, such as multi-pronged jets, or collimated leptons with or without displaced vertices, which require specific tools for their detection.
Besides, even though we are working in the framework of RPV models, stable DM particles due to a $Z_2$ symmetry of their couplings in the superpotential can be present in specific constructions. Such a DM is of the WIMP type, and can be formed by the bosonic or the fermionic components of the 
the singlet superfields under the SM gauge group. 
Thus, it has in general $Z'$ and Higgs mediated annihilations and interactions with the visible sector, that would be interesting to analyze.


\begin{acknowledgments}

The research of JAAS was supported by the Spanish Agencia Estatal de Investigaci\'on (AEI) through project PID2019-110058GB-C21.
The work of IL was funded by the Norwegian Financial Mechanism 2014-2021,
grant DEC-2019/34/H/ST2/00707.
The work 
of DL was supported by the Argentinian CONICET, and also acknowledges the support through PIP 11220170100154CO, and the Spanish grant FPA2015-65929-P (MINECO/FEDER, UE).
The research of CM was supported by the Spanish AEI 
through the grants 
FPA2015-65929-P (MINECO/FEDER, UE), PGC2018-095161-B-I00 and IFT Centro de Excelencia Severo Ochoa SEV-2016-0597.  
The authors acknowledge the support of the Spanish Red Consolider MultiDark FPA2017-90566-REDC.

\end{acknowledgments}

\clearpage

\appendix
\numberwithin{equation}{section}
\numberwithin{figure}{section}
\numberwithin{table}{section}

\numberwithin{equation}{subsection}
\numberwithin{figure}{subsection}
\numberwithin{table}{subsection}

\section{Values of the $U(1)'$ charges for the exotic quarks}

\label{Apendix:solution1}

Here we show solutions for the values of the $U(1)'$ charges for the exotic quarks in 
U$\mu\nu$SSM models, in addition to those solutions already shown in 
Sec.~\ref{Section:models}}.

\subsection{Solutions with exotic quarks $\hat{\mathbb{K}}_{1,2,3}$ and $\hat{\mathbb{K}}^c_{1,2,3}$
}
\label{SubAppendix:solution1}

\begin{table}[ht]
\begin{center}
\begin{tabular}{|c|c|c|c|c|}
\hline    
 {\bf Scenario 1}   &  Solution 1   &  Solution 2 & Solution 3 & Solution 4 
\tabularnewline
\hline 
$z({\mathbb{K}}_1)$ 
&  \ \ $\frac{25}{54}$ 
& $- \frac{11}{18}$ 
&  \ \ $ \frac{1}{27}$ 
& $-\frac{5}{27}$  
\tabularnewline
\hline 
$z({\mathbb{K}}_2)$ &  $-\frac{19}{108}$ & $-\frac{4}{9}$ & $ - \frac{65}{108}$ &  $ -\frac{71}{108}$ \tabularnewline
\hline 
$z({\mathbb{K}}_3)$ & 
$-\frac{1}{108}$ & $- \frac{5}{18}$ & \ \ $\frac{11}{54}$  &  \ \ $\frac{4}{27}$ \tabularnewline
\hline 
$z({\mathbb{K}}^c_1)$ &   $-\frac{77}{108}$ & \ \ $\frac{13}{36}$ & 
$ -\frac{31}{108}$  &  $ -\frac{7}{108}$\tabularnewline
\hline 
$z({\mathbb{K}}^c_2)$ &  $-\frac{4}{54}$& $ \ \ \frac{7}{36}$ & \ \ $\frac{19}{54}$  & \ \ $\frac{11}{27}$ \tabularnewline
\hline 
$z({\mathbb{K}}^c_3)$ &  $- \frac{13}{54}$ & \ \ $\frac{1}{36}$ & \ \ $ \frac{49}{108}$  &  $ -\frac{43}{108}$\tabularnewline
\hline 
\end{tabular}
\caption{
Values of the $U(1)'$ charges for the exotic quarks in 
U$\mu\nu$SSM models
using solution~(\ref{exotic1}) with the hypercharges
of
Table~\ref{table:ND1-11}.
For each column,
the $U(1)'$ charges of the rest of the fields are given in 
Table~\ref{table:general} and Scenario 1 of Table~\ref{table:NK3-general}
of Sec.~\ref{Section:models}.
The number of singlets under the SM gauge group consistent with these charges is 
${\bf n_{\nu^c}=2}$, ${\bf n_{S} = 2}$ and
${\bf n_{N} = 1}$.
}
\label{table:NK3_NReven-2}
\end{center}
\end{table} 

\begin{table}[ht]
\begin{center}
\begin{tabular}{|c|c|c|c|c|}
\hline    
 {\bf Scenario 1}   &  Solution 1   &  Solution 2 & Solution 3 & Solution 4 
\tabularnewline
\hline 
$z({\mathbb{K}}_1)$ 
& \ \ $\frac{7}{18}$ 
&  $- \frac{29}{54}$
& \ \ $ \frac{4}{27}$ 
&  $ - \frac{8}{27}$ 
 \tabularnewline
\hline 
$z({\mathbb{K}}_2)$ & $- \frac{5}{18}$ & $-\frac{55}{108}$ & $ -  \frac{14}{27}$  & $ - \frac{17}{27}$ \tabularnewline
\hline 
$z({\mathbb{K}}_3)$ & \ \ $  \frac{1}{18}$ & $- \frac{19}{108}$ & \ \ $ \frac{19}{108}$ & \ \ $ \frac{7}{108}$ \tabularnewline
\hline 
$z({\mathbb{K}}^c_1)$ & $- \frac{23}{36}$ &\ \  $\frac{31}{108}$ & $- \frac{43}{108}$ & \ \ $ \frac{5}{108}$ \tabularnewline
\hline 
$z({\mathbb{K}}^c_2)$ & \ \  $ \frac{1}{36}$ &\ \  $ \frac{7}{27}$ & \ \ $ \frac{29}{108}$ & \ \  $ \frac{41}{108}$\tabularnewline
\hline 
$z({\mathbb{K}}^c_3)$ &   $- \frac{11}{36}$ & $- \frac{2}{27}$ & $- \frac{23}{54}$  & $- \frac{17}{54}$\tabularnewline
\hline 
\end{tabular}
\caption{The same as in 
Table~\ref{table:NK3_NReven-2}, but 
with the minimal number of singlets under the SM gauge group
${\bf n_{\nu^c}=4}$, ${\bf n_{S} = 3}$ and
${\bf n_{N} = 1}$.
}
\label{table:NK3_NReven-4}
\end{center}
\end{table}

\begin{table}[ht]
\begin{center}
\begin{tabular}{|c|c|c|c|c|}
\hline    
 {\bf Scenario 1}   &  Solution 1   &  Solution 2 & Solution 3 & Solution 4 
\tabularnewline
\hline 
 $z({\mathbb{K}}_1)$
& \ \ $\frac{7}{18}$ 
&  $- \frac{29}{54}$ 
&  \ \  $ \frac{1}{27}$ 
&  $- \frac{5}{27}$ 
  \tabularnewline
\hline 
$z({\mathbb{K}}_2)$ & $- \frac{7}{36}$ & $- \frac{23}{54}$ & $ - \frac{59}{108}$  & $ - \frac{65}{108}$ \tabularnewline
\hline 
$z({\mathbb{K}}_3)$ & - $\frac{1}{36}$ & $- \frac{7}{27}$ &  \ \ $ \frac{4}{27}$ &  \ \ $  \frac{5}{54}$  \tabularnewline
\hline 
$z({\mathbb{K}}^c_1)$ &   $- \frac{23}{36}$ & \ \  $\frac{31}{108}$ & $- \frac{31}{108}$ & $- \frac{7}{108}$\tabularnewline
\hline 
$z({\mathbb{K}}^c_2)$ &  $- \frac{1}{18}$& \ \  $  \frac{19}{108}$ & \ \  $ \frac{8}{27}$ & \ \  $ \frac{19}{54}$ \tabularnewline
\hline 
$z({\mathbb{K}}^c_3)$ &  $- \frac{2}{9}$ &  \ \ $\frac{1}{108}$ & $- \frac{43}{108}$ & $- \frac{37}{108}$  \tabularnewline
\hline 
\end{tabular}
\caption{The same as in Table~\ref{table:NK3_NReven-2}, 
but 
with the minimal number of singlets under the SM gauge group
${\bf n_{\nu^c}=6}$, ${\bf n_{S} = 4}$ and
${\bf n_{N} = 1}$.
}
\label{table:NK3_NReven-6}
\end{center}
\end{table}

\begin{table}[ht]
\begin{center}
\begin{tabular}{|c|c|c|c|c|}
\hline    
  {\bf Scenario 1}   &  Solution 1   &  Solution 2 & Solution 3 & Solution 4 
\tabularnewline
\hline 
$z({\mathbb{K}}_1)$
& \ \  $\frac{55}{108}$ 
&  $-\frac{71}{108}$
& $-\frac{1}{36}$ 
&  $-\frac{7}{54}$ 
\tabularnewline
\hline 
$z({\mathbb{K}}_2)$ & $- \frac{25}{216}$ & $- \frac{11}{27}$ & $-  \frac{47}{72}$  & $-  \frac{29}{108}$  \tabularnewline
\hline 
$z({\mathbb{K}}_3)$ & $- \frac{5}{108}$ & $- \frac{73}{216}$ & \ \ $ \frac{2}{19}$  &  $- \frac{23}{108}$ \tabularnewline
\hline 
$z({\mathbb{K}}^c_1)$ &  $- \frac{41}{54}$ & \ \ $\frac{11}{27}$ & $- \frac{2}{9}$ & $- \frac{13}{108}$ \tabularnewline
\hline 
$z({\mathbb{K}}^c_2)$ & $ -\frac{29}{216}$& \ \  $ \frac{17}{108}$ & \ \  $ \frac{29}{72}$ & \ \ $ \frac{1}{54}$ \tabularnewline
\hline 
$z({\mathbb{K}}^c_3)$ &  $- \frac{11}{54}$ & \ \ $\frac{19}{216}$ & $- \frac{17}{36}$ & $- \frac{1}{27}$ \tabularnewline
\hline 
\end{tabular}
\caption{The same as in Table~\ref{table:NK3_NReven-2}, 
but 
with the minimal number of singlets under the SM gauge group
${\bf n_{\nu^c}=1}$, ${\bf n_{S} = 1}$,
${\bf n_{N} = 1}$ and ${\bf n_{\xi} = 2}$.
}
\label{table:NK3_NR-Odd-1}
\end{center}
\end{table} 

\begin{table}[ht]
\begin{center}
\begin{tabular}{|c|c|c|c|c|}
\hline    
  {\bf Scenario 1}   &  Solution 1   &  Solution 2 & Solution 3 & Solution 4 
\tabularnewline
\hline 
$z({\mathbb{K}}_1)$ 
 &  \ \ $\frac{17}{36}$ 
&  $- \frac{67}{108}$ 
&  \ \  $  \frac{1}{108}$ 
& $- \frac{17}{108}$ 
 \tabularnewline
\hline 
$z({\mathbb{K}}_2)$ & $- \frac{11}{72}$ & $- \frac{23}{54}$ & $-  \frac{133}{216}$ & $-  \frac{71}{108}$  \tabularnewline
\hline 
$z({\mathbb{K}}_3)$ & $- \frac{1}{36}$ & $- \frac{65}{216}$ &  \ \ $  \frac{11}{54}$ & \ \  $ \frac{35}{216}$ \tabularnewline
\hline 
$z({\mathbb{K}}^c_1)$ &  $- \frac{13}{18}$ & \ \  $\frac{10}{27}$ & $- \frac{7}{27}$ & $- \frac{5}{54}$ \tabularnewline
\hline 
$z({\mathbb{K}}^c_2)$ & $- \frac{7}{72}$&  \ \ $  \frac{19}{108}$ & \ \  $ \frac{79}{216}$ &  \ \ $ \frac{11}{27}$\tabularnewline
\hline 
$z({\mathbb{K}}^c_3)$ &  $- \frac{2}{9}$ & \ \  $\frac{11}{216}$ & $- \frac{49}{108}$ & $- \frac{89}{216}$\tabularnewline
\hline 
\end{tabular}
\caption{The same as in 
Table~\ref{table:NK3_NReven-2}, 
but 
with the minimal number of singlets under the SM gauge group
${\bf n_{\nu^c}=3}$, ${\bf n_{S} = 2}$,
${\bf n_{N} = 1}$ and ${\bf n_{\xi} = 2}$.
}
\label{table:NK3_NR-Odd-3}
\end{center}
\end{table} 

\begin{table}[ht]
\begin{center}
\begin{tabular}{|c|c|c|c|c|}
\hline    
  {\bf Scenario 1}   &  Solution 1   &  Solution 2 & Solution 3 & Solution 4 
\tabularnewline
\hline 
 $z({\mathbb{K}}_1)$  
& \ \  $ \frac{43}{108}$ 
&  $- \frac{59}{108}$ 
& \ \   $ \frac{5}{36}$ 
& $- \frac{31}{108}$ 
 \tabularnewline
\hline 
$z({\mathbb{K}}_2)$ & $- \frac{29}{108}$ & $-  \frac{109}{216}$   & $- \frac{19}{36}$ & $-  \frac{137}{216}$  \tabularnewline
\hline 
$z({\mathbb{K}}_3)$ & \ \   $ \frac{11}{216}$ & $- \frac{5}{27}$  & \ \  $ \frac{13}{72}$ & \ \  $ \frac{2}{27}$  \tabularnewline
\hline 
$z({\mathbb{K}}^c_1)$ &  $- \frac{35}{54}$ & \ \  $ \frac{8}{27}$ & $- \frac{7}{18}$ & \ \   $ \frac{1}{27}$ \tabularnewline
\hline 
$z({\mathbb{K}}^c_2)$ & \ \   $\frac{1}{54}$ & \ \   $ \frac{55}{216}$ & \ \   $ \frac{5}{18}$ & \ \  $ \frac{83}{216}$ \tabularnewline
\hline 
$z({\mathbb{K}}^c_3)$ &  $- \frac{65}{216}$ & $- \frac{7}{108}$ & $-\frac{31}{72}$ & $- \frac{35}{108}$ \tabularnewline
\hline 
\end{tabular}
\caption{The same as in 
Table~\ref{table:NK3_NReven-2}, 
but 
with the minimal number of singlets under the SM gauge group
${\bf n_{\nu^c}=5}$, ${\bf n_{S} = 3}$,
${\bf n_{N} = 1}$ and ${\bf n_{\xi} = 2}$.
}
\label{table:NK3_NR-Odd-5}
\end{center}
\end{table}

\begin{table}[ht]
\begin{center}
\begin{tabular}{|c|c|c|c|c|}
\hline    
 {\bf Scenario 2}   &  Solution 1   &  Solution 2 & Solution 3 & Solution 4 
\tabularnewline
\hline 
$z({\mathbb{K}}_1)$
&  \ \ $\frac{29}{36}$ 
&  $- \frac{31}{108}$
 &   \ \  $ \frac{37}{108}$
&   \ \  $ \frac{19}{108}$
\tabularnewline
\hline 
$z({\mathbb{K}}_2)$ & $- \frac{59}{72}$ & $- \frac{59}{54}$ & $-  \frac{277}{216}$  & $-  \frac{143}{108}$ \tabularnewline
\hline 
$z({\mathbb{K}}_3)$ &  $-\frac{25}{36}$ & $- \frac{209}{216}$ & $- \frac{25}{54}$  & $- \frac{109}{216}$ \tabularnewline
\hline 
$z({\mathbb{K}}^c_1)$ &   $-\frac{19}{18}$ &  \ \  $\frac{1}{27}$ & $- \frac{16}{27}$ & $- \frac{23}{54}$\tabularnewline
\hline 
$z({\mathbb{K}}^c_2)$ &  \ \  $\frac{41}{72}$&  \ \  $ \frac{91}{108}$ &  \ \  $ \frac{223}{216}$ &   \ \ $ \frac{29}{27}$\tabularnewline
\hline 
$z({\mathbb{K}}^c_3)$ &   \ \  $\frac{4}{9}$ &  \ \  $\frac{155}{216}$ &  \ \  $ \frac{23}{108}$ &   \ \ $ \frac{55}{216}$ \tabularnewline
\hline 
\end{tabular}
\caption{
Values of the $U(1)'$ charges for the exotic quarks in 
U$\mu\nu$SSM models
using solution~(\ref{exotic1}) with the hypercharges
of
Table~\ref{table:ND1-11}.
For each column,
the $U(1)'$ charges of the rest of the fields are given in 
Table~\ref{table:general} and Scenario 2 of Table~\ref{table:NK3-general}.
The minimal number of singlets under the SM gauge group consistent with these charges is
${\bf n_{\nu^c}=3}$, ${\bf n_{S} = 2}$,
${\bf n_{N} = 1}$ and ${\bf n_{\xi} = 2}$.
}
\label{table:NK3_NR3-Nolepto}
\end{center}
\end{table}

\begin{table}[ht]
\begin{center}
\begin{tabular}{|c|c|c|c|c|}
\hline    
 {\bf Scenario 3}   &  Solution 1   &  Solution 2 & Solution 3 & Solution 4 
\tabularnewline
\hline 
 $z({\mathbb{K}}_1)$
& \ \  $\frac{5}{36}$
&  $- \frac{103}{108}$
 &  $- \frac{35}{108}$
& $- \frac{53}{108}$ 
\tabularnewline
\hline 
$z({\mathbb{K}}_2)$ & \ \   $\frac{37}{72}$ & \ \   $\frac{13}{54}$ & \ \   $\frac{11}{216}$  & \ \   $ \frac{1}{108}$ \tabularnewline
\hline 
$z({\mathbb{K}}_3)$ &  \ \   $\frac{2}{36}$ & \ \   $\frac{79}{216}$ &  \ \  $\frac{47}{54}$  & \ \   $\frac{179}{216}$ \tabularnewline
\hline 
$z({\mathbb{K}}^c_1)$ &  \ \    $\frac{7}{18}$ & \ \   $\frac{19}{27}$ & \ \   $\frac{2}{27}$ & \ \   $\frac{13}{54}$\tabularnewline
\hline 
$z({\mathbb{K}}^c_2)$ & $- \frac{55}{72}$& $- \frac{53}{108}$ & $- \frac{65}{216}$ & $- \frac{7}{27}$\tabularnewline
\hline 
$z({\mathbb{K}}^c_3)$ &  $- \frac{8}{9}$ & $- \frac{133}{216}$ & $- \frac{121}{108}$ & $- \frac{233}{216}$ \tabularnewline
\hline 
\end{tabular}
\caption{
Values of the $U(1)'$ charges for the exotic quarks in 
U$\mu\nu$SSM models
using solution~(\ref{exotic1}) with the hypercharges
of
Table~\ref{table:ND1-11}.
For each column,
the $U(1)'$ charges of the rest of the fields are given in
Table~\ref{table:general} and Scenario 3 of Table~\ref{table:NK3-general}.
The minimal number of singlets under the SM gauge group consistent with these charges is
${\bf n_{\nu^c}=3}$, ${\bf n_{S} = 2}$,
${\bf n_{N} = 1}$ and ${\bf n_{\xi} = 2}$.
}
\label{table:NK3_NR3-Nolepto-Ze1}
\end{center}
\end{table} 

\clearpage

\subsection{Solutions with exotic quarks 
$\hat{\mathbb{K}}$, $\hat{\mathbb{K}}^c$
and $\hat{\mathbb{D}}, \hat{\mathbb{D}}^c$
}
\label{SubAppendix:solution2}

\begin{table}[ht]
\begin{center}
\begin{tabular}{|c|c|c|}
\hline  
 {\bf Scenario 5}   &  Solution 1   &  Solution 2  
\tabularnewline
\hline 
$z({\mathbb{K}})$   &  
$ \ \ \frac{1}{9} $& $- \frac{5}{36} $
\tabularnewline
\hline 
 $z({\mathbb{K}}^{c})$ &     $- \frac{13}{36} $  &  $-  \frac{1}{9} $ 
 \tabularnewline
\hline 
 $z({\mathbb{D}})$ &     $- \frac{7}{18} $    &  $- \frac{11}{18} $ \tabularnewline
\hline 
 $z({\mathbb{D}}^{c})$   &    \ \   $ \frac{5}{36} $   &  \ \   $ \frac{13}{36} $ 
 \tabularnewline
\hline 
\end{tabular}
\caption{
Values of the $U(1)'$ charges for the exotic quarks in 
U$\mu\nu$SSM models
using solution~(\ref{exotic2})
with the hypercharges of Case 1 of Table~\ref{table:ND1-12}.
The $U(1)'$ charges of the rest of the fields are given in  
Table~\ref{table:general} and Scenario 5 of Table~\ref{table:NK1-ND1-general-lepto}.
The minimal number of singlets under the SM gauge group consistent with these charges is
${\bf n_{\nu^c}=2}$. 
}
\label{table:ND1-1-nolepto-Zeneg}
\end{center}
\end{table}

\begin{table}[ht]
\begin{center}
\begin{tabular}{|c|c|c|}
\hline  
 {\bf Scenario 5}   &  Solution 1   &  Solution 2  
\tabularnewline
\hline 
$z({\mathbb{K}})$  &  $ -\frac{7}{9} $    &    $ - \frac{83}{108} $
\tabularnewline
\hline 
 $z({\mathbb{K}}^{c})$  &   \ \   $ \frac{19}{36} $   & \ \   $ \frac{14}{27} $ 
 \tabularnewline
\hline
$z({\mathbb{D}})$ &   $- \frac{5}{18} $     &  $ - \frac{8}{27} $ 
\tabularnewline
\hline
$z({\mathbb{D}}^{c})$  &   \ \   $ \frac{1}{36}$   &  \ \ $ \frac{5}{108} $ 
 \tabularnewline
\hline
\end{tabular}
\caption{
The same as in Table~\ref{table:ND1-1-nolepto-Zeneg},
but for 
the hypercharges of Case 2 of Table~\ref{table:ND1-12}.
}
\label{table:ND1-3-Zeneg}
\end{center}
\end{table}

\begin{table}[ht]
\begin{center}
\begin{tabular}{|c|c|c|}
\hline  
{\bf Scenario 6}   &  Solution 1   &  Solution 2  
\tabularnewline
\hline 
 $z({\mathbb{K}})$  &   $- \frac{1}{9} $   &   $- \frac{5}{36} $ 
 \tabularnewline
\hline 
 $z({\mathbb{K}}^{c})$  &     $- \frac{5}{36} $ 
 &  $-  \frac{1}{9} $
 \tabularnewline
\hline 
 $z({\mathbb{D}})$  &  \ \  $ \frac{7}{18} $    & \  \  $\frac{7}{18} $ 
 \tabularnewline
\hline 
 $z({\mathbb{D}}^{c})$ &    $- \frac{23}{36} $  &  $ - \frac{23}{36} $   \tabularnewline
\hline 
\end{tabular}
\caption{
Values of the $U(1)'$ charges for the exotic quarks in 
U$\mu\nu$SSM models
using solution~(\ref{exotic2}) with 
the hypercharges of Case 1 of Table~\ref{table:ND1-12}.
The $U(1)'$ charges of the rest of the fields are given in 
Table~\ref{table:general} and Scenario 6 of Table~\ref{table:NK1-ND1-general-lepto}.
The minimal number of singlets under the SM gauge group consistent with these charges is
${\bf n_{\nu^c}=2}$.
}
\label{table:ND1-1-nolepto-Ze1}
\end{center}
\end{table}

\begin{table}[ht]
\begin{center}
\begin{tabular}{|c|c|c|}
\hline  
{\bf Scenario 6}   &  Solution 1   &  Solution 2  
\tabularnewline
\hline 
$z({\mathbb{K}})$  &  
 \ \  $\frac{5}{9} $  &   \ \   $  \frac{61}{108} $ 
 \tabularnewline
\hline 
$z({\mathbb{K}}^{c})$  &   $ - \frac{29}{36} $  &    $ - \frac{22}{27} $  
\tabularnewline
\hline 
$z({\mathbb{D}})$ &  \ \    $ \frac{1}{18} $  &   \ \    $  \frac{1}{27} $ 
\tabularnewline
\hline 
 $z({\mathbb{D}}^{c})$  &     $- \frac{11}{36}$  &   $- \frac{31}{108} $
 \tabularnewline
\hline
\end{tabular}
\caption{
The same as in Table~\ref{table:ND1-1-nolepto-Ze1},
but for 
the hypercharges of Case 2 of Table~\ref{table:ND1-12}.
}
\label{table:ND1-3-Ze1}
\end{center}
\end{table} 

\begin{table}[ht]
\begin{center}
\begin{tabular}{|c|c|c|}
\hline
 {\bf Scenario 7}   &  Solution 1   &  Solution 2  
\tabularnewline
\hline 
  $z({\mathbb{K}})$  &   $- \frac{1}{9} $   &  $- \frac{5}{36} $
  \tabularnewline \hline
  $z({\mathbb{K}}^{c})$&  $- \frac{5}{36} $  &
$-  \frac{1}{9} $
\tabularnewline
\hline
  $z({\mathbb{D}})$    & \ \ $ \frac{5}{36} $  & \ \ $\frac{5}{36} $ 
  \tabularnewline \hline
 $z({\mathbb{D}}^{c})$  &   $- \frac{7}{18} $   & $ - \frac{7}{18} $
 \tabularnewline \hline \end{tabular} \caption{
Values of the $U(1)'$ charges for the exotic quarks in 
U$\mu\nu$SSM models
using solution~(\ref{exotic2}) with 
the hypercharges of Case 1 of Table~\ref{table:ND1-12}.
The $U(1)'$ charges of the rest of the fields are given in  
Table~\ref{table:general} and Scenario 7 of Table~\ref{table:NK1-ND1-general-lepto}.
The minimal number of singlets under the SM gauge group consistent with these charges is
${\bf n_{\nu^c}=2}$.
}
\label{table:ND1-1-nolepto-Ze05}
\end{center}
\end{table}

\begin{table}[ht]
\begin{center}
\begin{tabular}{|c|c|c|}
\hline
 {\bf Scenario 7}   &  Solution 1   &  Solution 2  
\tabularnewline
\hline 
   $z({\mathbb{K}})$ & \ \ $ \frac{4}{9} $  & \ \ $ \frac{25}{108} $ 
   \tabularnewline \hline
 $z({\mathbb{K}}^{c})$  & $- \frac{25}{36} $ & $- \frac{13}{27} $ 
 \tabularnewline \hline
   $z({\mathbb{D}})$ & \ \ $ \frac{7}{36} $  &   $- \frac{5}{108} $ 
   \tabularnewline \hline
 $z({\mathbb{D}}^{c})$   &  $- \frac{4}{9} $  &  $- \frac{11}{54} $ 
 \tabularnewline \hline \end{tabular} \caption{
The same as in Table~\ref{table:ND1-1-nolepto-Ze05},
but for the hypercharges of Case 2 of Table~\ref{table:ND1-12}.
}
\label{table:ND1-3-Ze05}
\end{center}
\end{table}

\clearpage


\begin{table}[ht]
\begin{center}
\begin{tabular}{|c|c|c|}
\hline  
 {\bf Scenario 4}   &  Solution 1   &  Solution 2  
\tabularnewline
\hline 
 $z({\mathbb{K}})$  &  
  $ \frac{-23 + 2 \sqrt{7}}{216} $ 
 & $\frac{-23 - 2 \sqrt{7} }{216} $
 \tabularnewline
\hline 
 $z({\mathbb{K}}^{c})$ &     $ \frac{-31 - 2 \sqrt{7}}{216}  $   &   $ \frac{-31 + 2 \sqrt{7} }{216} $  \tabularnewline
\hline 
$z({\mathbb{D}})$  &     $\frac{-13 - 2 \sqrt{7}  }{108} $    &  $\frac{-13 + 2 \sqrt{7}}{108} $ \tabularnewline
\hline 
 $z({\mathbb{D}}^{c})$   &    $ \frac{- 7 +  \sqrt{7} }{54}  $   &   $ \frac{- 7 -  \sqrt{7}}{54} $   \tabularnewline
\hline 
\end{tabular}
\caption{
Values of the $U(1)'$ charges for the exotic quarks in 
U$\mu\nu$SSM models
using solution~(\ref{exotic2}) with the hypercharges of Case 2 of Table~\ref{table:ND1-12}
and irrational $U(1)'$ charges. 
The $U(1)'$ charges of the rest of the fields are given in  
Table~\ref{table:general} and Scenario 4 of Table~\ref{table:NK1-ND1-general-lepto}.
The minimal number of singlets under the SM gauge group consistent with these charges is
${\bf n_{\nu^c}=3}$ and ${\bf n_{\xi} = 2}$.
}
\label{table:ND1-2-irrational}
\end{center}
\end{table}

\begin{table}[ht]
\begin{center}
\begin{tabular}{|c|c|c|}
\hline
 {\bf Scenario 4}   &  Solution 1   &  Solution 2  
\tabularnewline
\hline 
$z({\mathbb{K}})$  &  $ \frac{-23 + 2 \sqrt{439}}{216} $   &  $\frac{-23 - 2 \sqrt{439} }{216}$ \tabularnewline
\hline 
$z({\mathbb{K}}^{c})$ &  $ \frac{-31 - 2 \sqrt{439} }{216}  $ &  $ \frac{-31 + 2 \sqrt{439} }{216} $  \tabularnewline
\hline 
 $z({\mathbb{D}})$  &   $\frac{-13 - 2 \sqrt{439} }{108}$    &  $\frac{-13 + 2 \sqrt{439}}{108} $  \tabularnewline
\hline 
$z({\mathbb{D}}^{c})$ &    $ \frac{- 7 +  \sqrt{439}  }{54}  $  &   $ \frac{- 7 -  \sqrt{439}}{54} $  \tabularnewline
\hline 
\end{tabular}
\caption{
The same as in Table~\ref{table:ND1-2-irrational}, 
but 
with the minimal number of singlets under the SM gauge group
${\bf n_{\nu^c}=3}$, ${\bf n_{S} = 2}$,
${\bf n_{N} = 1}$ and ${\bf n_{\xi} = 2}$.
}
\label{table:ND1-3-irrational}
\end{center}
\end{table} 

\clearpage

\bibliographystyle{utphys}
\bibliography{munussmbib-completo_v4}

\begin{thebibliography}{10}
\bibitem{CMS:2020qrs}
{\bfseries CMS} Collaboration, ``{Search for a spin-1 heavy resonance that
  decays to a Z boson and Higgs boson in the semileptonic final states with
  Run-2 data},''.
\end{thebibliography}
\end{document}